\documentclass[usegraphicx,usenatbib]{mn2e}

\usepackage{amssymb,bm,color,charter}

\newcommand{\e}[1]{\times 10^{#1}}                        % x 10^...
\renewcommand{\div}{\nabla\cdot}                          % divergence
\newcommand{\fig}[1]{Fig. \ref{#1}}

\pdfoutput=1

\font\tenbg=cmmib10 at 10pt
\def \rvecmu{{\hbox{\tenbg\char'026}}}

\title[Launching of Conical Winds]
{Launching of Conical Winds and Axial Jets from the Disk-Magnetosphere Boundary:
Axisymmetric and 3D Simulations}

\author[M. M. Romanova et al.]
{M. M. Romanova,$^1$\thanks{E-mail: romanova@astro.cornell.edu}, G.
V. Ustyugova$^2$\thanks{E-mail: ustyugg@rambler.ru}, A. V.
Koldoba$^3$\thanks{E-mail: koldoba@rambler.ru},  R. V. E.
Lovelace$^{1,4}$\thanks{E-mail:lovelace@astro.cornell.edu}\\
$^1$ Department of Astronomy, Cornell University, Ithaca, NY 14853, USA\\
$^2$ Keldysh Institute of Applied Mathematics, Russian Academy of
Sciences, Moscow, Russia \\
$^3$ Institute for Mathematical Modeling, Russian Academy of
Sciences, Moscow, Russia\\
$^4$ Departments of Applied and Eng. Phys. and Astronomy, Cornell University,
Ithaca, NY 14853}

\begin{document}
\maketitle \label{firstpage}

\begin{abstract}
\noindent We investigate the launching of outflows from the disk-magnetosphere boundary of slowly and rapidly rotating magnetized stars using axisymmetric
and exploratory 3D magnetohydrodynamic (MHD) simulations.
  We find long-lasting outflows in both cases.
(1)   In the case of {\it slowly rotating stars}, a new type of outflow, {\it a conical wind},  is found and studied in simulations.
  The conical winds appear in cases where the magnetic flux of the star is bunched up by the disk
  into an X-type configuration.
   The winds have
the shape of a thin conical shell with a half-opening angle
$\theta \sim 30^\circ-40^\circ$. About $10-30\%$ of the disk
matter  flows from the inner disk into the conical winds. The
conical winds may be responsible for episodic  as well as
long-lasting outflows in different types of stars.
 (2) In the case of {\it rapidly rotating stars} (the ``propeller regime''),
 a two-component outflow is observed.
   One component is similar to the conical winds. A significant fraction of
   the disk matter may be
   ejected into the winds. A second component is a high-velocity, low-density
magnetically dominated {\it axial jet} where matter flows along the opened polar
field lines of the star.
  The jet has a mass flux about $10\%$ that of
the conical wind, but its energy flux (dominantly
magnetic) can be larger than the energy flux of the conical wind.
  The jet's angular momentum flux (also dominantly
magnetic) causes the star to spin-down rapidly.
Propeller-driven outflows may be responsible for the jets in protostars and for their rapid
spin-down. The jet is collimated by the magnetic force while the conical
winds are only weakly collimated in the simulation region.
Exploratory 3D simulations show that conical winds are axisymmetric about the rotational axis
(of the star and the disk), even when the dipole field of the star is significantly misaligned.

\end{abstract}


\begin{figure*}
\centering
\includegraphics[width=7in]{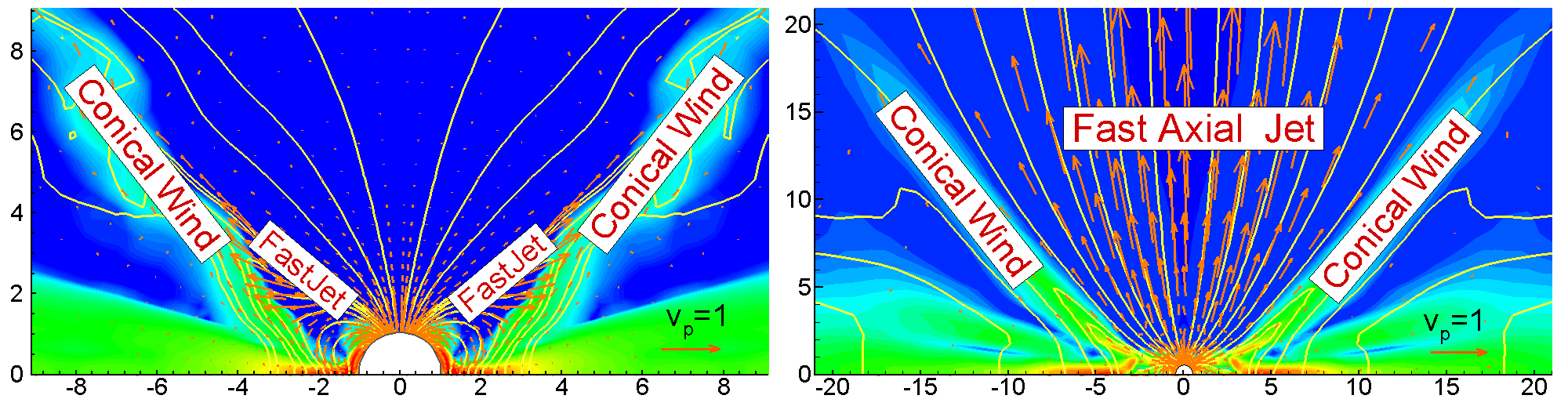}
\caption{Two-component outflows observed in slowly (left) and rapidly (right) rotating magnetized
stars for the reference runs described in this paper. The background shows
the poloidal matter flux $F_m=\rho v_p$, the arrows are the  poloidal velocity vectors, and the
lines are sample magnetic field lines. The labels point to the main outflow components.}
\label{p-x-label}
\end{figure*}

\section{Introduction}
Outflows or jets are observed from many disk accreting objects ranging from young stars to systems with white dwarfs, neutron stars, and black holes (e.g., Livio 1997).

%In addition  to steady outflows there are episodic outbursts
%associated with periods of enhanced accretion
A large body of observations exists for outflows
from young stars at different stages of their evolution, ranging from protostars, where powerful
collimated outflows are observed, to classical T Tauri stars (CTTSs), where the
outflows are weaker and often  less collimated (see review by Ray et al. 2007).
Correlation between the disk and jet power had been found in many CTTSs (e.g., Cabrit et al. 1990;
 Hartigan, Edwards \& Gandhour 1995).
A significant number of CTTSs show signs of outflows in spectral lines, in particular in
He I where two distinct components of outflows had been found (Edwards et al. 2003, 2006; Kwan, Edwards, \& Fischer 2007).
 Outflows are also observed from accreting compact stars such as accreting white dwarfs in
symbiotic binaries (e.g., Sokoloski \& Kenyon 2003),
or from the vicinity of neutron stars, such as from  Circinus X-1 (Heinz et al. 2007).

Different theoretical models have been proposed to explain the outflows from protostars and CTTSs
(see review by Ferreira, Dougados, \& Cabrit 2006).
The models include those where the outflow originates from a radially distributed disk
wind (K\"onigl \& Pudritz 2000;
Casse \& Keppens 2004; Ferreira et al. 2006) or  from the innermost region of the accretion disk (Lovelace, Berk \& Contopoulos 1991).
Further, there is the X-wind  model (Shu et al. 1994; 2007; Najita \& Shu 1994; Cai et al. 2008)
where most of
the outflow originates from the disk-magnetosphere boundary.  The maximum velocities in the outflows are usually of the order of the Keplerian velocity of the inner region of the disk (or higher).  This favors the models where the outflows originate from the inner disk region,
or from the disk-magnetosphere boundary (if the star has a dynamically important magnetic field).

Outflows from the disk-magnetosphere boundary were investigated in early simulations by
Hayashi, Shibata \& Matsumoto (1996) and  Miller \& Stone (1997). A one-time episode
of outflows from the inner disk and inflation of the innermost field lines connecting the star and the disk were observed
for a few dynamical time-scales.
Somewhat longer simulation runs were performed  by Goodson et al. (1997, 1999), Hirose et al. (1997),
Matt et al. (2002) and K\"uker, Henning \& R\"udiger (2003) where several episodes of field inflation and outflows were observed.
These simulations hinted at a possible long-term nature for the outflows.
 However,  the simulations were not sufficiently long to establish
the behavior of the outflows.
  Much longer simulation runs were obtained by treating the disk as a
  {\it  boundary condition} (e.g. Fendt \& Elsner 1999, 2000; Matsakos et al. 2008; Fendt 2009;  see also von Rekowski \& Brandenburg 2004; Yelenina, Ustyugova \& Koldoba 2006).
    These simulations help understand, for example, the roles of the disk wind and stellar wind components
in the outflow and collimation.
 However, for understanding the {\it launching mechanisms} it is important to have a {\it realistic},
low-temperature disk and to solve the full MHD equations in all of the disk and coronal space.

The goal of this work is to obtain long-lasting (robust) outflows from a realistic
low-temperature disk (not a boundary
condition) into a high-temperature, low-density corona.
    We obtained such outflows
in two main cases: (1) when the star rotates slowly but the field lines are bunched
up into an X-type configuration,  and (2) when the star rotates rapidly, in the propeller regime
(e.g., Illarionov \& Sunyaev 1975;
Alpar \& Shaham 1985; Lovelace, Romanova \& Bisnovatyi-Kogan 1999) and the condition for bunching is also satisfied.
In both cases, two-component outflows have been observed (see \fig{p-x-label}).
One component originates at the inner edge of the disk and has a narrow-shell conical shape,
and therefore we call it a ``conical wind". The other component is a magnetically
(or centrifugally) driven high-velocity
low-density wind which flows along stellar field lines. We call it a ``jet". The jet may be very powerful in the propeller regime.
  Below we discuss both regimes in detail (see \S 3 -\S 4)  after description of the numerical approach (see \S 2). In \S 5 we discuss different properties of outflows. In \S 6 we present exploratory 3D simulations of conical winds, and in \S 7 we compare conical winds and propeller
outflows with the $X$-wind model. In \S 8 we  apply the model to different types of
stars, and in \S 9 we present our conclusions.
{\it Appendixes A} and {\it B} clarify different aspects of the numerical model.
{\it Appendix C} summarizes results of different runs for a variety of parameters.

\begin{figure*}
\centering
\includegraphics[width=6in]{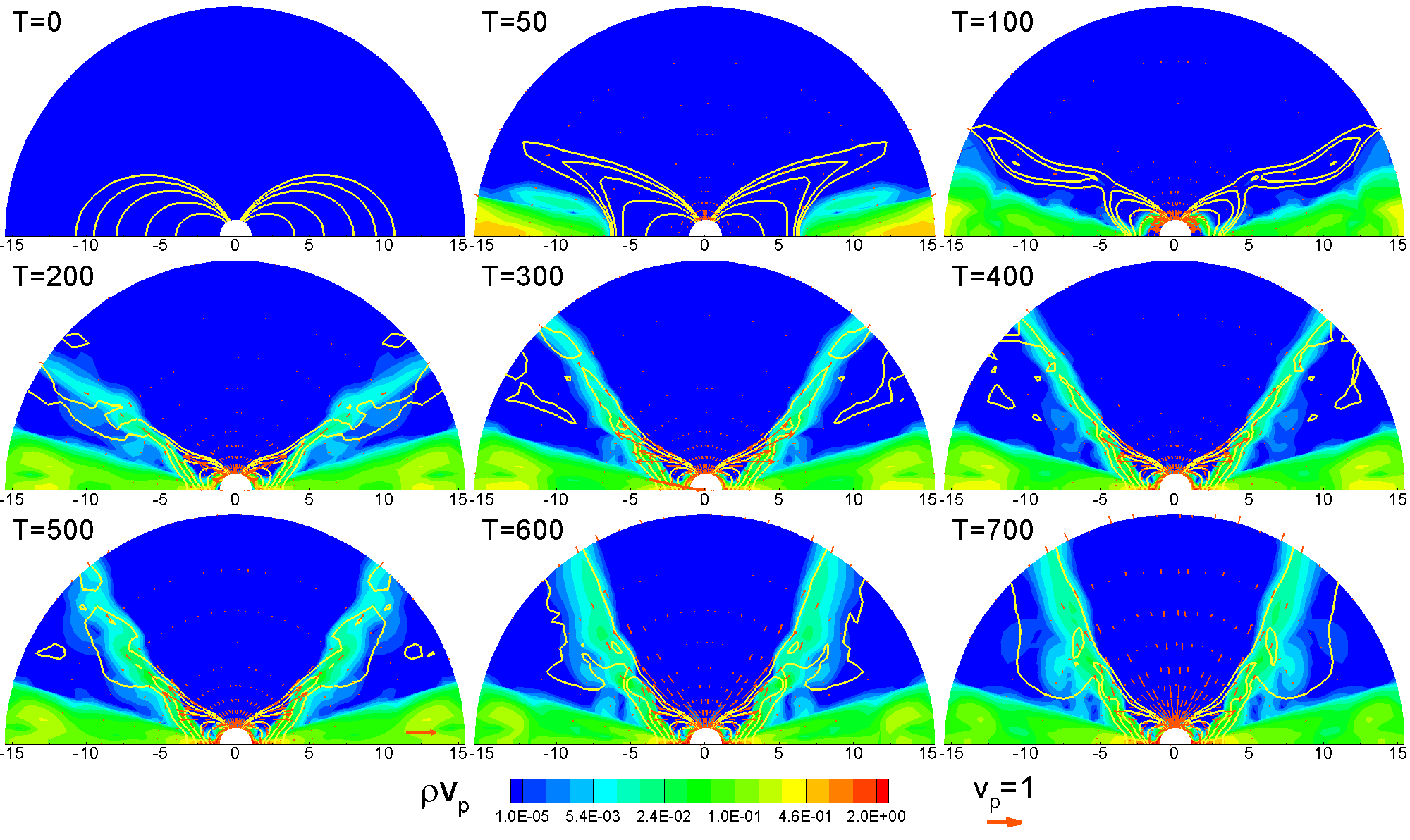}
\caption{Conical winds at different  times $T$.
The background shows
the poloidal matter flux $\rho v_p$ (with the scale below the plots), the arrows are the  poloidal velocity vectors,
and the lines are sample magnetic field lines (same set in all frames). Time $T$ is measured in Keplerian rotation periods at $r=1$.
For example, for CTTS $T_0=1.04$ days (see Table 1) and time $T=700$ corresponds to 2 years. The sample vector $v_p=1$ corresponds to $v_0=195$ km/s.}
\label{sym-x-9}
\end{figure*}

\begin{figure*}
\centering
\includegraphics[width=5in]{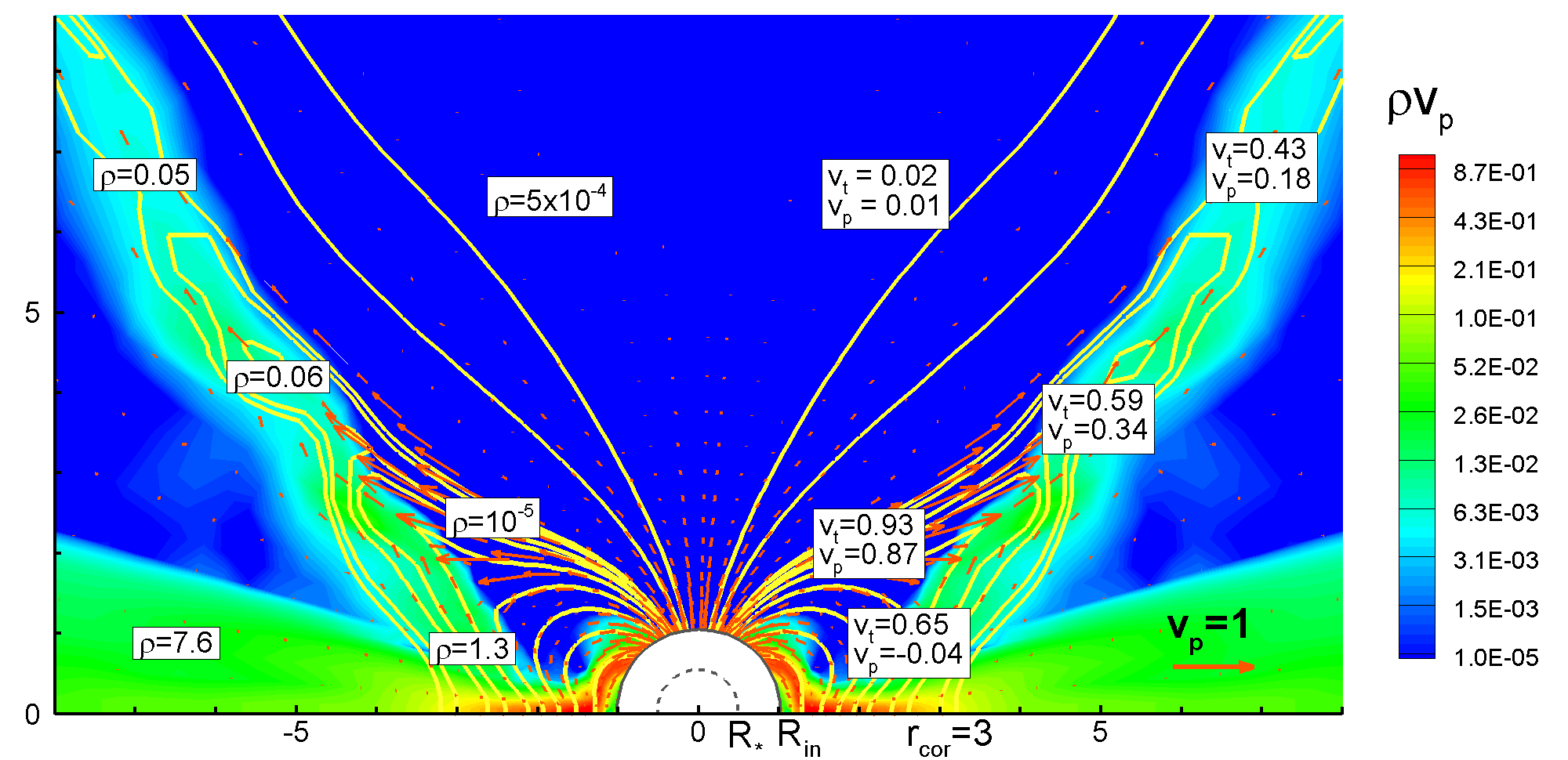}
\caption{The poloidal matter flux $\rho v_p$ (with the scale on the right-hand side), sample magnetic field lines, and
velocity vectors in the conical wind at time $T=500$.   Sample numbers are given for the dimensionless poloidal $v_p$ and total $v_t$ velocities, and for the density $\rho$. To obtain the  dimensional values one needs to multiply these
numbers by the  reference values given in the Table 1. For example, for application to CTTS: $v_p=1$ corresponds to $v_0=195$ km/s, $\rho=1$ corresponds to $\rho_0=4.1\times 10^{-13}$ g cm$^{-3}$, and the distance $r=1$ corresponds to $R_0=0.02$ AU.}
\label{con-numb}
\end{figure*}

\section{Numerical Model}

We simulate the outflows resulting from disk-magnetosphere interaction
using the equations of axisymmetric MHD  described below.  Axisymmetric simulations of the outflows
are similar to those performed earlier for the propeller regime
(e.g., U06), but differ in initial and boundary conditions. Below we give an outline
of the numerical model.

\subsection{Basic Equations}

Outside of the disk the flow is described by the
equations of ideal MHD.
     Inside the disk the flow is described by
the equations of viscous, resistive MHD.
In an inertial reference frame the equations are:
\begin{equation}\label{eq1}
\displaystyle{ \frac{\partial \rho}{\partial t} + {\bf \nabla}\cdot
\left( \rho
{\bf v} \right)} = 0~,
\end{equation}
\begin{equation}\label{eq2}
{\frac{\partial (\rho {\bf v})}{\partial t} + {\bf
\nabla}\cdot
{\cal T} } = \rho ~{\bf g}~,
\end{equation}
\begin{equation}\label{eq3}
{\frac{\partial {\bf B}}{\partial t} - {\bf
\nabla}\times ({\bf v} \times {\bf B}) + {\bf \nabla} \times\left(
\eta_t {\bf \nabla}\times {\bf B} \right)} = 0~,
\end{equation}
\begin{equation}\label{eq4}
{\frac{\partial (\rho S)}{\partial t} + {\bf
\nabla}\cdot ( \rho S {\bf v} )} =  Q~.\
\end{equation}
 Here, $\rho$ is the density and $S$ is the specific entropy; $\bf
v$ is the flow velocity; $\bf B$ is the magnetic field; $\cal{T}$ is the
momentum flux-density tensor; $Q$ is
the rate of change of entropy per unit volume;   and ${\bf g} = - (GM /r^{2})\hat{{\bf r}}$
is the gravitational acceleration due to the star, which has mass $M$. The total
mass of the disk is negligible compared to $M$.
     The  plasma is considered to be an
ideal gas with adiabatic index $\gamma =5/3$, and $S=\ln(p/
\rho^{\gamma})$. We use spherical coordinates $(r, \theta, \phi)$ with  $\theta$ measured
from the symmetry axis.  The condition for axisymmetry is $\partial /\partial
\phi =0$.
 The  equations in spherical coordinates are given in
U06.

The stress tensor $\cal T$ and the treatment of viscosity and diffusivity
are described in {\it Appendix A}.
     Briefly,  both the viscosity
and the magnetic diffusivity of the disk plasma are considered
to be due to turbulent fluctuations of the velocity and the magnetic field.
        We adopt the standard hypothesis
where the  molecular transport coefficients are  replaced by
 turbulent coefficients.
       To estimate the values of these coefficients, we
use the  $\alpha$-model of Shakura and Sunyaev (1973) where the
coefficient of the turbulent kinematic viscosity $\nu_t =
\alpha_v c_s^{2}/\Omega_K$, where $c_s$ is the isothermal sound
speed and $\Omega_K(r)$ is the Keplerian angular velocity.
        Similarly, the coefficient of the turbulent magnetic
diffusivity $\eta_t=\alpha_d c_s^{2}/\Omega_K$. Here,
      $\alpha_v$ and $\alpha_d$ are dimensionless coefficients which
are treated as parameters of the model.

\begin{figure}
\centering
\includegraphics[width=3.4in]{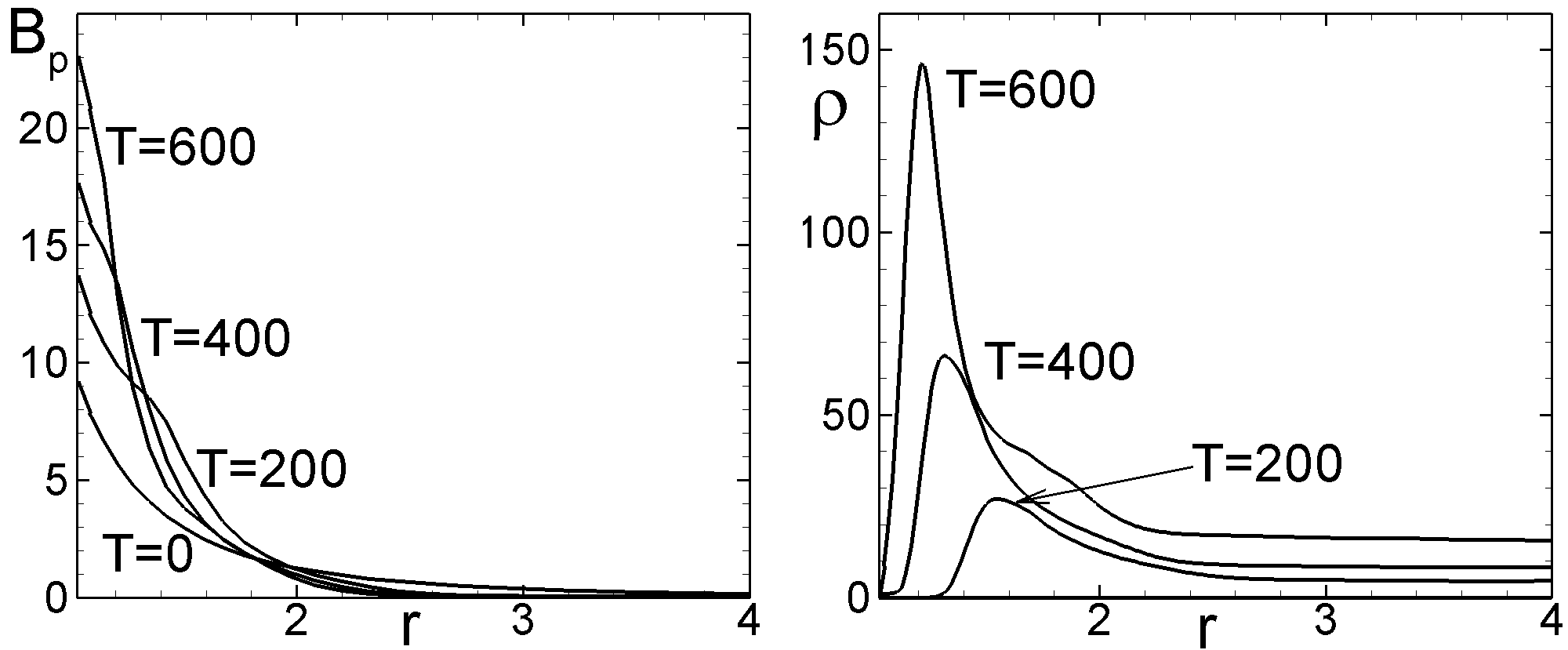}
\caption{{\it Left panel}: equatorial distribution of the poloidal magnetic
field in the inner part of the simulation region at different times $T$.
{\it Right panel}: same as left panel but for density.}
\label{bro-x-2}
\end{figure}

\begin{figure*}
\centering
\includegraphics[width=6in]{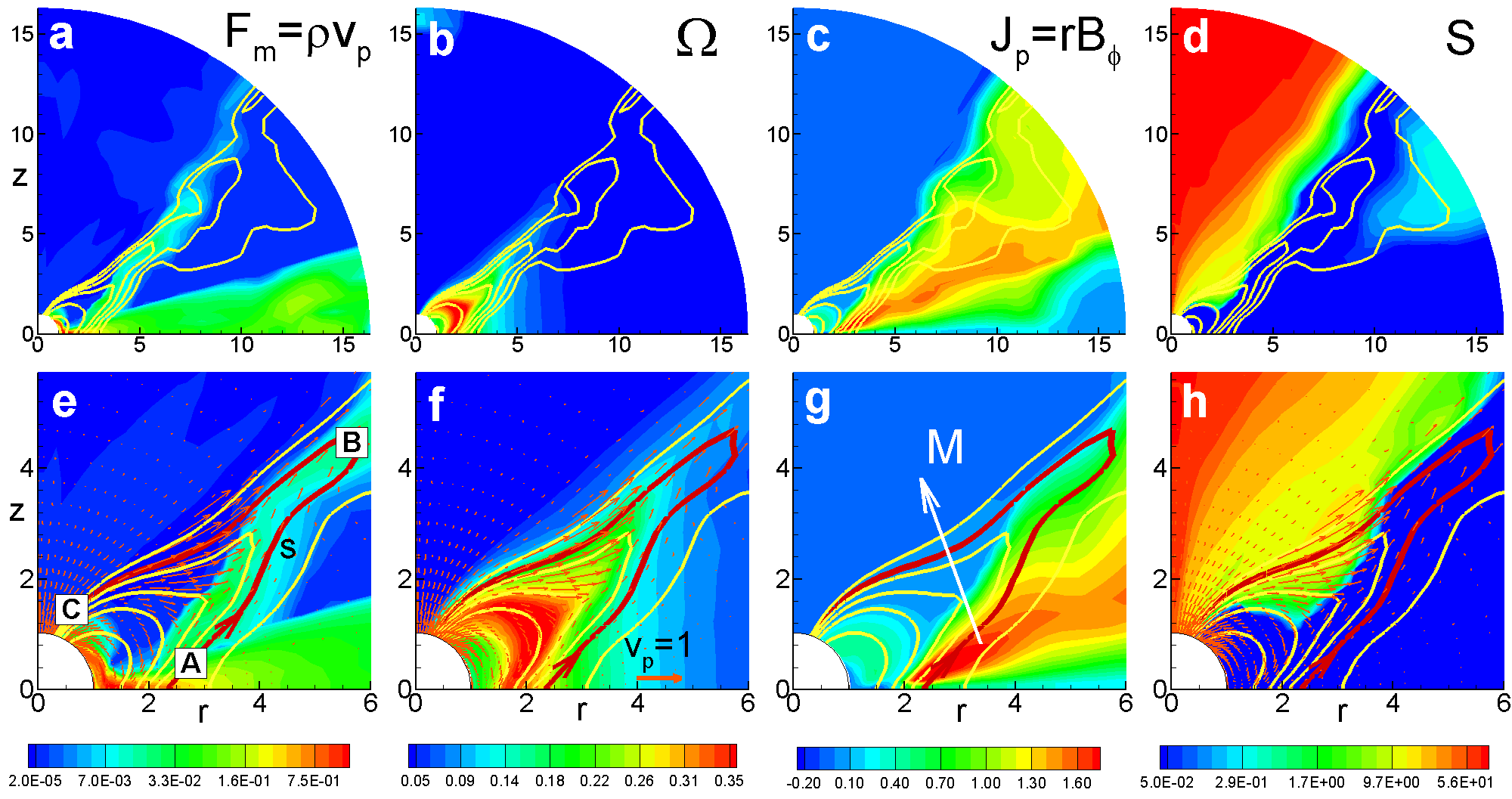}
\caption{The background shows different parameters of a conical wind
at  time $T=500$, in
the full simulation region, {\it panels a-d}, and near the star, {\it panels e-h}.  The
background shows (from left to right): the poloidal matter flux, $F_m=\rho
v_p$, the angular velocity $\Omega$, the poloidal current $J_p=r B_\phi$,
and the entropy $S$. Lines are sample magnetic field lines. Vectors are velocity vectors.
The thick red line and the marks {\bf A, B, C} show one of the field lines and positions along this line which are used for analysis of the forces and velocities
in \fig{for-vel-x-4}. The white vector in panel {\it g} shows schematically the direction of the magnetic force, M.}
\label{8-all}
\end{figure*}

\subsection{Reference Units}

The MHD equations are solved in
dimensionless form so that the results can be readily applied
to different accreting stars (see \S 7).
   We take the reference
mass $M_0$ to be the mass $M$ of the star.
  The reference radius is taken to be twice the radius
of the star, $R_0=2 R_*$.
    The surface
magnetic field $B_*$ is
%of course
different for different types of stars.
  The reference velocity is $v_0=(GM/R_0)^{1/2}$.
The reference time-scale $t_0=R_0/v_0$, and the reference angular velocity $\Omega_0=1/t_0$.
    We measure time
in units of $P_0=2\pi t_0$ (which is the Keplerian rotation period at $r=R_0$).
In the plots we  use the dimensionless time $T=t/P_0$.
    The reference magnetic field is $B_0=B_*(R_*/R_0)^{3}/{\tilde\mu}$,
    where $\tilde\mu$ is the dimensionless magnetic moment.
The reference density is taken to
be  $\rho_0 = B_0^{2}/v_0^{2}$.
The reference pressure is $p_0=B_0^{2}$.
The reference temperature is
$T_0=p_0/{\cal R} \rho_0 = v_0^{2}/{\cal R}$, where ${\cal R}$ is the gas constant.
   The reference accretion rate is $\dot M_0
= \rho_0 v_0 R_0^{2}$.
   The reference energy flux is $\dot
E_0=\dot M_0 v_0^{2}$.
The reference angular momentum flux is $\dot L_0=\dot M_0 v_0
R_0$.

The reference units are defined in such a way that the dimensionless MHD equations
have the same form as the dimensional ones, equations (1)-(4) (for such dimensionalization
we put $GM=1$ and ${\cal R}=1$).
Table 1 shows examples of reference variables for
different stars.
We solve the MHD equations (1)-(4) using normalized variables:
$\tilde\rho=\rho/\rho_0$, $\tilde v=v/v_0$, $\tilde B= B/B_0$, etc.
Most of the plots show the normalized variables (with the tildes
implicit). To obtain dimensional values one needs to multiply
values from the plots by the corresponding  reference values from
Table 1.

\begin{table*}
\begin{tabular}{l@{\extracolsep{0.2em}}l@{}lllll}

\hline
&                                              & Protostars      & CTTSs        & Brown dwarfs   & White dwarfs        & Neutron stars    \\
\hline

\multicolumn{2}{l}{$M(M_\odot)$}                  & 0.8            & 0.8            & 0.056             & 1                   & 1.4       \\
\multicolumn{2}{l}{$R_*$}                         & $2R_\odot$     & $2R_\odot$     & $0.1R_\odot$       & 5000 km             & 10 km     \\
\multicolumn{2}{l}{$R_0$ (cm)}                    & $2.8\e{11}$    & $2.8\e{11}$    & $1.4\e{10}$      & $1.0\e9$            & $2\e6$    \\
\multicolumn{2}{l}{$v_0$ (cm s$^{-1}$)}           & $1.95\e7$      & $1.95\e7$      & $1.6\e7$          & $3.6\e8$   & $9.7\e{9}$ \\
\multicolumn{2}{l}{$P_*$}                         & $1.04$ days    & $5.6$ days     & $0.13$ days        & $89$ s              & $6.7$ ms     \\
\multicolumn{2}{l}{$P_0$}                         & $1.04$ days    & $1.04$ days    & $0.05$ days       & $17.2$ s            & $1.3$ ms   \\
\multicolumn{2}{l}{$B_*$ (G)}                      & $3.0\e3$      & $10^{3}$      & $2.0\e{3}$            & $10^{6}$            & $10^{9}$    \\
\multicolumn{2}{l}{$B_0$ (G)}                     & 37.5           & 12.5          & 25.0                & $1.2\e4$            & $1.2\e{7}$  \\
\multicolumn{2}{l}{$\rho_0$ (g cm$^{-3}$)}         & $3.7\e{-12}$  & $4.1\e{-13}$  & $1.4\e{-12}$        & $1.2\e{-9}$         & $1.7\e{-6}$  \\
\multicolumn{2}{l}{$n_0$ (1/cm$^{-3}$)}            & $2.2\e{12}$   & $2.4\e{11}$   & $8.5\e{11}$         & $7.0\e{14}$         & $1.0\e{18}$          \\
\multicolumn{2}{l}{$\dot M_0$($M_\odot$yr$^{-1}$)}  & $1.8\e{-7}$  & $2.0\e{-8}$   & $1.8\e{-10}$      & $1.3\e{-8}$         & $2.0\e{-9}$  \\
\multicolumn{2}{l}{$\dot E_0$ (erg s$^{-1}$)}       & $2.1\e{33}$  & $2.4\e{32}$   & $2.5\e{30}$      & $5.7\e{34}$         & $6.0\e{36}$  \\
\multicolumn{2}{l}{$\dot L_0$ (erg s$^{-1}$)}       & $3.1\e{37}$  & $3.4\e{36}$   & $1.7\e{33}$      & $1.6\e{35}$         & $1.2\e{33}$  \\
\multicolumn{2}{l}{$T_d$ (K)}                       & $2293$       & $4586$        & $5274$            & $1.6\e{6}$          & $1.1\e{9}$  \\
\multicolumn{2}{l}{$T_c$ (K)}                       & $2.3\e{6}$   & $4.6\e{6}$    & $5.3\e{6}$        & $8.0\e{8}$          & $5.6\e{11}$  \\
\hline
\end{tabular}
\caption{Reference values for different types of stars. We choose the mass $M$, radius $R_*$,
equatorial magnetic field $B_*$ and  the period $P_*$ of the star and derive the other reference values (see \S 2.2).
To apply the simulation results to a particular star one needs to multiply the dimensionless values
from the plots by the reference values from this table.} \label{tab:refval}
\end{table*}

\subsection{Initial and Boundary Conditions}

We assume that the poloidal  magnetic field
of the star is an aligned dipole field
$
{\bf B}=[3(\rvecmu \cdot {\bf r}) {\bf r} -\rvecmu { r}^{2}]/ r^{5},~
$
where $\rvecmu$ is the star's magnetic moment.
The initial density and temperature distributions are different in cases
of conical winds and propeller-driven winds.

\subsubsection{Conical Winds}

{\it Initial Conditions}. At time $T=0$ the  simulation region is filled with a
low-density, high-temperature, isothermal  plasma  referred to as the corona.
Initially it is non-rotating. Thus
the density and pressure distributions in the corona are:
$$
\rho=\rho_c\exp[GM/({\cal R} T_c r)],
~~ ~p=p_c\exp[GM/({\cal R} T_c r)]~,
$$
where $T_c$ is the corona temperature, $\rho_c$ is the
coronal density at the external boundary, $p_c=\rho_c{\cal R} T_c$.

We divide the external boundary $R_{out}$ into a disk region, $\theta_d < \theta<\pi/2$,
 and a corona region, $0<\theta < \theta_d$, with $\theta_d\approx 65^\circ$.
 Initially, there is no disk in the simulation region.
When the simulations start, we permit high-density low-temperature matter
to enter the simulation region through the disk boundary region, $\theta >\theta_d$, with a fixed density $\rho=\rho_d$.
Matter continues to flow inward due to viscosity (see {\it Appendix A}).
We increase the spin of the star gradually from a
 small value corresponding to  $r_{cor}=R_{out}$ (where
$r_{cor} = (GM/\Omega^{2})^{1/3}$) up to a final value
$\Omega_*$.
Information about the stellar rotation propagates rapidly
(at the Alfv\'en speed) into the low-density corona.

We did simulations for a variety of parameters.
However, we take one case with typical parameters to be our {\it reference case}
and show the results for this case.
 In the reference case, the dipole moment of the star $\mu=10$;
the density in the corona  $\rho_c=0.001$,  the density in the disk $\rho_d=10$;
the corona is hot with temperature $T_c = 1$;
the disk is cold with temperature $T_d=(\rho_c/\rho_d) T_c = 10^{-4}$.
The angular velocity of the star corresponds to a corotation radius $r_{cor}=3$,
$\Omega_*=(GM/r_{cor}^3)=0.19$.
The coefficients of viscosity and diffusivity are
$\alpha_v=0.3$ and $\alpha_d=0.1$.
      The dependences of our results on different parameters
are discussed in {\it Appendix C}.

{\it The boundary conditions at the inner boundary $r=R_{in}$} are the following:
The frozen-in condition is applied to the poloidal component $\bf{B}_p$  of
the field, such that $B_r$  is fixed while $B_\theta$ and $B_\phi$ obey
 ``free" boundary conditions,  $\partial B_\theta/\partial r=0$
 and $\partial B_\phi/\partial r=0$.
The density, pressure, and entropy  also have free boundary conditions,
$\partial (...)/\partial r=0$.
The velocity components are calculated using  free boundary conditions.
    Then, the velocity vector is adjusted to be parallel to the magnetic field vector
    in the coordinate system rotating with a star.     Matter always flows inward at
the star's surface.
   Outflow to a stellar wind is not considered in this work.

{\it The boundary conditions at the external boundary $r=R_{out}$}
in the {\it coronal region}  $0<\theta <\theta_d$ are
 free for all hydrodynamic variables.
    However, we prevent
 matter from flowing into the simulation region from this part of the
boundary. We solve the transport equation for the flux function $\Psi$
so that the magnetic flux flows out of the region together with matter.
   If the matter has a tendency to flow back in, then we fix $\Psi$.
{\it In the disk region}, $ \theta_d < \theta <\pi/2 $, we fix the density at
$\rho=\rho_d$, and establish a slightly sub-Keplerian velocity, $\Omega_d=\kappa\Omega(r_d)$,
where $\kappa = 1 - 0.003$ so that matter flows into the simulation region through the boundary.
The inflowing matter has a fixed magnetic flux which is very small because $R_{out} \gg R_{in}$.

The boundary conditions on
the {\it equatorial plane} and on the {\it rotation axis} are symmetric and antisymmetric.

\begin{figure*}
\centering
\includegraphics[width=4.5in]{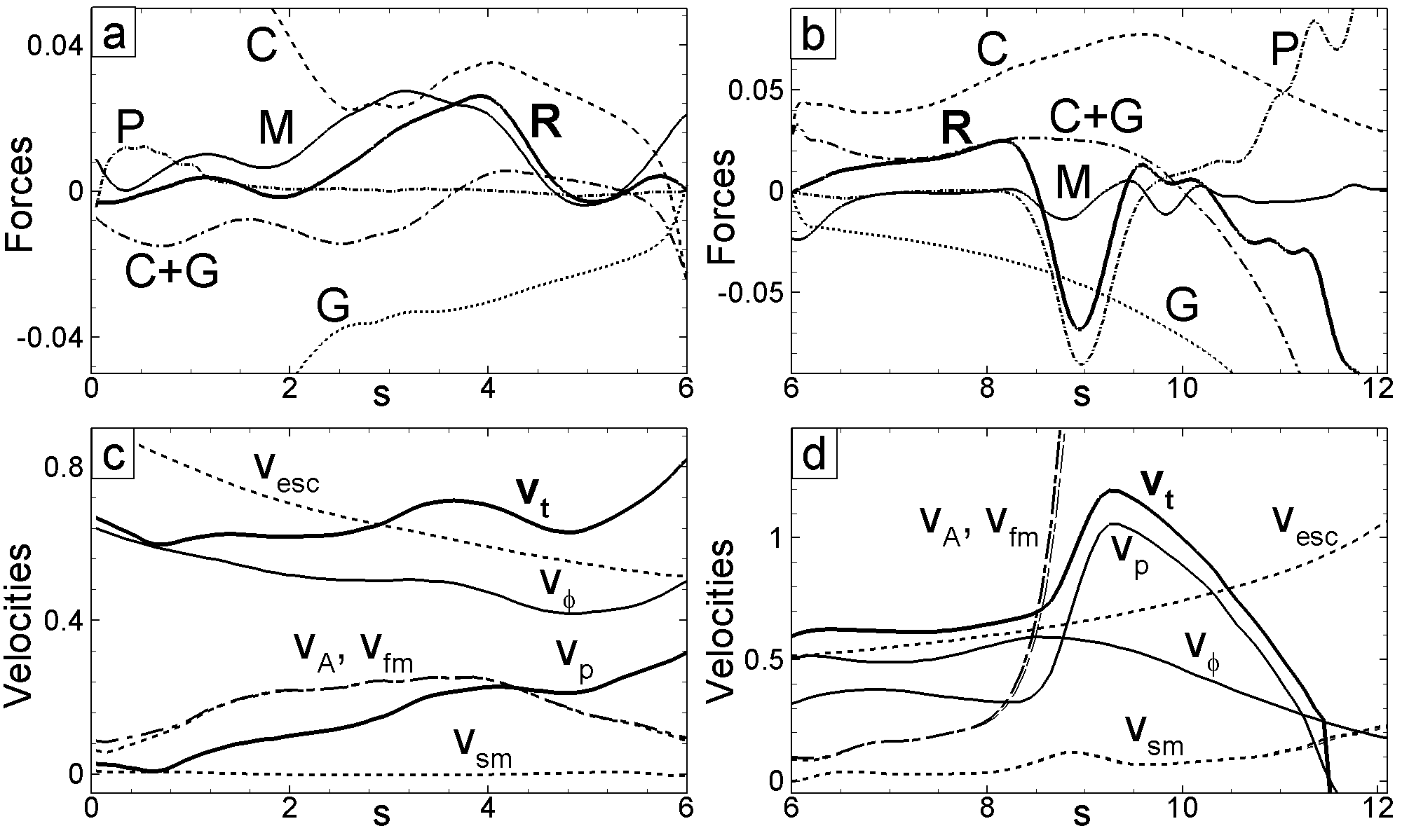}
\caption{{\it Panel a} shows the projection of forces onto part {\bf AB}
of the bold field line shown in
\fig{8-all}{\it e}.
   The labels are:
$G-$gravitational, $C-$centrifugal, $M-$magnetic, $P-$pressure gradient, and
$R-$sum of all forces. {\it Panel b} shows the projection of forces onto part
{\bf BC} of the field line. {\it Panel c} shows velocities along part {\bf AB}
 of the field line: $v_p-$poloidal velocity, $v_\phi-$azimuthal velocity, $v_t-$total velocity,
$v_A-$Alfv\'en speed, $v_{sm}-$slow-magnetosonic speed, $v_{fm}-$fast-
magnetosonic speed, and $v_{esc}-$escape velocity.  {\it Panel d} shows velocities along part {\bf BC}
of the field line.}\label{for-vel-x-4}
\end{figure*}

\begin{figure*}
\centering
\includegraphics[width=4.5in]{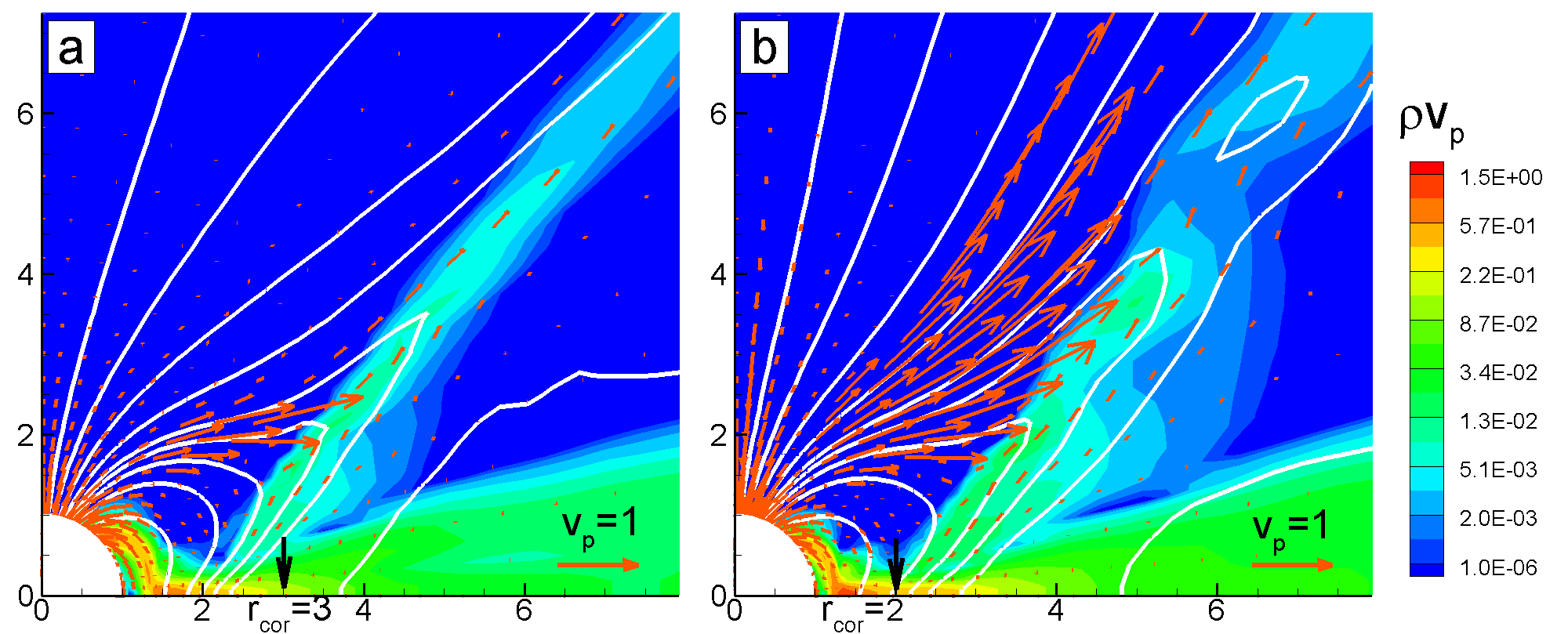}
\caption{{\it Panel a:} Two components of outflow in the reference case ($r_{cor}=3$).
{\it Panel b:} Same but for a more rapidly rotating star ($r_{cor}=2$). Both cases are shown at time $T=250$.}\label{rotation}
\end{figure*}

\subsubsection{\it Propeller Regime}

    The initial and boundary conditions for the propeller regime are the same as those
used in R05 and U06.   Here, we summarize these conditions.

{\it Initial Conditions}. We place both the disk and the corona into the simulation region.
We assume that the initial flow is barotropic with $\rho=\rho(p)$, and that
there is no pressure jump at the boundary between the disk and corona.
  Then the initial density distribution (in dimensionless units) is the following:
$$
\rho (p)= \left \{
      \begin{array}{lcl}
        p/{\cal R} T_d~,~~~ p>p_b~~~~ {\rm and}~~~~ r \sin \theta \geq
r_b~, \\[0.3cm]
        p/{\cal R} T_c~,~~~~ p<p_b ~~~~{\rm or}~~~~~~r \sin \theta \leq r_b~,
      \end{array}
\right.
$$
\noindent where $p_b$ is the pressure on the surface which separates the cold matter of the
disk from the hot matter of the corona. On this surface the density jumps
from $p_b/T_d$ to $p_b/T_c$. Here $r_b$ is the inner disk radius.
      Because the density distribution is barotropic,
the angular velocity is constant on coaxial cylindrical surfaces
about the $z-$axis.
        Consequently,
the pressure distribution may be determined from the Bernoulli equation,
$$
F(p) + \Phi + \Phi_c =E = {\rm const}~.
$$
Here, $\Phi = -GM /|{\bf r}|$  is gravitational potential, $\Phi_c
 = \int_{r \sin \theta}^\infty \Omega^{2} (\xi)\xi
d\xi$ is centrifugal potential, which depends only on the cylindrical
radius $ r\sin \theta$, and
$$
F(p)=\left \{
      \begin{array}{lcl}
        {\cal R} T_d \ln( p/p_b )~,~~~ p>p_b~~~~{\rm and}~~~r \sin \theta >r_b~,
\\[0.3cm]
        {\cal R} T_c \ln (p/p_b) ~,~~~~ p<p_b~~~~{\rm or}~~~~r\sin \theta <r_b~.
      \end{array}
\right.
$$
The angular velocity of the disk is slightly sub-Keplerian,
$\Omega(\theta=\pi/2)=\kappa \Omega_K$ ($\kappa = 1-0.003$), due to which the density and
pressure decrease towards the periphery.
      Inside the cylinder
$r\leq r_b$ the matter rotates rigidly with angular velocity
$\Omega(r_b)=\kappa (GM /r_b^{3})^{1/2}$.
      For a gradual start-up we change the angular velocity of the
star from its initial value $\Omega(r_b) =
5^{-3/2} \approx 0.09$ ($r_b=5$) to a final value of $\Omega_*=1$ over the course of three
Keplerian rotation periods at $r=1$.

For the propeller regime we use a slightly different set of parameters compared
with the conical wind case (in order to be consistent with our earlier simulations in R05, U06).
Below we describe the similarities and differences: the dipole moment of the star is the same, $\mu=10$.
The
angular velocity of the star in the propeller regime is larger, $\Omega_*=1$.
The initial density in the disk (at the
inner edge) is $\rho_d=1$, which is smaller than the external
density ($\rho_d=10$) in conical winds.  The initial temperatures are a
factor of two smaller than in conical winds, $T_c=(p/{\cal R}\rho)_c=0.5$, $T_d=0.0005$.

{\it Boundary conditions for the propeller regime} are similar to those
for conical winds with the following differences. At the external boundary (disk region)
 we take free conditions for all variables.
  There is no condition of fixed density at the disk part
  of the boundary.

The system of MHD equations (\ref{eq1}-\ref{eq4}) was integrated
numerically using the Godunov-type numerical scheme (see {\it
Appendix B}).
  The simulations were done
%spherical coordinates (see U06 for the equations)
in the region $R_{in} \leq r \leq R_{out}$, $0\leq \theta
\leq \pi/2$. The grid is uniform in the $\theta$-direction.
    The size steps in the radial direction were chosen
so that the poloidal-plane cells were
curvilinear rectangles with approximately
equal sides.
      A typical region for investigation of conical winds was
$1 \leq r \leq 16$,
with grid resolution $N_r\times N_\theta = 51\times 31$ cells. A typical region for
investigation of the propeller regime
was $1\leq r\leq 48$ with grid resolution $N_r\times N_\theta = 85\times31$ cells.
Test simulations at angular grids $N_\theta=51$ and $N_\theta=71$ were also performed.
Each simulation run at the lowest resolution takes about two months of computing
time on a single processor. We performed $40$ different simulation runs for
different parameters on our  local cluster of $20$ computers for the investigation of conical winds.
3D simulation runs were performed with ``Cubed sphere" parallel code on NASA high-performance facilities.

\begin{figure*}
\centering
\includegraphics[width=6in]{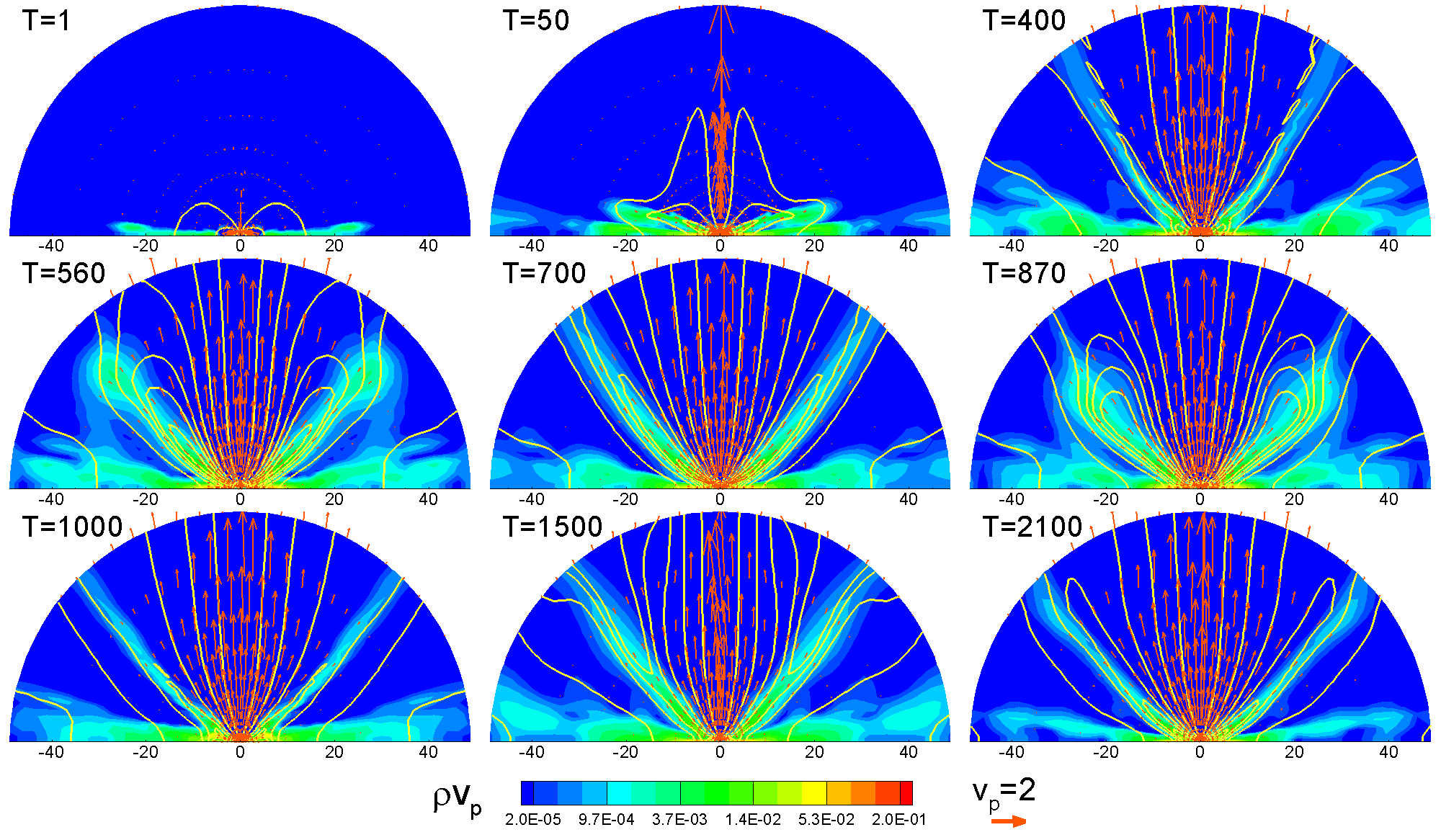}
\caption{Same as in \fig{sym-x-9} but for the propeller regime.
The simulation time $T=2200$ corresponds to 6 years. The sample vector $v_p=2$
corresponds to $v=2\times v_0= 390$ km/s.}\label{sym-p-9}
\end{figure*}

\begin{figure*}
\centering
\includegraphics[width=5in]{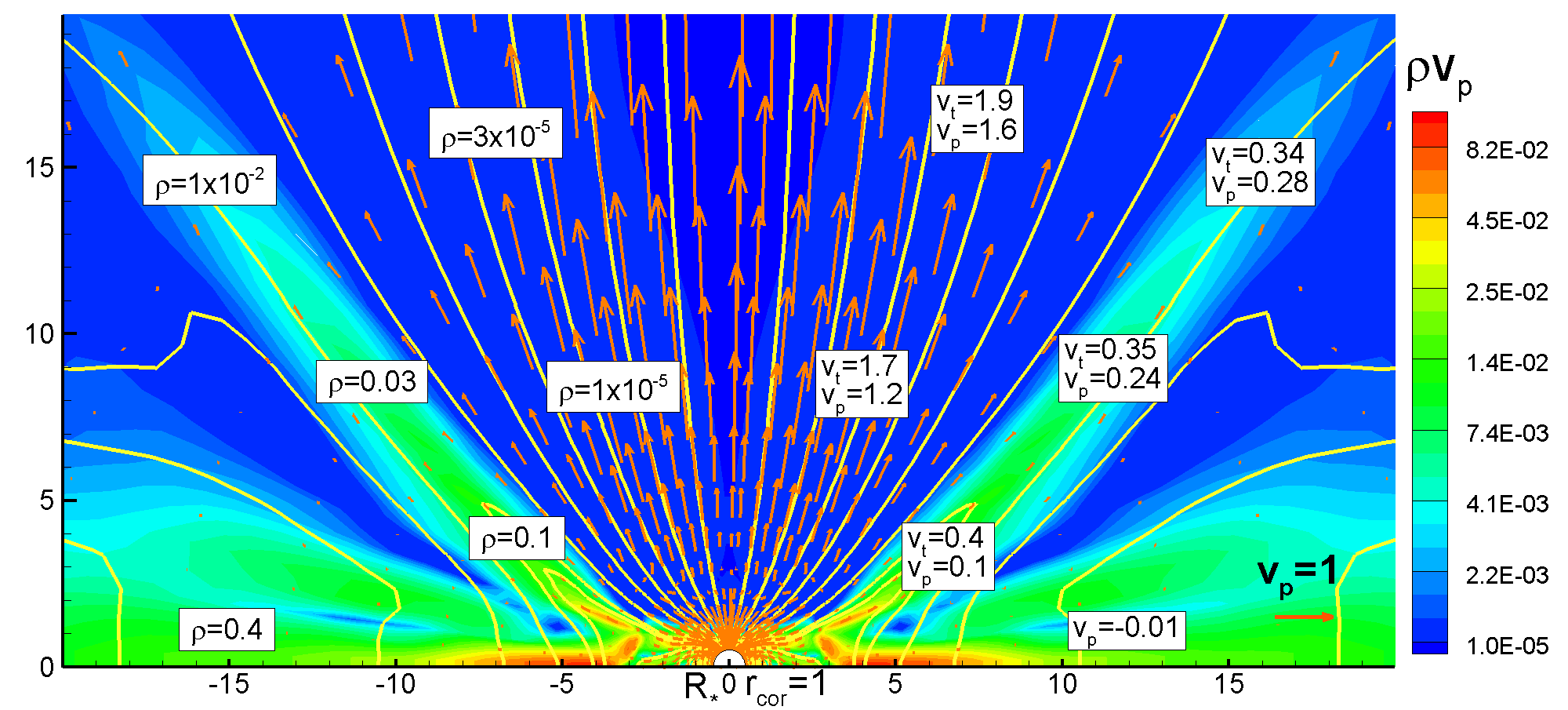}
\caption{Same as in \fig{con-numb} but for the propeller regime at time
$T=1400$.}
\label{prop-numb}
\end{figure*}

\section{Matter flow in conical winds and in the propeller regime}

\subsection{Matter flow, velocities, and forces in conical winds}

      A large number of simulations were done in order
to understand  the origin and nature of conical winds.
  All of the key parameters were varied in order   to ensure that
 there is no special  dependence on any parameter (see {\it Appendix C}).
  We observed that the formation of conical winds
is a common phenomenon for a wide range of parameters.
They are most persistent and strong in cases where the  viscosity and diffusivity coefficients are not very small, $\alpha_v\gtrsim 0.03$, $\alpha_d \gtrsim 0.03$.
Another important condition is  that $\alpha_v \gtrsim\alpha_d$;
that is, the magnetic Prandtl number
of the turbulence, Pr$_m=\alpha_v/\alpha_d \gtrsim 1$.
This condition favors the bunching of the stellar magnetic
field by the accretion flow.

    For a discussion of the
physics of conical winds we focus on
one set of parameters which serves as a reference case.
These parameters are:  $\alpha_v=0.3$ and $\alpha_d=0.1$;
 $\Omega_*=0.19$ ($r_{cor}=3$), $\mu=10$, $\rho_d=10$, and $\rho_c=10^{-3}$, $T_c=1$, $T_d=10^{-4}$.
    The simulations were done in dimensionless form and can be applied to
different stars  (see Table 1).
    However, for illustration we often show dimensional examples for
    CTTSs with parameters taken from Table 1. For example, in application
    to CTTSs, $\Omega_*=0.19$ corresponds to $P_*=5.4$ days and the unit of
    time used in the figures is $P_0=1.04$ days (see Table 1 for other reference values).

 \fig{sym-x-9} shows snapshots of simulations at different times $T$.
   One can see that the cold dense disk matter enters the simulation region
   from the external boundary and  moves inward towards the star on the viscous time-scale.
       The accretion flow bunches up the field lines of the dipole field to
       a relatively small region near the star.
    All field lines shown at $T=0$ are bunched up
 close to the star by $T=100$.
     The inclined configuration of the resulting poloidal field and inflation of
     external field lines create conditions favorable for matter
outflow from the inner disk.
     The outflow starts at $T \sim 120$ and gets stronger later.
     Matter flows from the inner disk into hollow, conical
     shaped winds with half-opening angle $\theta \sim 30^\circ -40^\circ$.
      The conical winds are non-stationary, showing variations
      associated with events of inflation and reconnection of the
      magnetic field lines (see animations at {\it http://www.astro.cornell.edu/$\sim$romanova/conical.htm}).
    The  simulation runs continue for a long time, about $T=740$,
    which is about 2 years for CTTSs.
     The outflows remain strong until the end of the simulation runs.
     It is reasonable to conclude that
these accretion-driven outflows into the conical winds will persist as long as matter is supplied from the disk.

 \fig{con-numb} shows the configuration at $T=500$.
     One can see that the disk matter comes close to the star and accretes
      onto the star through a small dense funnel.
     Some field lines are strongly inflated, and the conical wind flows from the disk along these lines.
       There is also a set of partially inflated field lines
 (a dead zone) where accretion does not occur (e.g.,
Ostriker \& Shu 2005; Spruit \& Taam 1990).
Matter in conical winds rotates with the Keplerian velocity at the base of outflow, $v_\phi\approx v_K$.
  It continues to rotate rapidly in the conical wind
at larger distances from the star.
    The poloidal velocity $v_p$ increases gradually from very small values at the beginning of the outflow, up to $v_p\approx 0.5 v_K$.
    The main contribution to the total velocity $v_t$ comes from the azimuthal component.
There is another, high-velocity component of the low-density matter which flows along the stellar field lines.
   In application to CTTSs the velocity is $ > 200$ km/s.

 \fig{bro-x-2} shows the variation of the density and the poloidal magnetic field
along the equator in the inner part of the simulation region
at different times $T$.
   One can see that the disk matter has approximately constant density at different radii,
   but there is a density peak closer to the star.
     The peak increases with time, but it does not
appreciably  influence the fluxes calculated  at the surface of the star (see \fig{flux-x-4}).
    \fig{bro-x-2} also shows that the poloidal magnetic field of the star is compressed, and the compression increases with time.
    Compression of the star's magnetic
field by the accretion flow is also assumed
 in the X-wind model (e.g., Najita \& Shu 1994).

\begin{figure}
\centering
\includegraphics[width=3.4in]{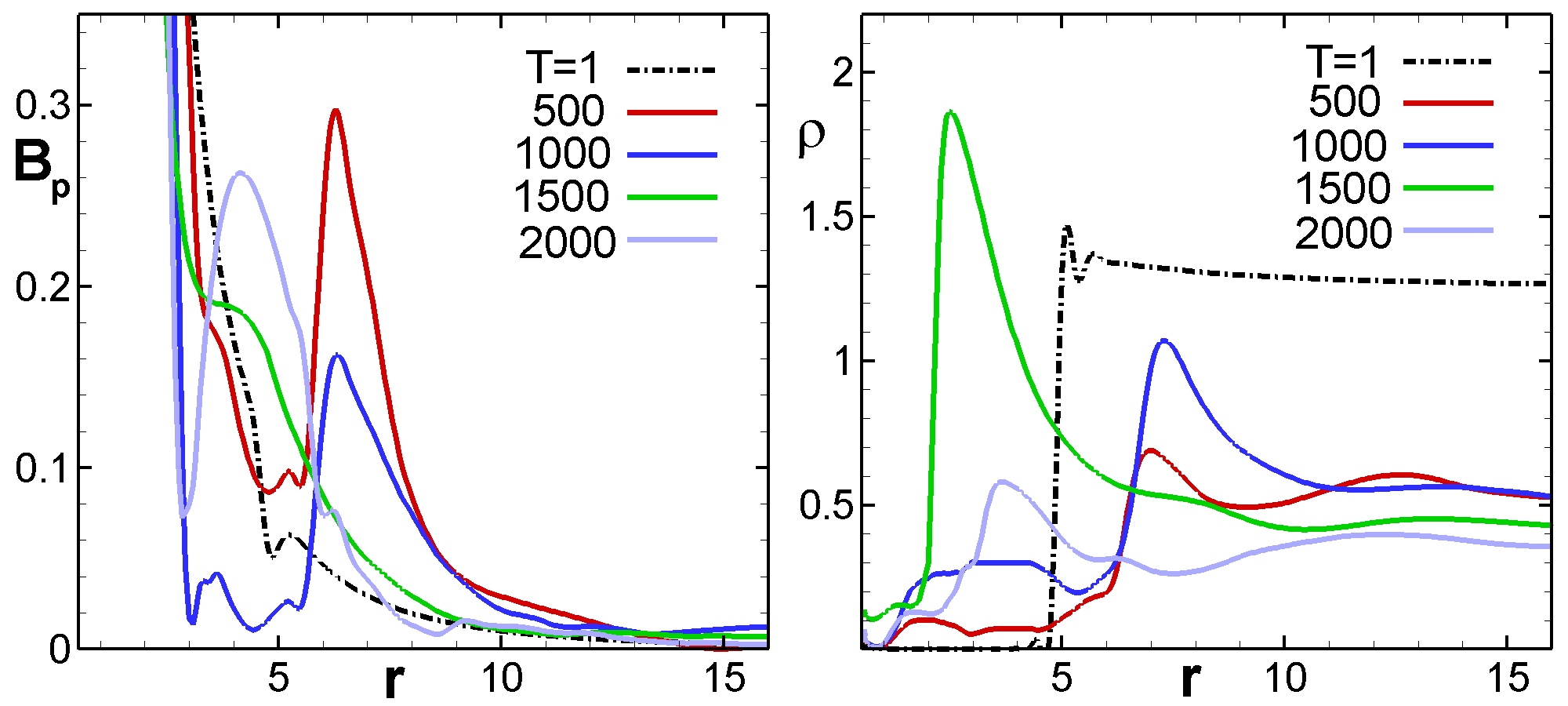}
\caption{Same as in \fig{bro-x-2} but in the propeller regime.}
\label{bro-p-2}
\end{figure}

\begin{figure*}
\centering
\includegraphics[width=4.5in]{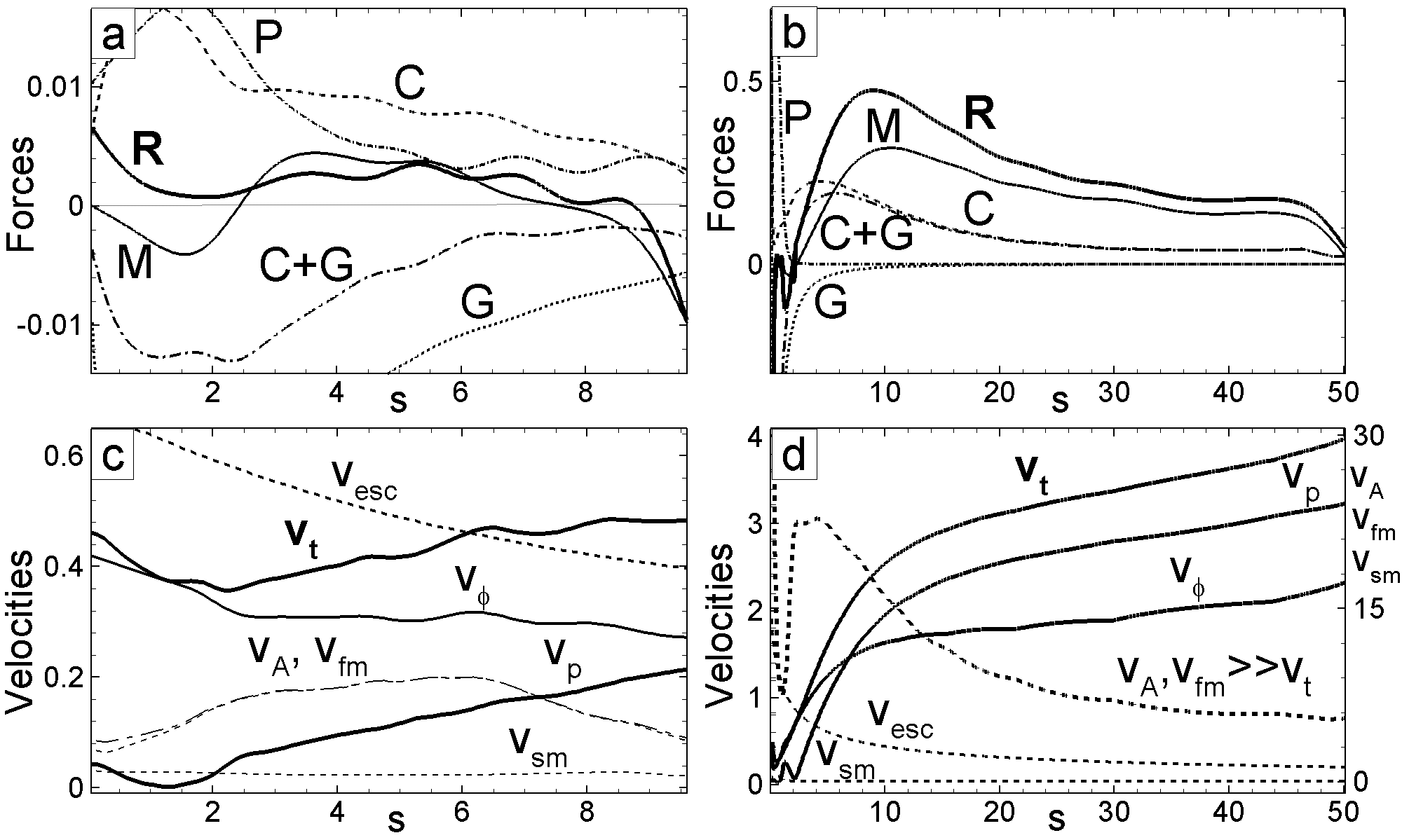}
\caption{
Forces ({\it top panels}) and velocities ({\it bottom panels}) along the field lines in the propeller regime at $T=1400$. {\it Left panels}: forces and velocities along the closed magnetic field
line which starts in the disk (at $r=4.3$) and ends on the star. We take only the part of the line up to the neutral point where the line curves towards the star($B_r=0$).
 Forces are projected onto the field line. Labels are the same as in \fig{for-vel-x-4}.  {\it Right panels}: forces and velocities along the open field line which starts on the surface of the star. Note that the scale for $v_A$, $v_{\rm sm}$,
$v_{\rm fm}$ is different from that for other velocities.}\label{for-vel-p-4}
\end{figure*}

To understand the physics of conical winds in greater detail we show in
 \fig{8-all} the distribution of different parameters  at time $T=500$.
Panels {\it b} and {\it f} show that the innermost region of the closed magnetosphere
rotates with the angular velocity of the star ($1.2 \leq r \leq 1.8$).
   At larger distances ($r > 2.5$), the corona above the disk rotates with the
angular velocity of the disk. Strongly inclined field lines which start
in the disk go through regions of lower and lower angular velocity and are strongly wound up owing to the difference in the angular rotation rates along the lines.
   This leads to a strong poloidal current flow $J_p \propto r B_\phi$
above the disk (see panels {\it c} and {\it g}) which gives rise
to the  magnetic force
 $F_p \sim - {\bf\nabla}[(r B_\phi)^2]$.
  This is the main force driving matter into the conical wind.
Driving of winds by the magnetic force from the inner disk was proposed earlier by Lovelace et al. (1991).
The direction of the magnetic force is shown schematically in \fig{8-all}{\it g}. It acts upwards and towards the axis, which is different from the centrifugal force.
      This determines three of the important properties of conical winds: (1) their small opening angle; (2) the fact that the wall of the cone is narrow, and (3).
      the gradual collimation of conical winds. If the centrifugal force were dominate (e.g. Blandford \& Payne 1982) then the cone would have a wider
opening angle and outflow would flow over a wide range of directions
(as in the X-wind model of Shu et al. 1994). Panels {\it d} and {\it h} show the distribution of entropy $S$
which shows that matter flowing from the disk into the wind is cold and it is not
thermally driven.
To analyze the forces driving matter into the conical winds we select one of the field lines, $s$, (see red bold line in
panels {\it e-h}) and we project forces onto this field line.
  We split the line into two parts (see panel {\it a}). Part ${\bf AB}$ starts from the disk and ends at the place where the line curves towards the star; part {\bf BC} continues from there to the surface of the star .

\fig{for-vel-x-4}{\it a} shows the projection of all forces onto part {\bf AB} of the field line.
   One can see that the main force accelerating matter into the conical wind is the magnetic force $M$.
     The centrifugal ($C$) and gravitational ($G$) forces approximately compensate each other and the sum $C+G$ is negative.
   The pressure gradient force $P$ is small.
   The  $\theta$-component of the
magnetic force leads to frequent forced reconnection events of the inflated field lines and to ejection of plasmoids into the conical wind.
  Panel {\it b} shows the projection of the forces onto segment {\bf BC} of the field line.
    One can see that it is chiefly the centrifugal force which accelerates the low-density matter to high velocities in this region.
   Panel {\it c} shows that in the conical wind the poloidal velocity $v_p$ (along part ${\bf AB}$)   gradually increases and crosses the slow magnetosonic ($v=v_{sm}$), Alfv\'en ($v=v_A$)
and fast magnetosonic ($v=v_{fm}$) surfaces.
   Matter rotates rapidly, therefore the azimuthal
component  $v_\phi$ is much larger than poloidal one, and the total velocity $v_t$ is  determined by the
azimuthal rotation of the flow.
Panel {\it d} shows that there is an interval of high velocity along the stellar part ({\bf BC}) of the field line.
Thus, we observe a two-component flow: (1) a high-density low-velocity conical wind which is the main component of the outflows, and (2) a low-density fast outflow along the stellar field lines which occupies a much smaller region.

%The region occupied by the second component increases in size with the spin of the star
%(see {\it Appendix C}), and it encompasses the whole coronal axial region in the case of propeller-driven outflows %(see \S 4).

We find that the region of the fast coronal flow increases in size with the star's rotation rate.
        As an example we decreased the corotation radius from $r_{cor}=3$ ($\Omega_*=0.19$) to $r_{cor}=2$ ($\Omega_*=0.35$)
        and observed that the region of  fast coronal flow increased significantly.
\fig{rotation} shows the difference. The region is even larger for smaller corotation radii when the star is closer
to the propeller regime.
  In the propeller regime (see \S 4) the fast jet component
occupies the entire region within the conical wind and is very powerful.
The star spins up for both $R_{cor}=2$ and $3$. Cases $R_{cor}=1, 1.5$ correspond to the propeller regime
where the star spins down due to the interaction with the disk and corona. We did not perform a refined search for the
rotational equilibrium state, in which the star has alternate spin-up and spin-down periods,
but zero torque on an average (e.g. R02, Long et al. 2005).
In this state we expect the jet component to occupy a large part of the region above conical winds.
Even a weak stellar wind
 (not considered in this paper) may enhance the jet component.

\begin{figure*}
\centering
\includegraphics[width=6in]{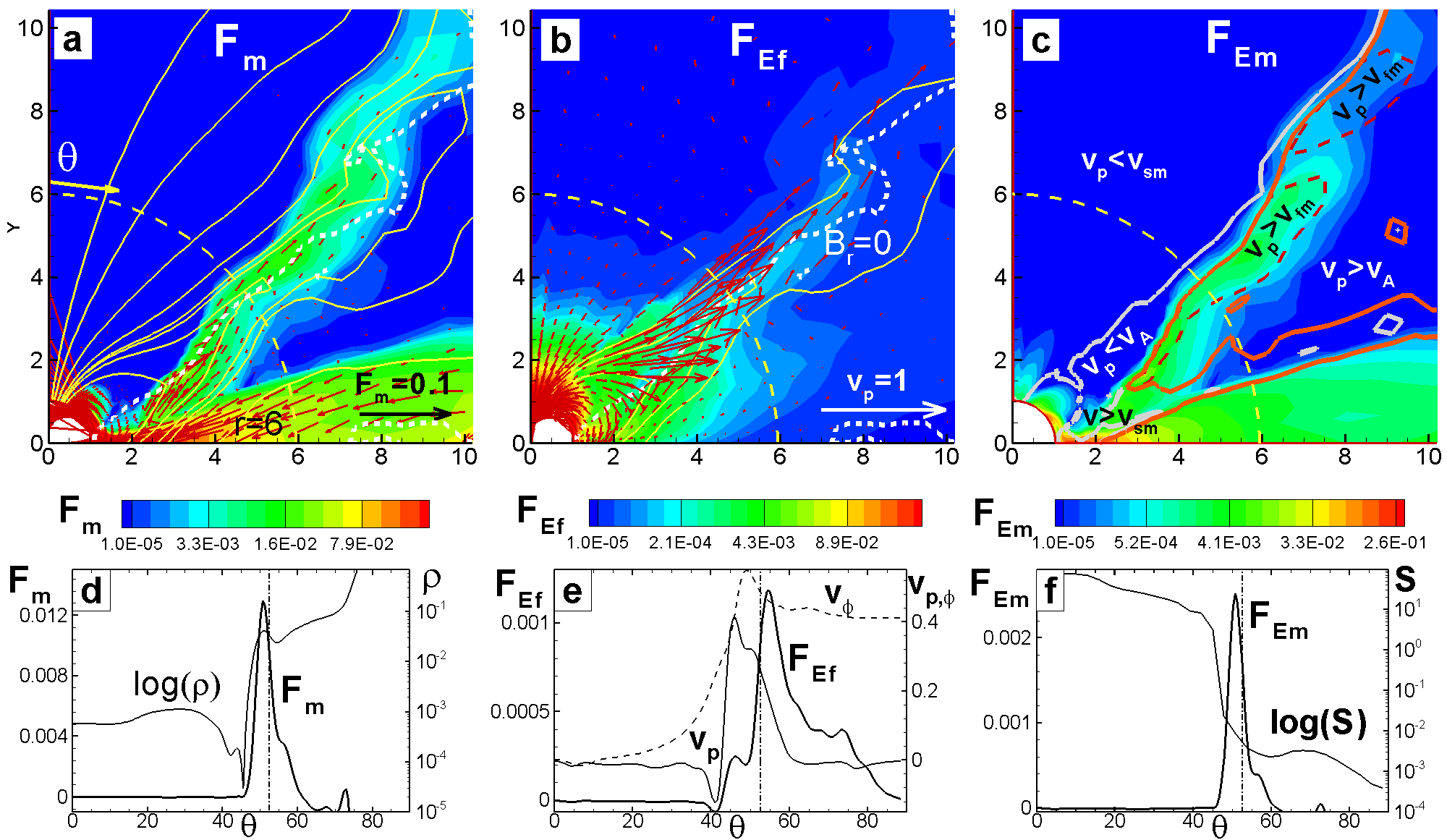}
\caption{{\it Top panels}: matter and energy fluxes at $T=500$.
{\it Bottom panels}: angular distribution of different quantities
at radius $r=6$ (starting from the axis).
{\it Panel a}: the background and vectors show matter flux $F_m$,
lines are poloidal field lines. The thick dashed line shows
the neutral line of the magnetic flux where $B_r=0$.
{\it Panel b}: the background shows the energy flux carried by the magnetic field, $F_{\rm Ef}$,
vectors are poloidal velocity vectors. {\it Panel c}:
the background shows the energy flux carried by the matter, $F_{\rm Ef}$. The solid red line is the line where
$v_p=v_A$, red dashed line corresponds to $v_p=v_{\rm fm}$, the solid white line corresponds to
$v_p=v_{\rm sm}$. {\it Panel d}: distribution of density $\rho$ and matter flux $F_m$ at $r=6$;
{\it Panel e}: same but for $F_{\rm Ef}$ and velocities $v_p$, $v_{\phi}$;
{\it Panel f}: same but for $F_{\rm Em}$ and entropy S. Dashed vertical line shows position of the neutral point, $B_r=0$.}
\label{en-x-6}
\end{figure*}

\begin{figure*}
\centering
\includegraphics[width=6in]{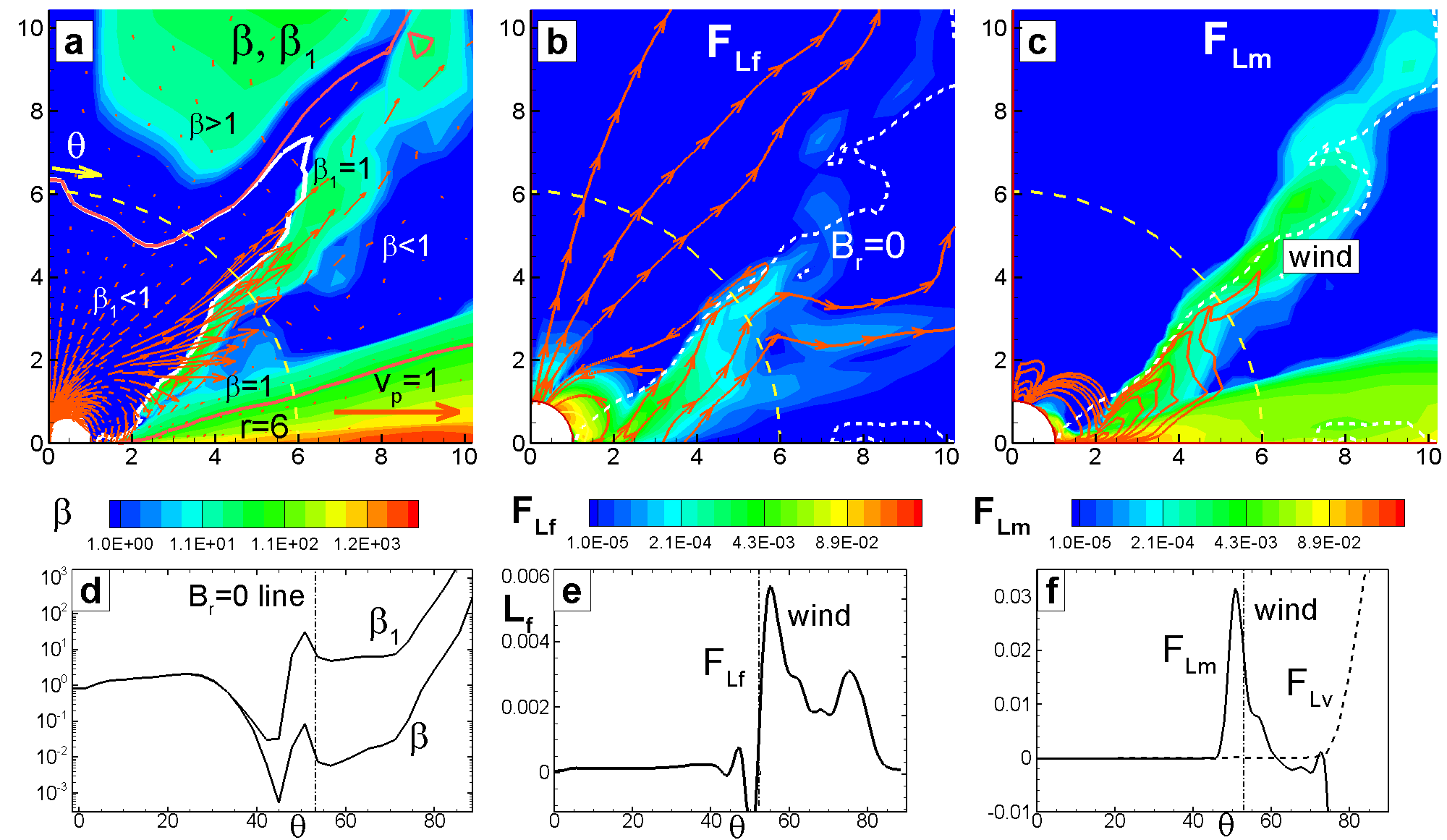}
\caption{The figure is similar to \fig{en-x-6}, but different quantities are shown.
{\it Panel a}: the background shows the distribution of the
kinetic beta parameter  $\beta_1$. The
solid white line is $\beta_1=1$ line,
the red line is
$\beta=1$ line. Arrows are velocity vectors.
{\it Panel b}: the background shows angular momentum flux carried by the magnetic field, $F_{\rm Lf}$,
streamlines show direction of this flux; the thick dashed line corresponds to $B_r=0$. {\it Panel c}:
the background shows angular momentum flux carried by the matter, $F_{\rm Lm}$.
The solid red lines show contours of the $F_{\rm Lf}$ flux; {\it Panel d}: angular distribution
of $\beta$ and $\beta_1$ at $r=6$. The dashed vertical line shows the position of the neutral line $B_r=0$.
{\it Panel e}: same but for $F_{\rm Lf}$. {\it Panel f}: same but for
$F_{\rm Lm}$ and the viscous flux $F_{\rm Lv}$ (dashed line).}
\label{ang-x-6}
\end{figure*}

\subsection{Matter flow, velocities and forces in the propeller regime}

Here we consider outflows from rapidly rotating
stars in the propeller regime. In earlier work
 we performed multiple simulation runs of the propeller stage
at a wide variety of parameters (R05; U06).
  Here we take the reference run shown in R05
and perform additional analysis.
The parameters are $P_*=1$ day, $\alpha_v=0.3$ and $\alpha_d=0.2$ (see \S 2.3.2 for the other parameter
values).

\fig{sym-p-9} shows snapshots of the matter flow
in the propeller regime at different times $T$.
One can see that  the outflow  appears at $T\approx 50$ and continues for a long time ($T=2200$ rotations, or about 6 years in application to protostars). These simulations are about $10$ times longer than previous simulations of outflows from a {\it real} disk (e.g. Goodson et al. 1997; Matt et al. 2002;
K\"uker et al. 2003). They are comparable in length with simulations of outflows from the disk as a {\it boundary condition} (e.g., Fendt 2009), although here we consider outflows form the ``real" cold disk to a hot low-density corona.

     \fig{sym-p-9} also shows that the outflow has two components.
 One is a conical-shaped wind similar to the conical winds of slowly rotating stars discussed earlier.   The other component is a fast flow of  matter interior to  the conical winds, which we term the {\it axial jet}.
     The disk-magnetosphere interaction is strongly non-stationary; the  magnetic field lines episodically inflate and the disk oscillates. The conical wind component seems to be weakly collimated inside the simulation region.
The jet component has stronger collimation.
   The jet collimation is stronger in the flow closer to the axis, and is enhanced during periods of strong inflation, like at times $T=560$ and $870$.  See animations  of propeller-driven outflows at {\it http://www.astro.cornell.edu/$\sim$romanova/propeller.htm}.

\fig{prop-numb} shows a typical snapshot from our simulations at time $T=1400$, with the dimensionless density and velocity at sample points (see Table 1 for reference values).
One can see that the velocities in the conical wind component are similar to those in conical winds around slowly rotating stars.
Matter launched from the disk has a velocity that is mainly azimuthal and approximately Keplerian.
It is gradually accelerated to poloidal velocities $v_p\sim (0.3-0.5) v_K$.
The flow has a high density and carries most of the disk mass into the outflows. The situation is the opposite in the axial jet component: the density is $10^2-10^3$ times lower, while the poloidal and total velocities are much
higher.
   Thus we find a  {\it two-component outflow}:  a dense,  slow conical wind and a low-density, fast axial jet.

\fig{bro-p-2} shows the time-variation of the equatorial density and the poloidal magnetic field
in the inner part of the simulation region. One can see that the density and the poloidal magnetic field are strongly enhanced at the inner edge of the disk, and the inner disk radius shows large
oscillation (see also R05, U06).

 \fig{for-vel-p-4}{\it a} shows the projection
of different forces onto a closed field line which starts in the disk at $r=4.3$ where the base of the conical wind (see \fig{prop-numb}). We take only the part of the
line from the disk to the neutral point where $B_r=0$
(this is the analog of part {\bf AB} of the line in \fig{8-all}{\it e}). One can see that the forces are large but more or less  compensate each other.
The magnetic force ($M$) seems to drive matter from the disk into the conical wind, though other forces, such as the centrifugal ($C$) and pressure gradient ($P$) forces are also important. It is interesting that conical winds in slowly rotating stars and in stars
in the propeller regime are similar, but that the distribution of forces is somewhat different.
In conical winds the winding of the field lines gives rise to a magnetic force in one localized region (above the inner disk) and this force dominates. In  the propeller regime the disk oscillates
strongly, and it is  important that the magnetosphere presents a centrifugal barrier for this matter,
and therefore the centrifugal and pressure gradient forces have  a larger role.  The magnetic force remains important.

Panel {\it b} shows the forces along the coronal field line which starts on the surface of the star. We consider the second line from the axis in \fig{prop-numb}, which is strongly inflated and is a representative line for the description of matter flow into the axial jet. One can see that the magnetic force $M$ is much larger than the other forces and is the main force accelerating matter into the jet.

Panel {\it c} shows velocities along the disk field line (as in  panel {\it a}). One can see that the azimuthal component $v_\phi$ dominates, while the poloidal velocity $v_p$ increases gradually from a very small value near the disk up to values comparable with $v_\phi$. It crosses the slow magnetosonic surface just above the disk, and later, the Alfv\'en
 and the fast magnetosonic surfaces.

Panel {\it d} shows that in the coronal region, the velocities are high and the poloidal velocity dominates. Matter crosses the slow magnetosonic surface but stays sub-Alfv\'enic. Both the
Alfv\'en, $v_A$, and the fast magnetosonic, $v_{\rm fm}$, velocities are about $10$ times larger than the flow velocity in the axial jet (note the scale at the right-hand side). The flow is in the Poynting flux regime found in simulations by Ustyugova et al. (2000) and analyzed theoretically by Lovelace et al. (2002).

\begin{figure*}
\centering
\includegraphics[width=5.0in]{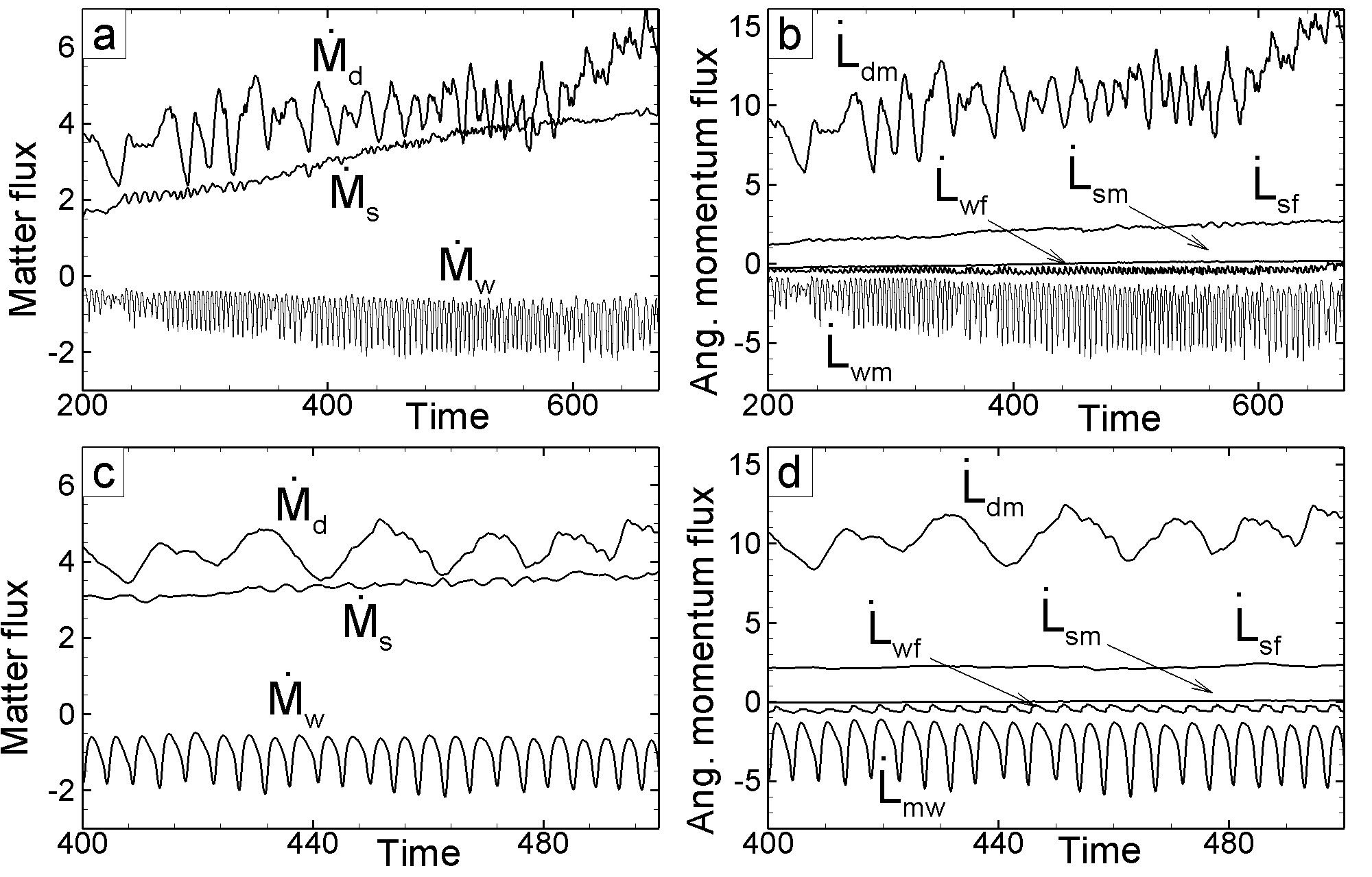}
\caption{{\it Panel a}: matter flux onto the star $\dot M_s$, into the
conical wind $\dot M_w$, and through the disk $\dot M_d$.
{\it Panel b}: angular momentum flux
carried by the disk $\dot L_d$, onto
the star carried by matter $\dot L_{\rm sm}$ and by the magnetic field
$\dot L_{\rm sf}$.
The angular momentum flux  of the conical wind
carried by matter is $\dot L_{\rm wm}$ and that carried
by the field is  $\dot L_{\rm wf}$.
{\it Panels c, d} show the time dependence of these fluxes.}\label{flux-x-4}
\end{figure*}

\section{Analysis of fluxes: matter, energy, angular momentum}

\subsection{Fluxes in Conical winds}

\fig{en-x-6}{\it a} shows the matter flux distribution $F_m$ and  a neutral line
of the magnetic field where $B_r=0$.
   This line separates the field lines starting
on the disk from those starting on the star.
    One can see that the conical wind flows along both sets of field lines.
Panel {\it d} shows that the matter flux $F_m$ has a sharp peak in its angular distribution (at $r=6$),
that is, the wall of the cone is narrow. The position
of the $B_r=0$ line in this panel shows that matter flows along both  the stellar and the disk field lines.
    For $\theta > 75^\circ$, the matter
flux is dominated by the disk and  is negative.
We also see that the density $\rho$ is low in the corona.
%   It increases by about a factor of 100 in the conical wind and continues to be large in the gap between the disk %and the conical wind. In this gap, matter flows from the disk to the wind with low velocity.
In the disk the density increases to much larger values, $\rho=1-10$.
There is also a low-density gap  at $\theta \sim  45^\circ$ where matter is accelerated to high velocities.

   The energy flux ${\bf F}_{\rm E}$ is the sum of the matter
component    ${\bf F}_{\rm Em}$ and the field component
${\bf F}_{\rm Ef}$ where
\begin{equation}\label{eq5}
{\bf F}_{\rm Em}=\rho {\bf v}_p \left(\frac{1}{2} v^2+w+\Phi_g\right), ~~{\bf F}_{\rm Ef}=\left(\frac {c}{4\pi}\right)({\bf E}\times{\bf B})_p,
\end{equation}
where $w$ is the enthalpy and $\Phi_g$ is the  gravitational potential.

\fig{en-x-6}{\it b}  shows that the magnetic
field energy flux  $F_{\rm Ef}$
is high near the star, at the base of the conical wind and in the region of fast flow.
Panel {\it e} shows that the energy flux $F_{\rm Ef}$ at $r=6$ has a peak in the region of the conical wind.
%The energy flux is also quite high at larger $\theta$, towards the disk. This indicates that the disk wind  also %contributes to this flux.

\fig{en-x-6}{\it c} shows that the distribution of the matter energy flux, $F_{\rm Em}$, is similar to the matter flux distribution.
  The panel also shows that in the conical wind,
matter crosses all the critical surfaces.
   It ends up flowing
with a super-fast-magnetosonic velocity.
In contrast, in the corona, away from the regions of outflow, the flow is sub-slow-magnetosonic.
Panel {\it f} shows that the entropy $S$ is high in the corona and low in the conical wind and the disk.

 \fig{ang-x-6}{\it a} and {\it d} shows the ratio of the gas and magnetic pressures,
 which is the conventional $\beta$ parameter.  We also use what we term  the
 {\it kinetic} parameter $\beta_1$, where
\begin{equation}\label{eq6}
\beta=\frac{p}{B^2/8\pi}~,\quad ~ \beta_1=\frac{p + \rho v^2}{B^2/8\pi}~.
\end{equation}
   The flow region is magnetically dominated when the $\beta$ or
$\beta_1$  is less than unity.
   The magnetic pressure  dominates only in the region
near the star in the conical wind case.   The situation is
different in the propeller regime where the axial
region is magnetically dominated (see \S 4).

 We calculate the {\it angular momentum flux} distribution
 which consists of three components,  ${\bf F}_L={\bf F}_{\rm Lm}+{\bf F}_{\rm Lf}+{\bf F}_{\rm Lv}$,
where ${\bf F}_{\rm Lm}$, ${\bf F}_{\rm Lf}$ and ${\bf F}_{\rm Lv}$ are the angular momentum fluxes  carried by the matter,   the magnetic field,  and   the viscosity:
\begin{eqnarray}\label{eq7}
{\bf F}_{\rm Lm} &=&  r \sin \theta \rho v_\phi  {\bf v}_p~, \quad\quad
{\bf F}_{\rm Lf}= - r \sin \theta \frac{B_\phi {\bf B}_p}{4 \pi}~, ~
\nonumber\\
{\bf F}_{\rm Lv}&= &- \nu_t \rho (r \sin \theta)^2 {\bf \nabla}\Omega~ .
\end{eqnarray}
\fig{ang-x-6}{\it b} shows that the magnetic component $F_{\rm Lf}$ of the flux, dominates
near the star and in the part of the conical wind close to the disk. The streamlines show that angular
momentum flows from the disk onto the star, from the disk into the conical wind, and from the star into the corona.
  Panel {\it c} shows that the conical winds carry
   away angular momentum associated with matter, $F_{\rm Lm}$.
     The magnetic component is also high at the base of conical wind.
However, at larger distances this angular momentum is converted into angular momentum carried by matter.
Comparison of panels {\it e} and {\it f} shows that at $r=6$, the angular momentum carried by the matter is much larger than that carried by the field.
Panel {\it e} also shows that some angular momentum flows into the disk wind (at $\theta > 60^\circ$).
Panel {\it f} shows that the angular momentum carried by viscosity  is significant. The disk viscous component is much larger than the matter
component flowing into the conical wind, so most of the angular momentum flows outward along the disk.

We also calculated the matter and angular momentum fluxes flowing through
the surface of the star, and through a spherical surface of radius $r=6$.
\begin{equation}\label{eq8}
\dot M = \int d {\bf S}\cdot {\bf F}_m ~, \quad\quad {\bf F}_m=\rho{\bf v}_p~,
\end{equation}
\begin{equation}\label{eq9}
\dot{L} = \int d {\bf S}\cdot ({\bf F}_{\rm Lm} +
{\bf F}_{\rm Lf} +
{\bf F}_{\rm Lv})~,
\end{equation}
where $d {\bf S}$ is the surface area element directed outward.
 \fig{flux-x-4} (left panels) shows that about $2/3$ of the
 incoming disk matter flows to the  star,
 and the rest going into the conical wind.
    Fluxes into the wind oscillate due to  magnetic field inflation and reconnection events.
The right-hand panels show that  angular momentum flows inward with the disk matter, $\dot L_{\rm dm}$.
 Part of this angular momentum is carried away by the conical wind. Angular momentum is carried mainly by matter,   $\dot L_{\rm wm} >> \dot L_{\rm wf}$.
 Part of the disk angular momentum flows to the star and spins it up (the star rotates slowly in that  $r_{cor}=3$).
   The angular momentum carried to the star by the matter, $\dot L_{\rm sm}$, is converted into  angular momentum
   carried by
   the field, $\dot L_{\rm sf}$, and hence on the surface of the star $\dot L_{\rm sf} >> \dot L_{\rm sm}$
   (see also R02). One can see from \fig{flux-x-4} that all these fluxes are smaller than the flux carried by the disk matter. This means that the main part of the angular momentum of the disk flows outward to larger distances due to viscosity.

\subsection{Fluxes in the propeller regime}

\begin{figure*}
\centering
\includegraphics[width=6in]{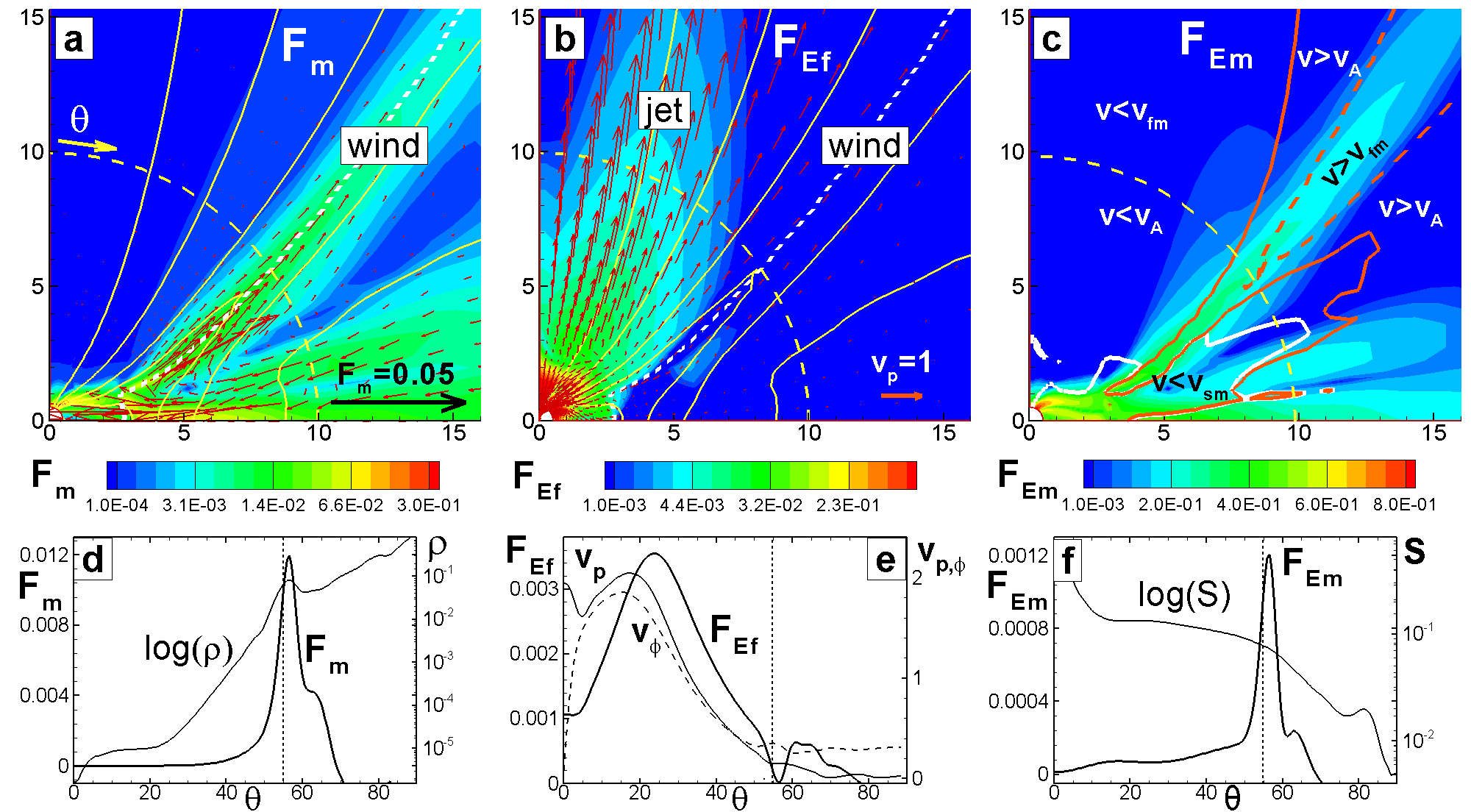}
\caption{
Same as in \fig{en-x-6} but for the propeller regime at time
$T=1400$.}
\label{en-p-6}
\end{figure*}

We analyze fluxes in the propeller regime in a similar manner to that for the conical winds.
\fig{en-p-6}{\it a} shows the distribution of the  poloidal matter flux $F_m$ in the background with the vectors ${\bf F}_m$ on top. The neutral line (dashed white line) separates the field lines which start in the disk from those which start on the star.
    %The conical wind flows along both sets of lines with the neutral line approximately in the middle.

Panel {\it d} shows that  most of the matter flows along field lines threading the disk, although some matter flows along the stellar field lines.
    For $\theta > 80^\circ$ (in the disk region) the matter flux becomes much larger and negative (we exclude this part of
the plot to show the conical wind part more clearly).
The plot of the density shows that the density is very low on the axis but  gradually increases toward the region of the conical wind and continues to grow towards the disk.

Panel {\it b}  shows the distribution of the magnetic energy flux  $F_{\rm Ef}$. One can see that a strong flux of magnetic energy  (Poynting flux) flows into the corona.
      This is the region where matter is accelerated to high velocities (see velocity vectors).
Panel {\it e} shows that at $r=10$ the magnetic energy flux is very large,
 and is distributed over a range of angles with a maximum at $\theta \sim 25^\circ$  (not on the axis).  The plot also shows that the poloidal velocity is slightly larger than the azimuthal velocity and both velocities are high, up to
 $v_p=2$ (which is $400$ km/s for protostars).
 The jet component is smaller in the case of slowly rotating stars.

Panel {\it c} shows the energy flux associated with the matter flow.
  One can see that matter in the conical wind crosses the slow magnetosonic surface $v=v_{\rm sm}$ just above the disk, and soon crosses the Alfv\'en surface $v=v_A$,  and  the fast magnetosonic surface $v=v_{\rm fm}$.
  Panel {\it f} shows that the matter energy flux distribution has a sharp peak in the region of the conical wind and that the entropy $S$ is high in the corona but drops towards the disk.

Comparison of panels {\it e} and {\it f} shows that the maximum of the energy flux carried by the magnetic field
into the jet, $F_{\rm Ef}$, is about 3 times larger than that carried by the matter into conical winds,  $F_{\rm Em}$.
In addition, the integrated flux carried by the magnetic field is a few times larger. Therefore, the cumulative energy flux carried by the magnetic field into the jet (the Poynting flux) is about
10 times larger than that carried by matter. This means that the jet component is 10 times more powerful.
Part of this energy is converted into  kinetic energy of the fast component
inside the simulation region. However, most of the magnetic energy may be transferred to
particles or converted into radiation at larger distances from the star.

Next, we analyze the angular momentum flow.
\fig{ang-p-6}{\it a} shows the $\beta$-parameter in the background and the
$\beta=1$ and $\beta_1=1$ surfaces in the foreground (see equation 6).
One can see that the magnetic energy dominates in the whole axial coronal region
interior to the conical wind ($\beta_1<1$ and $\beta<1$).
    If one uses only the standard  criterion $\beta=1$, then one can see that
    the region above the disk is also magnetically dominated ($\beta < 1$).
Panel {\it d} shows the angular distribution of $\beta$ and $\beta_1$ at
$r=10$.

Panel {\it b} shows the distribution of the  angular momentum
flux carried by the magnetic field, $ F_{\rm Lf}$, and the streamlines associated with this flux.
   One can see that {\it a significant amount of angular momentum flux flows from the star into the corona} along the stellar field lines.
Panel {\it e} shows that the angular momentum flows out along the set of field lines
 between the axis and the neutral field line with a maximum right above the conical wind component.
    Some of them thread the low-density corona, while others thread the upper part of the conical wind above the neutral line.
These panels also show that a significant amount of angular momentum flows from the inner part of the disk.   See U06 for detailed analysis of the different
components  of the angular momentum flow.

Panel {\it c} shows that the angular momentum flux carried by the matter is also large and is carried by the conical wind.
Panel {\it f} shows that most of this angular momentum flows along the disk field lines  while some angular momentum flows along the stellar field lines.

 \fig{flux-p-6}{\it a} shows the matter fluxes onto the star $\dot M_s$,
 and into the outflows, $\dot M_w$, integrated over a surface with radius $r=10$ (any flow with $v_r > 0$ is taken into account).
One can see that the matter flux into the wind is much larger than that onto the star, $\dot M_w>>\dot M_s$,
that is,
almost all disk matter is ejected from the system into the outflows. Here we should note that we
consider the ``strong propeller" case, $r_{cor}<<r_m$. If the star rotates slower, then the fraction of the matter
flux going into the wind decreases, and a larger portion of the matter may accrete onto the star (see U06 for dependences of matter fluxes on $\Omega_*$, $B_*$, $\alpha_v$ and $\alpha_d$.)
Both fluxes are strongly variable and show episodic enhancement of accretion and outflows. Simulations show that an interval between the strongest outbursts increases when diffusivity coefficient $\alpha_d$ decreases (R05, U06).

Panel {\it b} shows the  integrated angular momentum fluxes through the same $r=10$ surface.
   Here we calculate separately the angular momentum fluxes carried by the field
    and by the matter.
One can see that the star spins down due to the angular momentum carried by the magnetic field, $\dot L_{\rm sf}$, while  the angular momentum carried by the matter flow, $\dot L_{\rm sm}$, is negligibly small.
   The angular momentum outflow from the star, $\dot L_{\rm sf}$, almost coincides with the
   angular momentum carried by magnetic field lines into the magnetically-dominated jet, $\dot L_{\rm wf}$.
       This indicates that angular momentum flows from the star into the
magnetically dominated axial jet.
   Thus a star in the propeller regime is expected to spin down rapidly due to angular momentum flow into the magnetically-dominated axial jet.
    Analysis of U06 shows that this jet angular momentum is approximately equally split between the flux carried into the corona along open field lines, and the flux which flows along partially inflated field lines which close inside the simulation region and are connected  with both the star and the disk.
Panel {\it c} shows that the angular momentum carried by matter into the conical winds,
 $\dot L_{\rm wm}$   is approximately equal to that carried by the field to the corona.
 The bottom panels show  the same plots at higher time resolution.
Therefore, the star-disk system loses its angular momentum through both the wind and jet components,
 via the inner disk and star respectively. So, there is no problem
 with excess angular momentum in the star-disk-system; it flows into the jet/wind.

\section{Other properties of outflows}

\subsection{Inflation of Field Lines and Disk Oscillations}

The field lines connecting the disk and the star have the tendency to
inflate (e.g., Lovelace,  Romanova \& Bisnovatyi-Kogan 1995).
Quasi-periodic reconstruction of the magnetosphere due to
inflation and reconnection has been discussed theoretically (Aly
\& Kuijpers 1990; Uzdensky, Litwin \& K\"onigl 2003) and has been
observed in a number of axisymmetrtic simulations (Hirose et al.
1997; Goodson et al. 1997, 1999; Matt et al. 2002; Romanova et al.
2002 - hereafter R02; von Rekowski \& Brandenburg 2004).
Goodson \& Winglee (1999) discuss the physics of inflation cycles.
They have shown that each cycle of inflation consists of a period of matter accumulation near
the magnetosphere, diffusion of this matter through the magnetospheric field,
inflation of the corresponding field lines,
accretion of some matter onto the star, and outflow of some matter as winds,
 with subsequent expansion of the magnetosphere. There simulations show $5-6$
 cycles of inflation and reconnection.

Our simulations show $30-50$ cycles of inflation and reconnection in the propeller regime.
   We chose one outburst from our simulations
and plotted the density and a fixed set of magnetic field lines
at different times.
\fig{flare-p-6} shows
   that at $T=890$, the magnetosphere is relatively expanded,
although some matter accretes around the expanded field lines (see also Romanova et al. 2004a).
   At $T=900$, the disk matter comes closer to the star and some field lines inflate or partially inflate, thus blocking accretion. At $T=910$ even more field lines inflate and accretion is blocked. However, outflow is permitted at both of these moments of time.
      At $T=916$, the internal field lines reconnect, permitting accretion onto the star.
At $T=921$, the magnetosphere expands and accretion onto the star is  again
prevented. Later, at $T=927$, the field lines reconnect and some matter accretes along a longer path
-- around the expanded magnetosphere.
This picture is similar to that described by Goodson \& Winglee (1999).

\fig{flux-p-6}{\it d} shows that the time interval between the strongest outbursts
in the propeller regime is $\Delta T\approx 50-70$.
In application to
protostars and CTTSs ($P_0=1.04$ days) this time corresponds to $(\Delta t)_{\rm outb}=52-73$ days.
In some young stars, like CTTS HH30 (XZ Tau), for example, the outbursts into the jet
occur at intervals of a few
months, which hints that episodic inflation of field lines may be responsible for some outbursts.
During the outbursts, the matter flux into outflows increases several times
and the velocities also increase.
This may lead to the formation of new blobs or to the generation of shock waves
in the outflow.
This mechanism may be relevant for formation of blobs or shocks in protostars and rapidly rotating CTTSs.
In  slowly rotating stars, the time-interval between outbursts is smaller,
$(\Delta t)_{\rm outb}\approx 5$ days (see \fig{flux-x-4}), so the outbursts have a smaller
amplitude but are more frequent.
The interval depends on the diffusivity in the disk, $\alpha_d$.
At very small diffusivity the time-interval between outbursts may be much larger.

Diffusivity is important for reconnection processes in the corona.
We have diffusivity only in the disk. We choose a certain density
level $\rho_d=0.3$ below which the diffusivity is absent, so that
high-density regions, $\rho>0.3$, which correspond to the disk and
the funnel streams, have diffusivity, and low-density regions do
not.
     In the corona and the conical outflows,
the diffusivity has only a numerical origin and is small. Namely,
we observe in simulations that in conical winds, the two layers of
plasma with an oppositely directed magnetic field reconnect only
slowly. Similar behavior has been observed in ideal MHD
simulations by Fendt \& Elstner (2000). An anomalous (high)
diffusivity was added by Hayashi et al. (1996) to a part of the
simulation region to enhance the reconnection process in the
inflating plasmoids. Diffusivity had been added into the whole
simulation region by Fendt \& Cemelji\'c (2002). They observed
that at higher diffusivity the level of collimation by the
magnetic field and the Lorentz force decrease, while the
centrifugal force increases.  We performed exploratory simulations
with non-zero diffusivity in the corona. We added to the corona
the same diffusivity as in the disk with $\alpha_d=0.1-0.2$, which
operates at different density levels, $\rho>\rho_d$, where
$\rho_d=0.1, 0.03, 0.01$ and in the whole simulation region
(formally, $\rho_d=0$). We observed that in case of conical winds
(slowly rotating stars) the diffusivity in the corona does not
change the result. However, in case of propeller-driven winds, we
observed that propeller becomes weaker. We believe that the
difference is in the fact that in the case of conical winds, the
wind and the neutral line of the inflated magnetic field have
approximately the same position in space, leading to slower
reconnection. In the other case, in the propeller regime the inner
disk and the region of outflows strongly oscillates, and so the
position of the neutral line varies, and hence the reconnection is
forced (that is, the plasma layers with oppositely directed fields
are pushed towards each other by an external force).

\begin{figure*}
\centering
\includegraphics[width=6in]{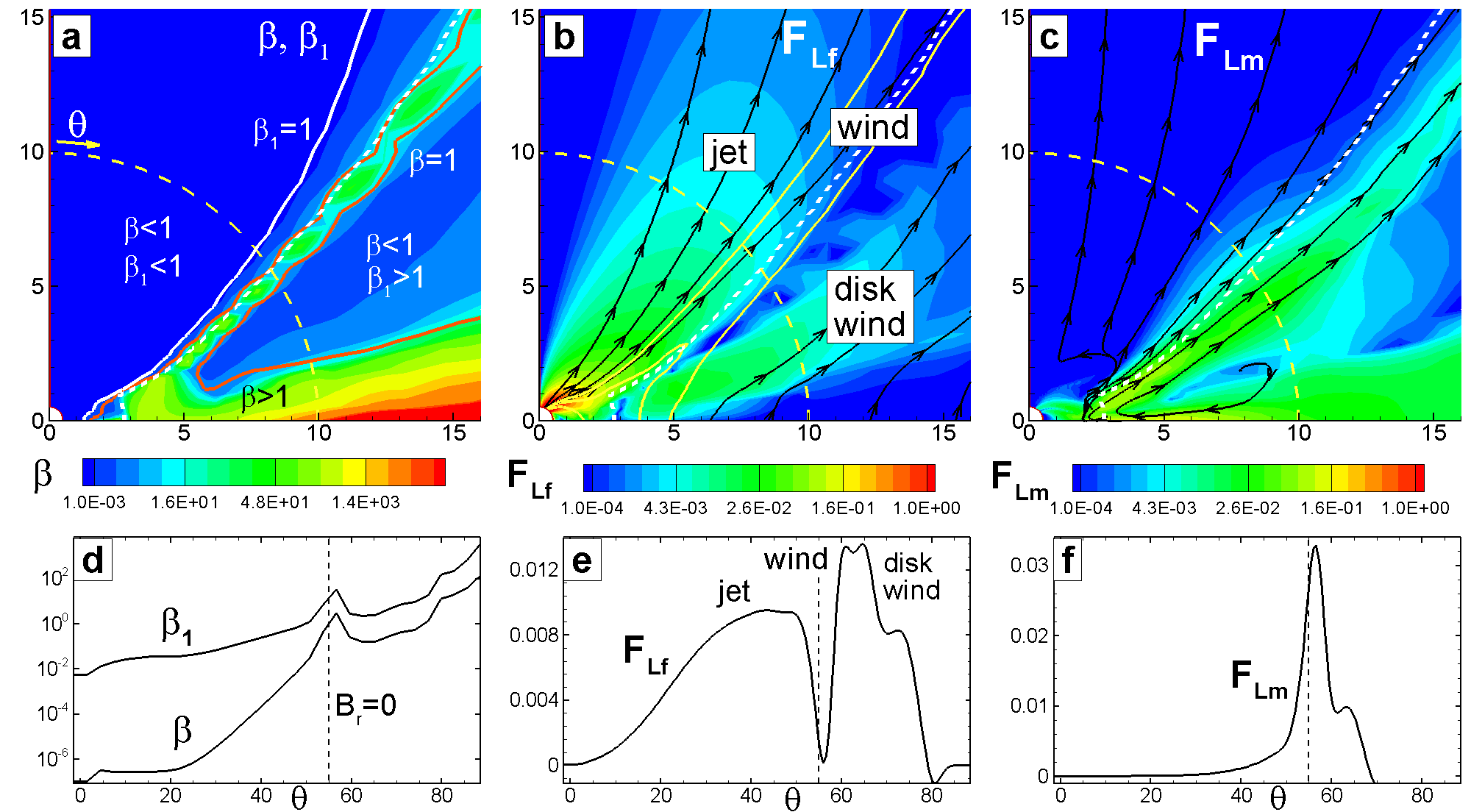}
\caption{Same as in \fig{ang-x-6} but for the propeller regime at time
$T=1400$.}
\label{ang-p-6}
\end{figure*}

\begin{figure*}
\centering
\includegraphics[width=6in]{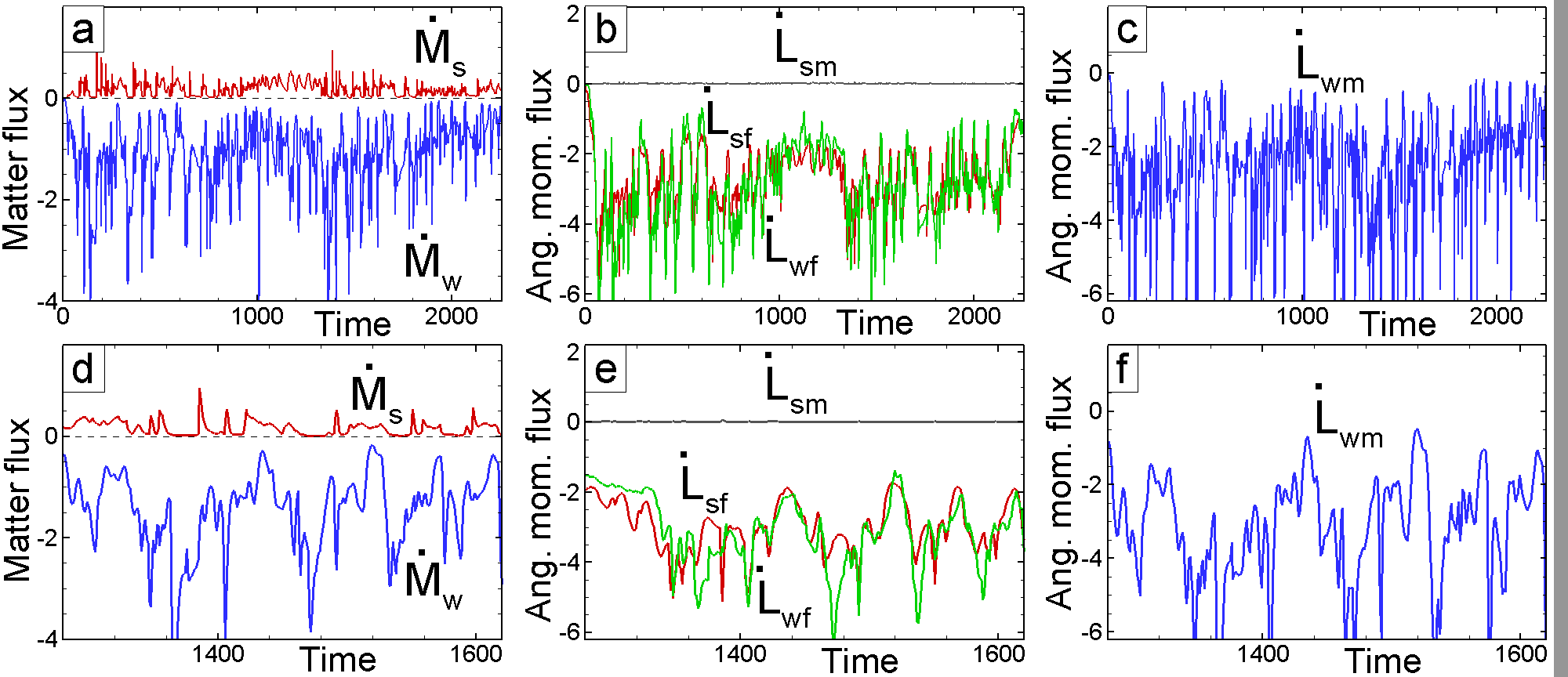}
\caption{{\it Panel a}: matter fluxes onto the surface of the star $\dot M_s$,
and into the outflows (both wind and jet) through the surface $r=10$ versus time $T$.
{\it Panel b}: angular momentum fluxes
carried to (or out from)  the star by the matter, $\dot L_{\rm sm}$  (gray line), and by the magnetic field,
$\dot L_{\rm sf}$ (red line).  The green line shows  the angular momentum flux
carried by the magnetic field through the surface $r=10$.
{\it Panel c}: angular momentum flux carried by the matter component
 $\dot L_{\rm wm}$ through the surface $r=10$.
The bottom panels show the same fluxes but during a part of the simulation
time.}
\label{flux-p-6}
\end{figure*}

\begin{figure*}
\centering
\includegraphics[width=6in]{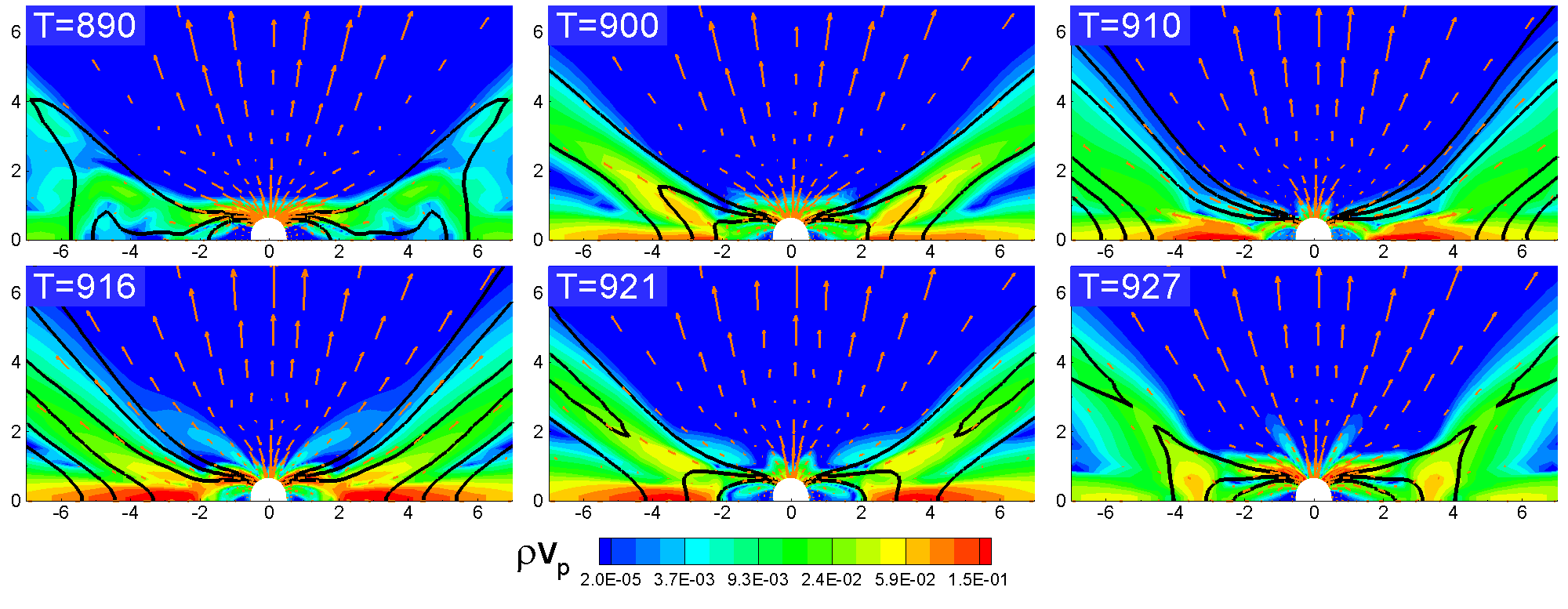}
\caption{Several snapshots from simulations show the mechanism of
enhanced accretion and outflows observed in the light-curves of propeller-driven accretion and outflows.
The background shows the matter flux, lines are the sample magnetic field
lines, the same set of which is shown in all plots.}\label{flare-p-6}
\end{figure*}

\subsection{Matter loading onto stellar field lines, and possible role of stellar wind}

\begin{figure}
\centering
\includegraphics[width=3.4in]{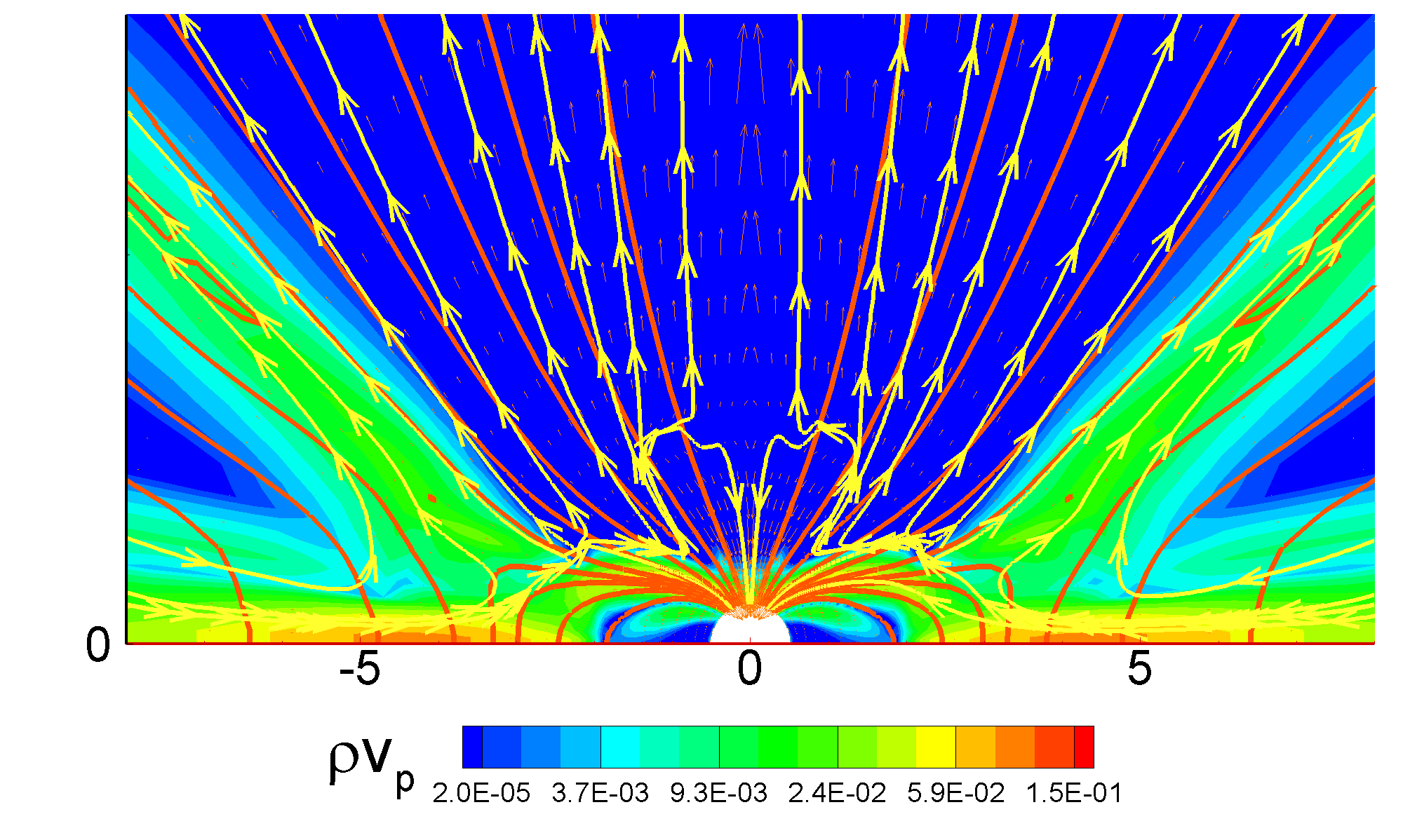}
\caption{White lines with arrows show streamlines of matter flow in the propeller regime at $T=1500$. Red lines show sample magnetic field lines. Only the inner part of the simulation
region is shown.}\label{stream}
\end{figure}

Here we discuss how the disk matter gets loaded onto the stellar
field lines and then flows into the jet in the propeller regime
(where the jet is strong). It is important to have diffusivity in
the disk, so that the matter of the disk threads the field lines
of the star and flows onto the star in funnel streams. When a
sufficient amount of matter is accumulated in the inner disk, the
field lines connecting the star and the disk inflate. During and
after inflation, part of the disk matter ends up on the field
lines connecting the disk with corona (usually most of the matter
flows along these field lines). Another, smaller part of the
matter ends on the field lines connecting the star with the
corona. For example, Fig. 15a demonstrates the result of such
inflation, where the neutral line dividing the stellar and disk
lines is in the middle of the conical wind component. On the other
hand, when matter flows in a funnel stream, most of it accretes
onto the star. However, part of it is stripped away by the
magnetic  and centrifugal forces and  flows into the jet along the
stellar field lines.  \fig{stream} shows that there is a dividing
line running through the upper part of the  funnel stream,
separating the regions from which matter flows onto the star (most
of it) from those from which it flows into the jet along the
stellar field lines (a small fraction). In the funnel region the
density is usually high enough so that the diffusivity which works
in our disk also works in the funnel stream. This diffusivity
helps launch matter from the funnel stream field lines onto the
coronal, jet field lines. Both processes are consistent with the
strong decrease in coronal matter density  along the axis.  This
region is ``matter-starved''.

Our simulations do not take into account possible {\it stellar winds}. Even a weak wind from the star may have a significant influence on the axial region of the jet in the propeller regime and the ``matter-starved" jet region
in slowly rotating stars.
The existence of powerful stellar winds was  suggested by Matt \& Pudritz (2005, 2008) in order to
explain the loss of angular momentum by young stars.
The spectra of many  CTTSs (e.g. Edwards et al. 2003; Dupree et al. 2005) require up to 10\% of the disk
mass flowing out as winds, in order to explain different spectral lines (Edwards 2009).
No such winds are observed in diskless, weak-line T Tauri stars.  Hence, the winds must be
accretion-driven  (e.g., Edwards et al. 2006; Kwan et al. 2007).
The physics of these accretion-driven stellar winds is not understood yet.
In the standard approach it is suggested that matter falling onto the surface of the star through the funnel stream forms a shock
near the surface and is heated by this shock.
However, it cools rapidly in the radiative zone behind the shock wave, and no
reverse flow into the wind  is expected (e.g., Lamzin 1998; Koldoba et al. 2008).
In another investigation, however, Alfv\'en waves and other processes at the stellar surface help  accelerate up
to $1\%$ of the accreting disk matter into the wind (e.g. Cranmer 2008).
We did not incorporate stellar winds into the present simulations. Weak winds may help supply
matter to the magnetically-accelerated
axial jets and the ``matter-starved" region of fast flow in slowly rotating stars. On the other hand, if the wind is very strong,
 say $\dot M_{sw}\sim 0.1 \dot M_d$,
it will probably be matter-dominated at moderate
distances from the star (say $\gtrsim 10$ stellar radii) and
will have a decollimating effect on the outflows (Fendt 2009).  In summary, a weak stellar wind will contribute matter to the jet component.

\subsection{Collimation of outflows}

\begin{figure*}
\centering
\includegraphics[width=7.0in]{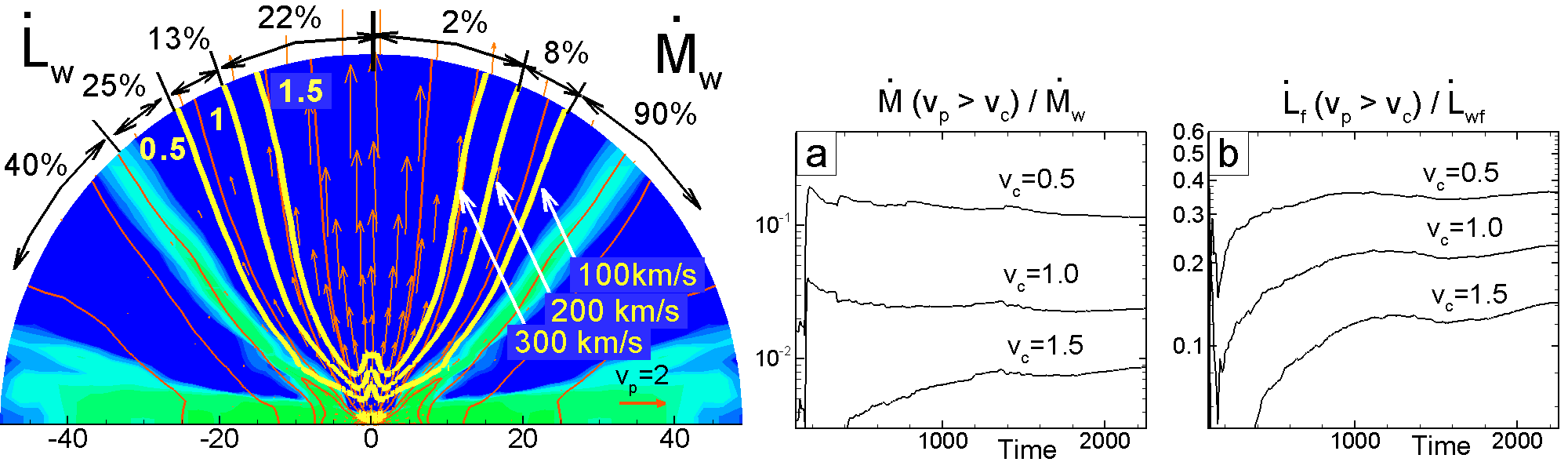}
\caption{Analysis of matter and angular momentum distribution between the jet and wind components in the propeller
regime at $T=1400$. {\it Left panel}: The yellow lines show surfaces of constant poloidal velocity, $v_c=0.5, 1, 1.5$ (
in application to protostars they are approximately $100$, $200$ and $300$ km/s).
Numbers on top of the plot show fractions of total matter flux $\dot M_w$
into different sectors of the region, and the same for angular momentum carried by the field, $\dot L_{\rm wf}$. {Panel \it b}:
Matter flux into the jet through an external boundary, $r=R_{out}$, with velocities $v_p > v_c$, $\dot M (v_p > v_c)$ versus total matter flux $\dot M_w$. {Panel \it c}:
Same as in panel b, but for the angular momentum flux carried by the magnetic field.}\label{jw-v-3}
\end{figure*}

\begin{description}
\item[\bf Collimation of conical winds.] We observe conical winds in both slowly and rapidly rotating stars.
In both cases, matter in the conical winds passes through the Alfv\'en surface,
beyond which the flow becomes matter-dominated.
We note that in slowly rotating stars, the distribution of the poloidal current $J_p$ (see \fig{8-all}{\it c}) is such that  the corresponding magnetic force  has a component towards the axis. This may explain why conical winds show some collimation (see \fig{sym-x-9}).
The conical wind component of the propeller-driven outflows shows stronger collimation during periods of inflation and outbursts (see \fig{sym-p-9}). However, this collimation may not be sufficient to explain well-collimated jets.

Conical winds may be further collimated at larger distances from the star either by the pressure
of the external medium (Lovelace et al. 1991;  Frank \& Mellema 1996), or by disk winds (K\"onigl \& Pudritz 2000; Ferreira et al. 2006;  Matsakos et al. 2008; Fendt 2009). In addition, Matt, Winglee, \& B\"ohm (2003) have shown that a weak axial magnetic field ($B<<0.1$ G) associated with the disk,  may collimate the winds at a distance of a few AU.

\item[\bf Collimation of the jet.] In the propeller regime,
the jet  component is self-collimated by the magnetic hoop-stress. The level of collimation
increases towards the axis.
The poloidal velocity in the jet also increases towards the axis,
 and varies between $v_p\approx 2$ near the axis and $v_p\approx 0.2$
near the conical wind.
That is why we choose a few typical velocity levels $v_p = v_c$, with $v_c = 0.5, 1$ and $1.5$,
 and plot lines of equal velocity (\fig{jw-v-3}).
In application to protostars, the fast component of the jet, $v_p\gtrsim 200$ km/s,
carries $\sim 2\%$ of the mass and $\sim 22\%/0.6\approx 37\%$ of the angular momentum
flux out of the star.
At the lower velocity limit, $v_p\gtrsim 100$ km/s,
these numbers are $10\%$ and $35\%/0.6\approx 60\%$.
\end{description}

It is of interest to know the dependence of the mass outflow rate on the
poloidal velocity $v_p$.
We calculate the matter flux $\dot M (v_p > v_c)$  through the external boundary  $r=R_{out}$
at poloidal velocities above a
certain value $v_c$ for different values of $v_c$.
Panel {\it a} shows that the jet component with  $v_p > 0.5$ carries
about $10\%$ of the total outflowing mass $\dot M_w$, while the very fast components, $v_p > 1.5$ and $v_p > 1$
carry only about $1\%$ and $2\%$ correspondingly. So, about $10\%$ of the total mass flows into the collimated jet.
The fractions of matter flowing into different parts of the jet and into the conical
wind are shown in the left panel.

The jet carries angular momentum out of the star along different field lines corresponding to different $v_p$.
It is of interest to know which part of the jet carries most of the angular momentum.
We calculate the angular momentum flux  carried by the magnetic field $\dot L_f (v_p > v_c)$ (this component dominates in the jet) through the external
boundary and normalize it to the total magnetic flux through this boundary,  $\dot L_{\rm wf}$.
Panel {\it b} shows that the high-velocity part of the jet, $v_p > 1.5$, carries about $13\%$ of the total angular momentum flux,
while the entire inner part of the jet in the velocity intervals $v_p>1$ and $v_p>0.5$ carry $22\%$ and
$35\%$ of the flux correspondingly.
Another fraction ($12\%$) flows out along stellar field lines threading the conical wind component
and the low-velocity area above it. All this flux is
responsible for spinning down the star (which is about $60\%$ of the total flux). The rest of the flux
($40\%$) flows along the disk field lines threading the conical winds and the disk.
We conclude that the jet component above the conical wind carries a relatively small
mass but has a significant contribution to the angular momentum outflow from the star.
Note that only about half the star's angular momentum flows into the jet. The other
 half is associated with star-disk interaction through the field lines which are closed inside
 the simulation region and were not taken into account in this analysis (see U06 for details).
In application to protostars, the fast component of the jet, $v_p\gtrsim 200$ km/s,
carries $\sim 2\%$ of the mass and $\sim 22\%/0.6\approx 37\%$ of the angular momentum
flux out of the star.
At the lower velocity limit, $v_p\gtrsim 100$ km/s,
these numbers are $10\%$ and $35\%/0.6\approx 60\%$.

\section{3D Simulations of Conical Winds}

We did exploratory simulations of conical winds in global 3D
simulations.
We chose a case where the dipole magnetic  field of the star is misaligned
with the rotation axis (of the star and disk) by an angle $\Theta=30^\circ$.
One  question is what  the
direction of the conical wind is in the case of an inclined dipole.
We used the Godunov-type 3D MHD ``cubed sphere'' code
developed by Koldoba et al. (2002). In the past we have used this code to study magnetospheric
accretion close to the star (Romanova et al. 2003, 2004b).
Compared with that work we decreased the
density in the corona by a factor of
$10$ to $\rho_c=0.001$ and created conditions suitable for bunching of the
field lines.
We used a grid resolution of $N_r\times N^2= 120\times 51^2$ in each of 6 blocks of
the sphere.
    We took the density in the
disk to be $\rho_d=2$ which is $5$ times lower than in the  axisymmetric
case shown above.
      At the same time, we chose a smaller magnetic moment for the star,
$\mu=2$ compared to $\mu=10$ in the axisymmetric case (to reduce
the computing time).
We start the disk flow not from large
distances but  from $r=5$, to limit the  computing time.
The bunching of field lines is achieved by having a sufficiently high
viscosity, $\alpha_v=0.3$. We do not have diffusivity in the 3D
code, but at the grid resolution we use, the estimated numerical diffusivity at the
disk-magnetosphere boundary is at the level of $\alpha_d\sim 0.01-0.02$,
and hence the main condition for conical wind formation, Pr$_m \gtrsim 1$, is satisfied
(see also {\it Appendix C}).

\begin{figure}
\centering
\includegraphics[width=3.4in]{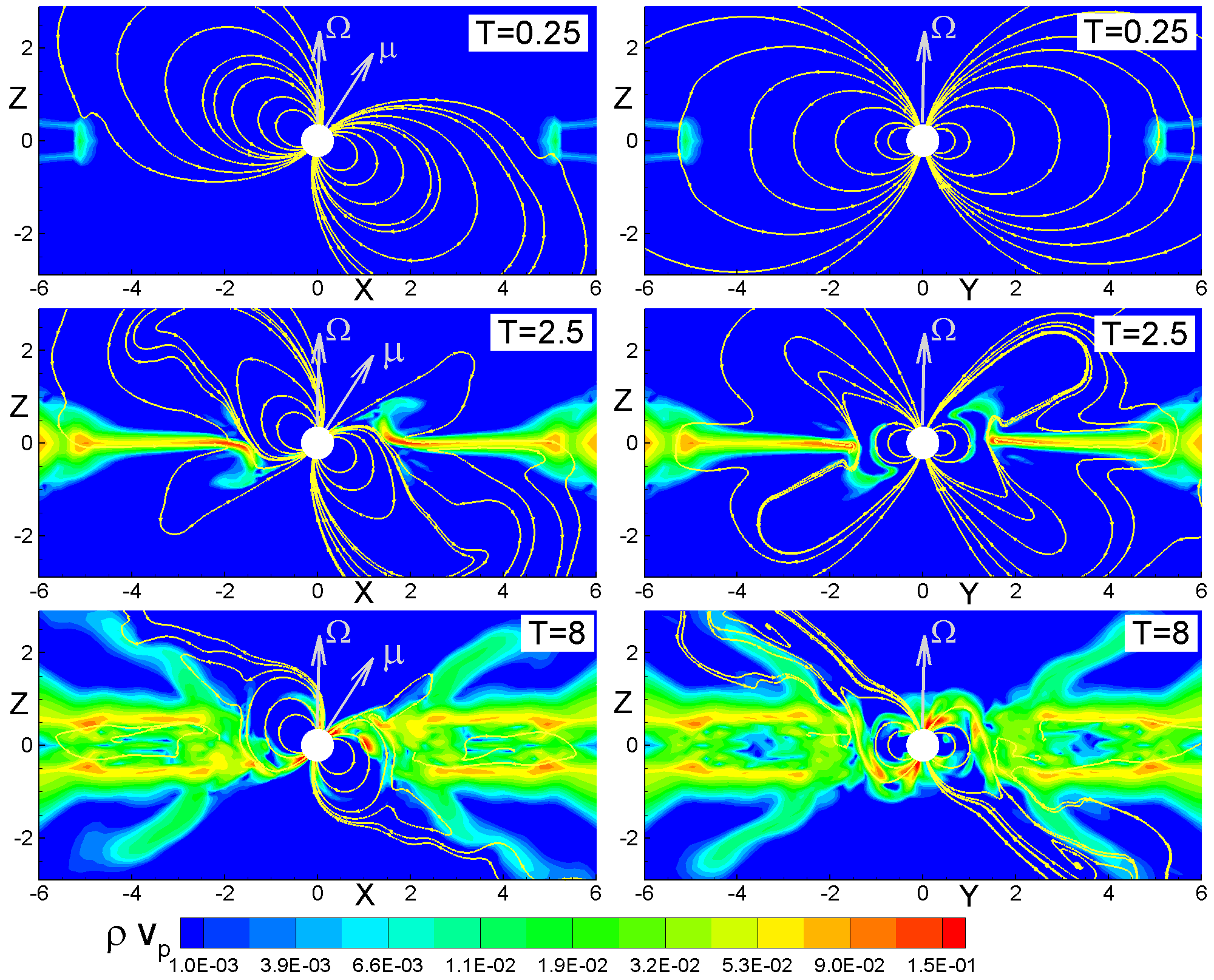}
\caption{Projections show formation of conical winds obtained in 3D
MHD simulations. {\it Left panels} show the distribution of
matter flux (background) and sample magnetic field lines in the XZ plane at different
moments of time. {\it Right panels} show corresponding YZ slices. Arrows
show the direction of the magnetic moment of the star $\mu$ and
angular velocity of rotation $\Omega_*$.}\label{3d-6}
\end{figure}

Simulations show that the accreting matter bunches up field lines and
some matter flows out as a conical wind.  \fig{3d-6} shows that
the wind is geometrically symmetric about the rotation
axis. However, the density distribution in the wind shows a spiral structure
which rotates with the angular velocity of the star, $\Omega_*$, and
represents a one-armed spiral wave from each side of the outflow.

Note that for high $\alpha_v$ ($0.3$ in this case),  the disk-magnetosphere boundary
may exhibit the magnetic interchange instability
(Romanova, Kulkarni \& Lovelace 2008; Kulkarni \& Romanova 2008).
     In these simulations
we do observe some accretion due to this instability in addition to
the main funnel stream accretion that dominates at
high misalignment angles, $\Theta \gtrsim 30^\circ$ (Kulkarni \& Romanova 2009).
However, the conical wind originates at larger radii compared with
 the inner disk radius where accretion through instability dominates.
      We believe  that both processes can ``peacefully" co-exist for
 $\Theta \gtrsim 30^\circ$.
   However, in other situations the conical wind
may be influenced by the interchange instability.
    For example, we did not try to investigate
outflows at small $\Theta$ where accretion through
instability often dominates.
Accretion through instability opens up a new path for penetration of matter
through the magnetosphere, and thus may possibly decrease the
bunching of field lines and consequently the strength of conical winds.
This interrelation between instabilities and conical
winds  needs to be investigated  in  future 3D simulations.
Longer simulations should be performed, and accretion to
 rapidly rotating stars should also be examined.

\section{Comparison with the X-wind model}

Winds from the disk-magnetosphere boundary
have been proposed earlier by Shu and collaborators and referred to as
X-winds (e.g., Shu et al. 1994).
In this model, X-winds originate from a small region near the corotation radius $r_{cor}$,
while the disk truncation radius $r_t$ (or, the magnetospheric radius $r_m$) is only slightly smaller than
$r_{cor}$ ($r_m\approx 0.7 r_{cor}$, Shu et al. 1994).
   It is suggested that excess angular
momentum flows  from the star to the disk and from there into the X-winds.
  The model aims to explain the slow rotation of the star and the formation of jets.
In the simulations discussed here we have  obtained outflows from both
slowly and rapidly rotating stars.
Both have conical wind components which are reminiscent of  X-winds.
What, then, is the difference between X-winds, conical winds and propeller-driven winds?

 In some respects conical/propeller winds {\bf are similar} to
X-winds: {\bf (1)} They both require {\it bunching} of the poloidal field lines and show outflows from the inner disk; {\bf (2)} They both have high rotation and show gradual poloidal acceleration (e.g., Najita \& Shu 1994).~

{\bf The differences} are the following:
{\bf (1)} The conical/propeller outflows have {\it two components}: a slow high-density  conical wind (which can be considered as an analogue of the X-wind), and a fast low-density jet.
No jet component is discussed in the X-wind model.
{\bf (2)} Conical winds form around stars with {\it any rotation rate} including very slowly rotating stars. They do not require fine tuning of the corotation and truncation radii.
For example, bunching  of field lines is often expected during periods of enhanced or unstable accretion when the disk comes closer to the surface of
the star and  $r_m<<r_{cor}$. Under this condition conical winds will form. In contrast, X-winds require $r_m\approx r_{cor}$.~ {\bf (3)}
The base of the conical wind component in both slowly and rapidly rotating stars is associated with the region where the field lines are bunched up, and not with the corotation radius.
{\bf (4)} X-winds are driven by the {\it centrifugal force} (Blandford \& Payne 1982), and as a result matter flows over a wide range of directions below the ``dead zone"   (Shu et al. 1994; Ostriker \& Shu 1995).
In conical winds the matter is driven by the {\it magnetic force} (Lovelace et al. 1991) which acts such that
the matter flows into a {\it thin shell} with a cone angle $\theta\sim 30^\circ$. The same force acts to partially collimate the flow.
{\bf (5)} In the X-wind model it is suggested that angular momentum flows from the star to the disk in spite of the fact that the truncation radius of the disk is located at $r_m\approx 0.7 r_{cor}$ and the disk rotates faster than the star (Shu et al. 1994). Simulations show that
if the funnel stream starts at $r_m < r_{cor}$, then angular momentum flows from the disk to the star along magnetic field lines of the funnel stream
which form a leading spiral, and the star spins up (R02, Romanova et al. 2003; Bessolaz et al. 2008).
The star may transfer its angular momentum to the disk if $r_m>r_{cor}$, like
in the propeller case considered above.
~{\bf (6)}  The X-wind regime is somewhat similar to the propeller regime,
where the
star transfers part of its angular momentum to the disk, and this excess angular momentum may
flow into the conical component of the wind. However, in the propeller regime,
 angular momentum also flows from the star into the jet.
%{\bf (7)} Here we considered only the strong propeller regime ($r_t>>r_{cor}$), or slowly rotating stars ($r_t<<r_{cor}$).
%In reality most of stars are expected to be in the rotational equilibrium state, $r_t\approx r_{cor}$, where variation of the %accretion rate will lead to a alternation between periods of weak spinning-up and spinning-down with zero torque on average
%(R02; Long et al. 2005). Simulations of conical winds in this state (which is closest to one discussed in the X-winds model) yet %should be done. ~
{\bf (7)} Conical and propeller-driven winds are {\it non-stationary}: the magnetic field constantly inflates and
reconnects. X-winds, on the other hand, are steady. This
difference, however, is not significant, and  models can be compared
using time-averaged characteristics.

\section{Application to different stars}

\subsection{Application to Young Stars}

Our simulation results can be applied to different types of young stars, including low-mass
protostars (class I YSOs) which
often show powerful outflows,  CTTSs (class II YSOs) which show less powerful outflows, EXors which show periods of
strongly enhanced accretion and outflows,  and  young brown dwarfs.

\subsubsection{Low-mass protostars (class I YSOs)}

Class I protostars are young stars  which are usually embedded inside a cloud of gas and dust. IR observations
show that protostars are surrounded by cold massive disks and that the
accretion rate is usually an order of magnitude larger than in CTTSs, that is, $\dot M\sim (10^{-6}-10^{-7}) M_\odot/$yr  (e.g., Nisini et al. 2005). The outflows are also more powerful than in CTTSs.
The stars are fully convective, and so rapid generation of a magnetic field that may even be larger than in
CTTSs is expected.

\begin{figure*}
\centering
\includegraphics[width=5in]{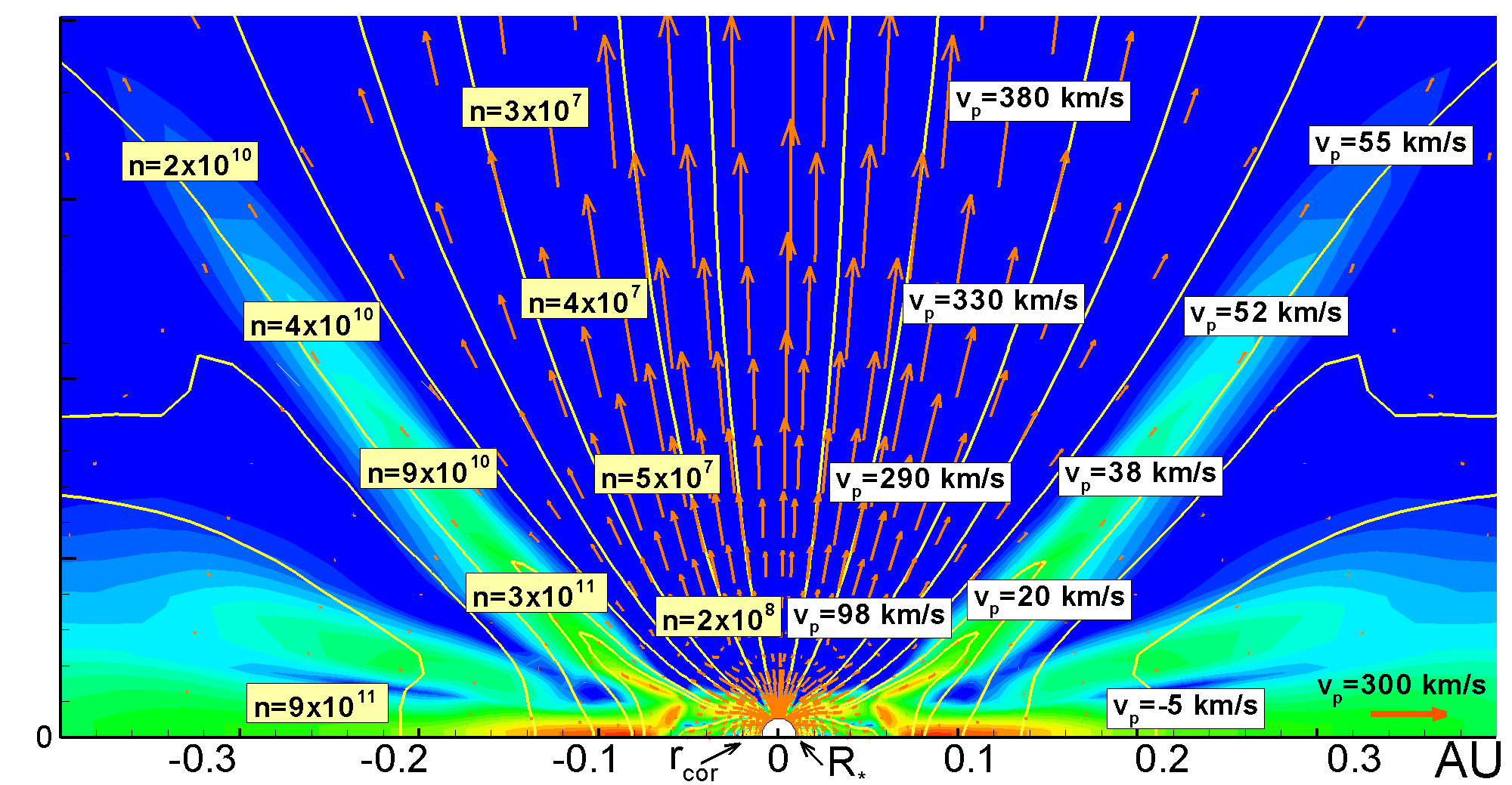}
\caption{A dimensional example of matter flow in the protostar regime shown in \fig{prop-numb}.
Time $T=1400$ corresponds to 3.8 years. Labels show the particles density $n$ in units of
$1/cm^3$ and the poloidal
velocity $v_p$. Azimuthal velocity is a few times larger that the poloidal velocity
in the beginning of the flow, but decreases at larger distances (see \fig{prop-numb}).}
\label{protostar}
\end{figure*}

We consider a protostar of mass $M_0=M=0.8 M_\odot$, radius $R_*=2 R_\odot$, and surface
magnetic field  $B_*=3\times 10^3~{\rm G}$.
The dimensionless radius of the star (the inner boundary) is $0.5$, and the unit radius $R_0=2 R_*=2.8\times 10^{11}~{\rm cm}$.
The velocity scale is $v_0=(GM/R_0)^{1/2}= 195$ km/s,  the time-scale is $t_0=R_0/v_0=0.16$ days,
and the period of rotation at $R_0=1$ is $P_0=1.04$ days. We take a rapidly rotating star with period
$P_*=1.04$ days (the corotation radius of $r_{cor}=1$).  The other reference variables are shown in Table 1.
For dimensionless temperatures in the disk and corona of
 $\tilde T_d=5\times 10^{-4}$ and $\tilde T_c=0.5$, we obtain corresponding initial dimensional
 temperatures: $T_d=2290$ K and $T_c=2.3\times 10^6$ K.
\fig{protostar} shows the distribution of density and velocity
around the protostar.
The age of protostars is $10^5-10^6$ years, and therefore
they may rotate more rapidly than CTTSs and it is likely that some of them are in the propeller regime.
If the propeller is strong enough (like in our simulations, where the period $P_*\approx 1$ day and $\alpha_v=0.3$)
then most of the disk matter will be ejected as slow conical winds with velocity $v_p\sim 50$ km/s,
which may be higher if the disk is closer to the star.
Most of the energy, however, flows into the magnetically-dominated axial jet, where a small fraction
 (about $10\%$) of the disk matter is accelerated up to $v_p\sim 100-400$ km/s inside the simulation region.
 A huge amount of angular momentum flows out of the star
through the same jet, and conical winds carry a comparable amount of angular momentum as well. This may solve the angular momentum problem of the system.
So, at this stage the outflows are powered by two things: the stellar rotational energy and the inner disk
winds (the conical winds).
\fig{flux-p-6}{\it b} shows that the outflow is strongly non-stationary with strong matter
ejection into jets/winds every $2-3$ months.
Ejection is accompanied by  larger than average matter flux and velocities, and hence formation
of new blobs or shock waves is expected.

A protostar in the propeller regime loses its angular momentum to an axial jet.
  From the right-hand panels of \fig{flux-p-6}, we obtain  the dimensionless value of the angular momentum
  loss: $\tilde{\dot L}_{sw}\approx 3$, which corresponds to a dimensional
  value of $\dot L_{sw}=\tilde{\dot L}_{sw} \dot L_0\approx 9.3\times 10^{37} ~{\rm g cm^2/s^2}$.
    The star's angular
velocity is $\Omega_*=2\pi/P_*\approx 7\times 10^{-5} {\rm s}^{-1}$,
its angular momentum is $J= k M r^2  \Omega_* =
 2.2\times  10^{51} k ~{\rm g cm}^2/{\rm s}$, where $k<1$.
   Taking $k=0.4$, the spin-down
time-scale is $\tau = J/{\dot L}_{sw} \approx 3\times 10^5$ years.
Note that this time-scale is calculated for  $B_*=3\times 10^3
{\rm G}$. The time-scale decreases with the magnetic
field of the star as $\sim B_*^{-1.1}$ (see U06) and will be $\tau \approx
3\times 10^6$ years for $B_*=10^3 {\rm G}$. If the magnetic field is weaker, then the
protostar will continue to spin rapidly even in the CTTSs stage.
U06 present the dependence of the spin-down time-scale on the magnetic field, the spin of the star and other parameters.

\subsubsection{Classical T Tauri Stars (class II YSOs)}

CTTSs and their jets have been extensively studied in recent years.
High-resolution observations of CTTSs show that the outflows often have an
``onion-skin" structure, with better-collimated, higher-velocity outflows in the axial
region, and less-collimated, lower-velocity outflows at a larger distance from the axis (Bacciotti et al. 2000).
In other observations, high angular resolution
[FeII] $\lambda~ 1.644\mu$m emission line maps taken along the jets of
DG Tau, HL Tau and RW Aurigae reveal two components: a
high-velocity well-collimated extended component with
velocity $v\sim 200-400$ km/s, and a low-velocity, $v\sim 100$ km/s, uncollimated
component closer to the star (Pyo et al. 2003, 2006).
High-resolution observations of molecular hydrogen in HL Tau have shown that at small distances from the star, the flow shows a conical structure with outflow velocity $\sim 50-80$ km/s (Takami et al. 2007).
In XZ Tau, two-component outflows are observed: one component is a powerful but low-velocity conical wind with an opening angle of about $1$ radian, and the other is a fast well-collimated axial jet (e.g., Krist et al. 2008).
The origin of these outflows is not known, but we can suggest that at least the lower-velocity component may be explained by the conical winds suggesting that the condition for bunching,
Pr$_m>1$, is satisfied. If a CTTS rotates rapidly (in the propeller regime) then the jet component may originate from the propeller effect.

\begin{figure}
\centering
\includegraphics[width=3.4in]{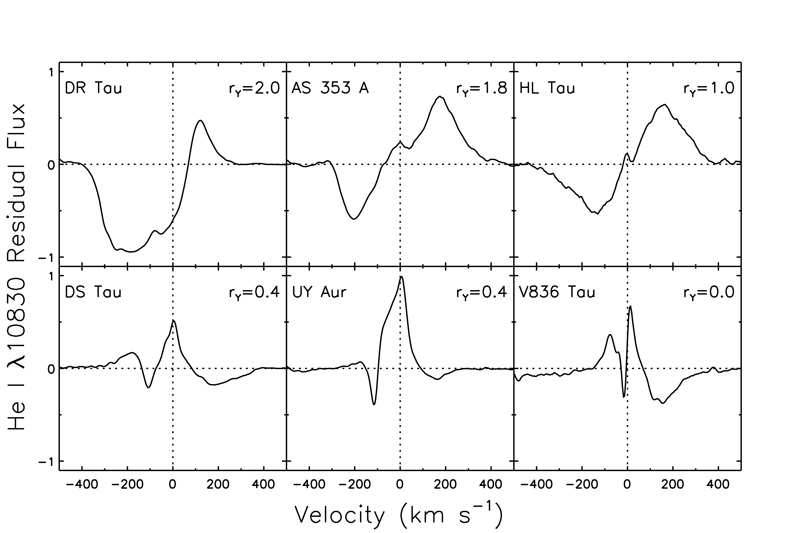}
\caption{Examples of HE I $\lambda$10830 residual profiles and corresponding
1$\mu$ veiling, $r_\gamma$, for 6 accreting CTTSs. The upper row shows high veiling
objects with P Cygni profiles characterized by deep and broad blue absorption which is a sign of the
high-velocity outflow. The lower row shows low veiling objects with narrow blue absorption which is
a sign of a low-velocity outflow. The latter also show red absorption from magnetospheric infall
(from Edwards 2009).}\label{heline}
\end{figure}

\begin{figure}
\centering
\includegraphics[width=3.4in]{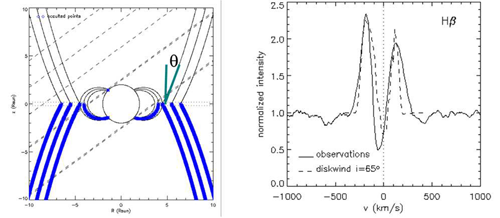}
\caption{Modeling of the  $H_\beta$ line in RW Aurigae led to the
conclusion that a cone-shaped wind with half-opening angle
$30-40^\circ$ and a narrow annulus gives the best match to the
observations of this line (from Alencar et al. 2005).}\label{rw_aur}
\end{figure}

Spectral observations of the He I (10830A) line show clear evidence of two-component outflows (Edwards et al. 2003, 2006; Kwan et al. 2007). Observations show (see \fig{heline}) that at smaller
accretion rates only the relatively low-velocity component, $v\lesssim 100$ km/s, appears. At higher accretion rates there is evidence of
a very fast component with
$v\sim$ 200-400 km/s  which requires an outflow rate of up to $\dot M_w \sim 0.1 \dot M_d$
(Edwards et al. 2006; Edwards 2009).
 We suggest that the low-velocity component may be a conical wind.
However, it is not clear what can explain the high-velocity component. Even if a star is in the propeller regime, then
at the required velocities $v>200$ km/s, only (2-3)\% of the disk matter flows into the jet.
Possibly, additional matter influx from the wind of the star may  enhance the matter flux into the jet component.
In another example, observations of the $H_\beta$ spectral line in RW Aurigae, and comparison
of possible outflow geometries led to the conclusion that a thin cone-shaped wind with a
 half-opening angle of $30^\circ-40^\circ$  gives the best fit
to the observations (from Alencar et al. 2005; see \fig{rw_aur}).

The strongest outbursts supplying CTTS jets are usually episodic or quasi-periodic
(e.g., Ray et al. 2007). For example, blobs are ejected every few months in HH30 (XZ Tau),
and every 5 years in DG Tau
(Pyo et al. 2003).  Both of these may be connected with episodes of enhanced accretion
and formation of conical winds. The velocity and density in the outflow are larger during periods of enhanced accretion,
because the disk comes closer. If the CTTS is in a binary system, then the accretion rate may be episodically enhanced due to interaction
with the secondary star, and this may explain the longer intervals of a few years) between outbursts observed in other CTTSs.  Events of fast, implosive accretion are also possible due to thermal or global magnetic instabilities (e.g., Lovelace et al. 1994).
Alternatively, a  period of a few months may be connected with long-term episodes
of oscillations of the magnetosphere. In the propeller regime the time-interval between oscillations is 1-2 months
even for mild parameters. Bouvier et al. (2007) have shown that
magnetospheric expansion in the CTTS AA Tau may occur with a period of a few weeks.
Multi-year observations of variability in CTTSs show that they are strongly variable on different time-scales (e.g., Herbst et al. 2004; Grankin et al. 2007) which is probably connected with periods of enhanced accretion.

For CTTSs we suggest the same parameters as for protostars but take
a weaker magnetic field, $B=10^3$ G, so that for the same dimensionless runs we obtain lower accretion rates (see Table 1).
Taking from  \fig{flux-x-4}{\it c} the dimensionless values of the matter flux onto the star, $\tilde{\dot M_s}\approx 3.8$,
and into the conical winds, $\tilde {\dot M_w}\approx 1.3$, and taking the value of
$\dot M_0$  Table 1, we obtain an accretion rate onto the star of
$\dot M_s=\tilde{\dot M_s} \dot M_0\approx 7.6\times 10^{-8}~{\rm M_\odot/yr}$ and
into the wind of $\dot M_w\approx 2.6\times 10^{-8}~{\rm M_\odot/yr}$.
For a corotation radius of $r_{cor}=3$, the period of the CTTS is $P_*=$ 5.6 days.
In typical simulation run the truncation radius $r_m\approx 1.2$ is much smaller than the corotation radius $r_{cor}=3$.  This situation corresponds to
the case of enhanced accretion when the star spins up, and which corresponds to ejection of conical winds.

Many CTTSs are expected to be in the rotational equilibrium state, when $r_m\approx r_{cor}$. Without
the bunching condition, and at small viscosity and diffusivity parameters, no significant outflows had been observed in simulations
(R02; Long et al. 2005).  On the other hand, if the bunching condition is satisfied and/or the
accretion rate is enhanced, then conical winds are expected. It is possible that the jet component is also powerful
enough in this state so as to produce the fast
jet component that is observed. Additional simulations are needed for better understanding of outflows in this important state.

\subsubsection{Periods of enhanced accretion and outflows in EXors}

\begin{figure}
\centering
\includegraphics[width=3.4in]{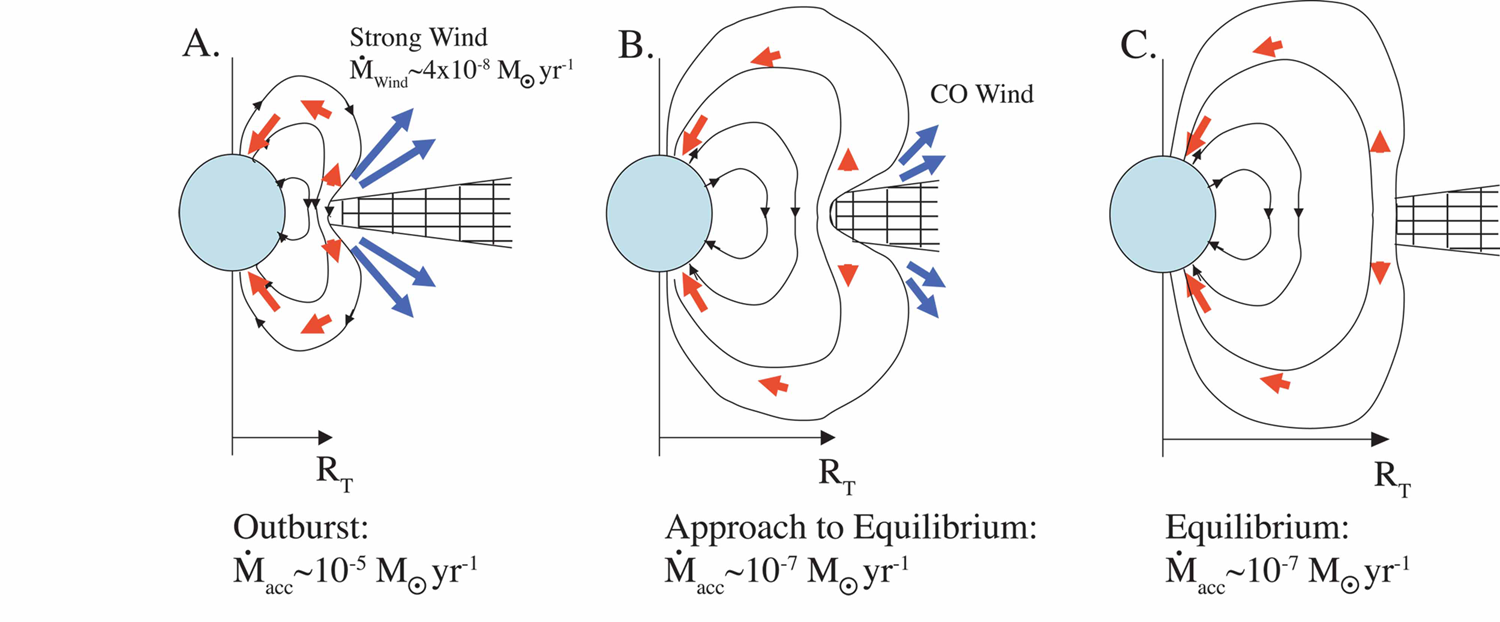}
\caption{Schematic model of an Exor V1647 Ori. During the outburst
the accretion rate is enhanced, and the magnetospheric radius
$r_m$ decreases and the magnetic field lines are bunched up (A). This
results in a fast, hot outflow. As the accretion rate decreases, the
disk moves outward and this results in a slower, cooler CO outflow
(B). Further decrease in the accretion rate leads to a quiescent
state where the production of warm outflows stops (C). From Brittain
et al. (2007)}\label{brittain}
\end{figure}

EXors represent an interesting stage of evolution of young stars where the accretion rate is strongly enhanced and powerful outflows are observed  (e.g., Coffey, Downes \& Ray 2004; Lorenzetti et al. 2006; Brittain et al. 2007).
 Brittain et al. (2007) reported on the
outflow of warm gas  from the inner disk around EXor V1647, observed
in the blue absorption of the CO line during the decline of the EXor
activity. They concluded that this outflow is a continuation of
activity associated with early enhanced accretion and bunching of the
magnetic field lines (see \fig{brittain}).
The EXor stage may correspond to the initial stage of our simulations, during which a
 significant amount of matter comes into the region. Or, it is more probable that initially there is
  weak outflow
at the level of that in CTTSs, but later the accretion rate increases by a few
orders of magnitude, leading to a powerful outburst which produces conical winds.
For conversion into dimensional values, we suggest that the disk comes close to the
stellar surface, which is at $r=1$ (as opposed to 0.5 in the previous examples),
and the disk stops much closer to the star ($r_m=1.2 R_*$). Then all velocities are higher by a factor of
$\sqrt{2}\approx 1.4$, densities by a factor of $32$, and
matter fluxes by a factor of $11$
than in the main example relevant to CTTSs.

\subsubsection{Outflows from Brown Dwarfs}

Recently outflows were discovered from a few
brown dwarfs (BDs) (e.g., Mohanty, Jayawardhana \& Basri 2005;
Whelan, Ray \& Bacciotti 2009).
Clear signs of  CTTS-like  magnetospheric accretion (broad spectral
lines with full-widths of $v > 200$ km/s) were reported earlier for a number of
young BDs (e.g., Natta et al. 2004).
BDs are fully or partially convective and the generation of a
strong magnetic field is expected (Chabrier et al. 2007).
Magnetic fields of the order of $0.1-3$ kG may explain the observed
properties of magnetospheric accretion (Reiners, Basri \& Christensen 2009).
Recently, radio pulses were discovered from the L dwarf binary 2MASSW J0746425+200032
with period $P\approx 124$ minutes, which point to a magnetic field of
$B\approx 1.7$ kG (Berger et al. 2009).
The accretion rates in young BDs are smaller than in CTTS:
$\dot M = 10^{-11}-10^{-9}M_\odot/$yr,  and are often strongly variable.
For example, in 2MASSW J1207334-393254, $\dot{M}$ varied by a factor of $5-10$
during a $6-$ week period (Scholz, Jayawardhana \& Brandeker 2005).
We suggest that outflows may form in BDs during periods of enhanced accretion or
in the propeller regime if the BD is rapidly rotating.

As an example we consider a BD with mass $M_{BD}=60 M_J=0.056 M_\odot$, radius $R_{BD}=0.1 R_\odot$,
and surface magnetic field $B_{BD}=2$ kG and obtain the reference parameters shown in the Table 1. The period of the star is another independent parameter. Here we suggest $r_{cor}=2$ which corresponds to $P_*=0.13$ days, which is a typical period  for a BD.
We also suggest that in \fig{con-numb} the star's radius is $r=0.5$, that is, the disk is truncated at $r_m=1.2/0.5=2.4 R_*$.
For these parameters we obtain an accretion rate of $\dot M_{BD}\approx 1.8\times 10^{-10}{\rm M_\odot/yr}$.
For a smaller magnetic field, $B=1$kG, the  same truncation radius will correspond to a smaller accretion rate $\dot M_{BD}\approx 4.6\times 10^{-11}{\rm M_\odot/yr}$.
The reference velocity $v_0=210$ km/s is not different from the CTTSs case, and therefore the
poloidal velocity of matter in the conical wind is $v_p\lesssim (40-60)$ km/s.
The higher-velocity component of the outflow, $v_p\sim 200$ km/s, can be easily explained if the BD is in the propeller regime.
It is also possible that in the rotational equilibrium state the jet component is strong enough to drive jets.

\subsection{Application to Compact Stars}

\subsubsection{Symbiotic stars --- white-dwarf hosting binaries}

Outflows are observed in some white-dwarf hosting systems. One class of them is
the  symbiotic stars (SSs). SSs  are binary stars in which a white dwarf orbits a red giant star and captures material from the wind of the red giant. Collimated outflows have been observed from more than $10$ (out of $\sim 200$)  symbiotic binaries.
Most of them are transient and appear during or after an optical outburst that indicates an enhanced accretion rate (Sokoloski 2003).
If SSs have a magnetic field then enhanced accretion may drive conical-type outflows from
the disk-magnetosphere boundary during periods of enhanced accretion.  The possibility of a
magnetic field $B \approx 6\times 10^6$G in the SS Z And
is discussed by Sokoloski \& Bildsten (1999) where flickering with a definite frequency was observed.
In other SSs the magnetic field has not been estimated, but present observations do not rule it out
  (Sokoloski 2003). The flickering in many SSs
does not show a definite period, but the presence  of a weak magnetic field is not excluded
(Sokoloski, Bildsten \& Ho 2000). For a typical SSs accretion rate of
$\dot M \approx 10^{-8} {\rm M_{\odot}/yr}$, a magnetic field as small
as $B \sim 3\times 10^4$ G will be dynamically important for the disk-star interaction.
Thus it is possible that outflows are launched from the vicinity of the SS as accretion-driven conical winds.
Collimation may be connected with a disk wind, disk magnetic flux and/or the
 interstellar medium as discussed in \S 5.3.

\subsubsection{Circinus X-1 - the neutron-star hosting binary}

Circinus X-1 represents one of a few cases where a jet is seen from the vicinity
of an accreting neutron star.
   The system is unusual because
Type I X-ray  bursts as well as twin-peak X-ray QPOs are observed.
    The neutron star is estimated to have a
weak magnetic field (Boutloukos  et al. 2006).
    The binary system  has a high
eccentricity ($e\sim 0.4-0.9$)  and thus has periods of low and high
accretion rates (e.g., Murdin 1980).
Two-component outflows are observed.
Radio observations  show a
non-stationary jet with a small opening angle on both arcminute and arcsecond scales.
     At the same time  spectroscopic
observations in the optical (Jonker et al. 2007) and X-ray bands (Iaria
et al. 2008) show that outflows have a
conical structure with a half-opening angle of about $30^\circ$.
    Different explanations are possible for this conical structure,
such as precession of a jet (Iaria et al. 2008).
  However, this appears less likely
because the axis of the jet has not
changed in the last $10$ years (Tudose et al. 2008).
   This neutron star may be a good candidate for conical winds, because (1)
it has episodes of very low and very high accretion rates, and (2) a neutron star
has only a weak magnetic field which can be strongly compressed by the disk, favouring
 the formation of conical
winds. Table 1 shows possible parameters for neutron stars.
Episodic collimated radio jets are also observed from the neutron-star hosting system Sco X-1 (Fomalont, Geldzahler, \& Bradshaw 2001).

\subsubsection{Application to black-hole hosting systems}

Jets and winds are observed from accreting black holes (BHs)
including both
stellar-mass BHs and BHs in galactic nuclei.
The correlation between enhanced accretion rate and outflows has been
discussed extensively, and observational data are in favor of this correlation (e.g. Livio 1997).
Recently, a conical-shaped ionized outflow was discovered in the black-hole hosting X-Ray Binary LMC X-1 (Cooke et al. 2008).
It is not known what determines its shape, but the formation of conical winds is a possibility.
 Magnetic flux accumulation in the inner disk around the black hole
was discussed by Lovelace et al. (1994) and Meier (2005) and  observed in numerical simulations
(Igumenshchev, Narayan, \& Abramowicz 2003; Igumenshchev 2008).
Implosive accretion and outflows from black-hole hosting systems were analyzed by Lovelace et al.
(1994) where angular momentum flows from the disk into a magnetic disk wind, leading to a global magnetic instability and strongly enhanced accretion.
An accretion disk around a black hole may have an ordered magnetic field or loops threading the disk and corona. Fast accretion may lead to bunching of all field lines and possibly to conical winds.
The inward advection of a large scale weak magnetic field threading a
turbulent disk is strongly enhanced because the surface layers of the disk
are non-turbulent and highly conducting (Bisnovatyi-Kogan \& Lovelace 2007; Rothstein \& Lovelace 2008).
The mechanism of conical winds probably does not require a special magnetic field configuration (such as a dipole).
Mohanty and Shu (2008) have shown that the X-wind model works when the star has a complex magnetic field configuration  (see also Donati et al. 2006; Long, Romanova \& Lovelace 2007, 2008).

\section{Conclusions}

We have obtained long-lasting outflows of  cold disk matter into a hot low-density corona from the disk-magnetosphere boundary in cases of slowly and rapidly rotating stars.  The main results are the following:

\smallskip

\noindent{\bf Slowly rotating stars (not in the propeller regime):}
\noindent{\bf 1.}~ A new type of outflow --- a conical wind --- has been found and studied in our simulations.
Matter flows out forming a {\it conical wind} which has the shape of a thin conical shell with a half-opening angle $\theta \sim 30^\circ$.
The outflows appear in cases where the magnetic flux of the star is bunched up by the disk into an X-type configuration.
We find that this occurs when the turbulent magnetic Prandtl number
(the ratio of viscosity to diffusivity) Pr$_m > 1$, and when the viscosity is sufficiently high, $\alpha_v\gtrsim 0.03$.
In earlier simulations of funnel accretion (e.g., R02; Romanova et al. 2003; Long et al. 2005)
both viscosity and diffusivity were small and of the same order, and bunching of the
magnetic field did not occur.
\smallskip

\noindent {\bf 2.} The matter in the conical winds rotates with Keplerian velocity $v_K$ at the base of the wind and continues to rotate  higher up.
It gradually accelerates to poloidal velocities of $v_p \sim 0.5 v_K$.
The conical wind is driven by the magnetic force which acts upwards and towards the axis.
This is responsible for the small opening angle of the cone, the narrow shell shape of the flow, and
the gradual collimation of conical wind towards the axis inside the simulation region.

\smallskip
\noindent{\bf 3.} Conical winds form around stars with different, including very low, rotation rates.
   The amount of matter flowing into the conical wind depends on a number of parameters,
   but in many cases it is $\dot M_w\sim (10-30)\% \dot M_d$.
    It increases with the rotation rate of the star
and reaches almost $100\%$ in the propeller regime.
   For rapidly rotating stars the outflows become strongly non-stationary. The period
    between outbursts increases with the spin of the star.

\smallskip

\noindent{\bf 4.}  There is another
component of the outflow: a low-density, high-velocity component of gas flowing along the stellar field lines.
 The volume occupied by this component increases with the rotation rate of the star.
It occupies the entire region interior to the conical wind in the propeller regime.

\smallskip
\noindent{\bf 5.} A major part of the disk matter accretes to the star through the funnel flow and spins the star up.
 The conical winds carry away part of the disk angular momentum, but most of it is transported
radially outward by the viscous stress.

\smallskip
\noindent{\bf 6.} Conical winds can be further collimated at larger distances by the pressure of the surrounding medium
(Lovelace et al. 1991), disk winds (e.g. Fendt 2009), or by the magnetic flux threading the disk at large
distances (e.g., Matt et al. 2003).

\smallskip
\noindent{\bf 7.} Conical winds may appear during strong enhancements of accretion, as in EXors or
symbiotic variables.
At the same time our simulations indicate that relatively steady outflows can exist for a long time
($2$ years in application to young stars) if the conditions for the magnetic field bunching are maintained.

\smallskip

\noindent{\bf 8.} Exploratory 3D simulations of conical winds from accreting stars with
a significantly misaligned dipole field show that the conical winds are approximately symmetric about the rotational axis of the star (and the disk).

\medskip

\noindent {\bf  Propeller-driven outflows} appear around rapidly rotating stars for
conditions where
$r_m > r_{cor}$ and where the condition for bunching, Pr$_m>1$, is satisfied.
Their properties are the following:

\smallskip
\noindent{\bf 1.} Two distinct outflow components are found in the propeller regime: (1) a relatively low-velocity conical wind and (2) a high-velocity axial jet.
\smallskip

\noindent{\bf 2.} A significant part of the disk matter and  angular momentum flows into the conical winds.
   At the same time a significant part of the rotational energy of the star flows
   into the magnetically-dominated axial jet.
  Formation of powerful jets is expected.
 This regime is particularly relevant to protostars, where the star rotates rapidly and has a high
 accretion rate.

\smallskip
\noindent{\bf 3.} The star spins down rapidly due to the angular momentum flow into the axial jet
along the field lines connecting the star and the corona. For typical parameters a protostar
spins down in $3\times10^5$ years. The axial jet is powered by the spin-down of the star
rather than by disk accretion.

\smallskip
\noindent{\bf 4.} The matter fluxes into both components (wind and jet) strongly oscillate due to events of inflation and reconnection. Most powerful outbursts occur every $1-2$ months. The interval between outbursts is expected to be longer for smaller diffusivities in the disk.
Outbursts are accompanied by higher outflow velocities and stronger self-collimation of both components. Such outbursts may
explain the ejection of knots in some CTTSs every few months.
Enhanced accretion due to external factors will also lead to formation of a new blob/knot in the jet.

\smallskip

The values of the transport coefficients $\alpha_v$ and $\alpha_d$ in realistic accretion disks
remain uncertain, but it is widely thought that they are due to Magneto-Rotational
Instability (MRI)-driven turbulence (Balbus \& Hawley 1998).
MRI simulations suggest that the turbulence
may give values of $\alpha$ in the range: $\alpha_v = 10^{-2} -
0.4$ (e.g., Stone et al. 2000).

If in actual accretion disks the transport coefficients {\it are large}, $\alpha\sim 0.1$, then strong outflows
are expected during periods in which $\alpha_v \gtrsim \alpha_d$, or when the accretion rate is enhanced.
  The condition for magnetic field bunching is that the magnetic Prandtl number of the turbulence
Pr$=\alpha_v/\alpha_d >1$.  The effective Prandtl
number may be significantly increased owing to the
highly conducting surface layer of the disk
(Bisnovatyi-Kogan \& Lovelace 2007; Rothstein \& Lovelace 2008).
The field lines may be bunched up not only due to high viscosity, but also by many other
processes, e.g., due to thermal or magnetic instabilities, or due to the Rossby wave mechanism (Lovelace et al. 1999; Li et al. 2000).
   If the transport coefficients are {\it very small}, say
$\alpha_v\approx \alpha_d=0.01$, then quasi-stationary accretion through a funnel flow (with no outflows) is expected (R02; Long et al. 2005).
If the star rotates rapidly, it will be in the weak propeller regime in which a star spins down,
but no outflows are produced, Romanova et al. 2004a).
If at some point the accretion rate is enhanced due to one or another mechanism, then
conical winds will form in spite of the diffusivity being small.
%In case of slowly rotating star a new episode of enhanced accretion will lead to the bunching of %the field lines and to the formation of conical winds which will last as long as accretion rate %stays high. In the rapidly rotating stars enhanced accretion will lead to powerful outflows.

\section*{Acknowledgments}

The authors thank S. Edwards, S. Cabrit, E. Dougatos, J. Ferreira, W. Herbst, A. Kulkarni, S. Matt and F. Shu
for helpful discussions and the referee for multiple comments and questions which helped
improve the paper.
The authors were supported in part by NASA grant NNX08AH25G and by
NSF grants AST-0607135 and AST-0807129. MMR thanks NASA for use of
the NASA High Performance Computing Facilities. AVK and GVU were supported in
part by grant RFBR 09-02-00502a, Program 4 of RAS.

\appendix

\section{Viscosity and diffusivity}

Here we discuss the treatment of viscosity and diffusivity in the axisymmetric code in greater detail.
The stress tensor ${\cal T}_{ik}=T_{ik}+\tau_{ik}$ consists of an ideal part,
$$
{T}_{ik}=\rho v_i v_k +p \delta_{ik} +
\frac{B^2}{8 \pi} \delta_{ik} - \frac{B_iB_k}{4 \pi} ~,
$$
and a viscous part $\tau_{ik}$ which takes into account
small-scale  turbulent velocity and magnetic field fluctuations.

We assume that the stress due to the turbulent fluctuations
can be represented in the same way as the collisional
viscosity by substitution of the turbulent viscosity coefficient into it.
      Moreover, we consider that the viscous stress
is determined mainly by the gradient of the angular velocity because
the azimuthal velocity is the dominant component in the disk.
        The dominant components of the tensor
$\tau_{ik}$ in spherical coordinates are:
$$
\tau_{r \phi} = - \nu_t \rho r \sin \theta \frac{\partial
\Omega}{\partial r}~,\quad
\tau_{\theta \phi} = - \nu_t \rho \sin \theta \frac{\partial
\Omega}{\partial \theta}~.
$$
Here $\Omega = v_{\phi} /r \sin \theta$ is angular velocity of the
plasma and $\nu_t$ is the kinematic turbulent viscosity.

Separating out the viscous stress in the $\phi$ component of  eqn. (\ref{eq2}) gives
$$
\frac{\partial (\rho v_{\phi})}{\partial t} + \frac{1}{r^3}
\frac{\partial( r^3 T_{r \phi})}{\partial r} + \frac{1}{r \sin^2
\theta} \frac{\partial (\sin^2 \theta T_{\theta \phi})}{\partial
\theta} =
$$
\begin{equation}
\label{eq10}
     \frac{1}{r^3} \frac{\partial}{\partial r} \left( \nu_t \rho
r^4 \sin \theta \frac{\partial \Omega}{\partial r} \right) +
\frac{1}{r \sin^2 \theta} \frac{\partial}{\partial \theta} \left(
\nu_t \rho \sin^3 \theta \frac{\partial \Omega}{\partial \theta}
\right) ~,
\end{equation}
where $T_{r \phi}$ and $T_{\theta \phi}$ are components of the
inviscid part of the stress tensor.

      The viscosity leads to dissipation of the
kinetic energy and its conversion into thermal energy and to a
corresponding increase of the entropy.
       In both types of runs (propeller and conical winds)
we have neglected viscous heating. We have also neglected
radiative cooling.  Inclusion of heating and cooling
is a separate and complex physics problem
which is different for different types of stars.
   We suggest that the viscous heating is compensated by radiative cooling.
Thus, the main ``role" of viscous terms is the transport of angular momentum outward
which allows matter to accrete inward  to the disk-magnetosphere boundary.

We also assume that the plasma has a finite magnetic diffusivity.
 That is, the  matter may diffuse across the field lines.
 We assume that the finite diffusivity of the plasma is also
 due to  the small-scale turbulent
fluctuations of the velocity and the magnetic field.
      The induction equation averaged over
the small-scale fluctuations has the form
\begin{equation}
\label{eq11}
\frac{\partial {\bf B}}{\partial t} - {\bf
\nabla}\times ({\bf v}\times{\bf B}) + c {\bf \nabla}\times {\bf
E}^\dagger =0~.
\end{equation}
Here, $\bf v$ and ${\bf B}$ are the averaged velocity and magnetic
fields, and ${\bf E}^\dagger=- \left< {\bf v}'\times{\bf B}'
\right>/c$  is electromotive force connected with the fluctuating
fields.
      Because the turbulent electromotive
force ${\bf E}^\dagger$ is connected with the small-scale
fluctuations, it is reasonable to suppose that it has a simple
relation to the ordered magnetic field $\bf B$.
       If we neglect the magnetic dynamo
$\alpha$-effect (Moffat 1978), then $\left< {\bf v}'\times{\bf B}'
\right> = -\eta_t {\bf \nabla}\times {\bf B}$, where $\eta_t$ is the
coefficient of  turbulent magnetic diffusivity. Equation
(\ref{eq11}) now takes the form
\begin{equation}
\label{eq12} \frac{\partial {\bf B}}{\partial t} - {\bf
\nabla}\times ({\bf v} \times {\bf B}) + {\bf \nabla} \times\left(
\eta_t {\bf \nabla}\times {\bf B} \right) =0~.
\end{equation}
We should note that the term for ${\bf E}^\dagger$ formally
coincides with Ohm's law
$$
{\bf J} = \frac{c}{4\pi} {\bf \nabla}\times {\bf B} =
\frac{c^2}{4\pi \eta_t} {\bf E}^\dagger~.
$$
The coefficient of turbulent electric conductivity $\sigma = c^2
/4\pi \eta_t$. The rate of dissipation of magnetic energy per unit
volume is
$$
\frac{{\bf J}^2}{\sigma} = \frac{\eta_t}{4 \pi} ({\bf \nabla}\times
{\bf B})^2~.
$$

To calculate the evolution of the poloidal magnetic field it is
useful to calculate the $\phi$-component of the vector-potential
$\bf A$.
       Owing to the assumed axisymmetry,
\begin{equation}
\label{eq13}
B_r = \frac{1}{r \sin \theta} \frac{\partial (\sin \theta
A_{\phi})}{\partial \theta}~,\quad\quad B_{\theta} = - \frac{1}{r}
\frac{\partial (r A_{\phi})}{\partial r}~.
\end{equation}
Substituting  ${\bf B} = {\bf \nabla}\times {\bf A}$ into the
induction equation gives the equation for the $\phi$ component of the vector-potential
\begin{equation}
\label{eq14}
\frac{\partial A_{\phi}}{\partial t} - \eta_t \left( \frac{1}{r}
\frac{\partial^2 (r A_{\phi})}{\partial r^2} + \frac{1}{r^2}
\frac{\partial}{\partial \theta} \frac{1}{\sin \theta}
\frac{\partial (\sin \theta A_{\phi})}{\partial \theta} \right) =
[{\bf v}\times{\bf B}]_{\phi}~.
\end{equation}
The azimuthal component of the induction equation gives
$$
\frac{\partial B_{\phi}}{\partial t} - \frac{1}{r}
\frac{\partial}{\partial r} \left( \eta_t \frac{\partial (r
B_{\phi})}{\partial r} \right) - \frac{1}{r^2}
\frac{\partial}{\partial \theta} \left( \frac{\eta_t}{\sin \theta}
\frac{\partial (\sin \theta B_{\phi})}{\partial \theta} \right)
$$
\begin{equation}
\label{eq15} =\frac{1}{r} \left( \frac{\partial [r ({\bf
v}\times{\bf B})_{\theta}]}{\partial r} - \frac{\partial ({\bf
v}\times{\bf B})_r}{\partial \theta} \right)~.
\end{equation}
The Joule heating rate per unit volume is
\begin{equation}
\label{eq16}
\frac{\eta_t}{4\pi} ({\bf \nabla} \times{\bf
B})^2=\frac{\eta_t}{4\pi r^2} \times
\end{equation}
$$  \left[  \left( \frac{\partial( r B_{\theta})}{\partial r} -
\frac{\partial B_r}{\partial \theta} \right)^2 +
\left(\frac{\partial (r B_{\phi})}{\partial r} \right)^2 +
\frac{1}{\sin^2 \theta} \left( \frac{\partial( \sin \theta
B_{\phi})}{\partial \theta} \right)^2 \right]. $$
We included the Joule heating only in the propeller case runs, for completeness.
However, we observed that although this term led to some heating, it was not the reason for the production
the outflows in the propeller regime. Test simulations with no heating led to similar outflows, because
the main driving forces are magneto-centrifugal forces. In the conical wind simulations
we did not include Joule heating. Therefore we suggest that both Joule and viscous
heating are exactly compensated by the radiative cooling.

\section{Numerical Method}

For numerical integration of the MHD equations including the
magnetic diffusivity and viscosity in the disk,
we used a method of
splitting of the different physical processes.
      Our simulation algorithm
has a number of blocks: (1) an ``ideal MHD" block in which we
calculate the dynamics of the plasma and magnetic field with
dissipative processes switched off; blocks (2) and (3) for the
diffusion of the poloidal and azimuthal components of the magnetic
field calculated for frozen values of the plasma velocity and
thermodynamic parameters (density and pressure); and block (4) for
the calculation of viscous dissipation in which we took into
account only the $r \phi$ and $\theta \phi$ components of the viscous
stress tensor.

{\bf (1)} In the hydrodynamic block, the ideal MHD equations are
integrated numerically using an explicit conservative Godunov-type
numerical scheme. In our numerical code the dynamical variables
are determined in the cells, while the vector-potential of the
magnetic field, $A_\phi$, is determined on the corners.
       For calculation of fluxes between the
cells we use an approximate solution of the Riemann problem
analogous to the one described by Brio \& Wu (1988).

For better spatial resolution, the restricted antidiffusion terms
based on the MINMOD limiter are added to the fluxes (Kulikovskii,
Pogorelov \& Semenov 2001). The spacial splitting has not been
performed. Integration of the equations with time is performed
with a two-step
  Runge-Kutta method.
To guarantee the absence of magnetic charge, we calculate at each
time-step the $\phi$-component of the vector-potential $A_
{\phi}$, which is then used to obtain the poloidal components of
the magnetic field ($B_r$, $B_\theta$) in a divergence-free form
(Toth 2000). In other words, the condition $\div B=0$ is satisfied
with machine accuracy.

{\bf (2)} In the block where {\it the diffusion of the poloidal magnetic
field} is calculated, we numerically
integrate equation (\ref{eq14}) for
the $\phi$-component of vector-potential.
During this calculation we freeze the
values of  $A_\phi$ on the inner and outer boundaries of the
simulation region. In the equatorial plane we have the symmetry conditions
 $\partial A_\phi /\partial \theta=0$.
On the symmetry axis we have $A_\phi=0$.
Equation (\ref{eq14}) is approximated
by an implicit difference scheme.
The approximation is chosen so that the
operator on the implicit time-layer is symmetric and positive.
For solving the system of equations
on the implicit time-layer, we used ICCG (Incomplete
Cholesky Conjugate Gradient) method (Kershaw  1978).
Because the  size of the grid cells and
the coefficient of  magnetic diffusivity vary
strongly in space, the elements of the matrix of the system also
vary strongly. To remove this undesirable property,
we changed the matrix so that it has diagonal elements equal to unity.

{\bf (3)} In the block where {\it the
diffusion of the azimuthal component of the magnetic field} is
calculated, we numerically integrate equation (\ref{eq15}).
At the inner and outer boundaries,
$B_\phi$ is frozen in this computational block.
Along the rotation axis
and on the equatorial plane $B_\phi=0$.
Equation (\ref{eq15}) was approximated
by a numerical scheme with a
symmetric positive operator on the implicit time layer.
The corresponding system of linear equations is solved by the
ICCG method.

{\bf (4)} In the block where {\it the viscous stress} is calculated,  we
 integrate equation (\ref{eq10}) numerically for the angular velocity of matter $\Omega =
v_\phi / r \sin \theta$. At the inner boundary of the simulation
region we take $\Omega=\Omega_*$, the angular rotation
velocity of the star.
     At the outer boundary, $\Omega$ is taken to be fixed
and equal to the corresponding Keplerian value.
       On the axis and in the
equatorial plane we have the condition of zero stress for the
$\theta \phi$ -component of the viscous stress tensor.
Equation (\ref{eq10}) is approximated
by a numerical scheme with a symmetric  positive operator
on the implicit time-layer.
The corresponding system of linear equations is solved by the
ICCG method.

The code has passed all the  standard tests.
   In addition it has been  used
for the solution of a number of important
astrophysical MHD flow problems (e.g., Ustyugova et al. 1999; R02; R05; U06).
   In one problem,   stationary super-fast-magnetosonic MHD
outflows were  obtained for the first time with
the disk treated  as a boundary condition (Ustyugova et al. 1999).
   In that work all of the flux function integrals of motion were
calculated from the simulations and found to be constant as required
by the theory (e.g., Lovelace et al. 1986).   Additionally,
the cross-field force balance was checked numerically.
     Subsequently,  similar results were obtained by Krasnopolsky et al. (1999)
using the ZEUS code.
This test of the simulations against the  axisymmetric MHD theory
is an important test in that it involves
all three components of the flow velocity and the magnetic field.
Simulation results of accretion to a star with a dipole field (R02; Long et al. 2005)
were recently confirmed by Bessolaz et al. (2008).

\section{Tests of the Code}

We performed multiple tests of the code with different grids. We
plan to describe the whole set of tests in a separate paper.
However, in Appendix B we show two tests relevant to the ideal and
diffusive blocks of the code.
 Below we show two
examples of such tests.
 In the first test we checked the ideal MHD block of the code
(with viscosity and diffusivity switched off).
In the second test we checked the diffusion block of the code separately.

\begin{figure*}
\centering
\includegraphics[width=5in]{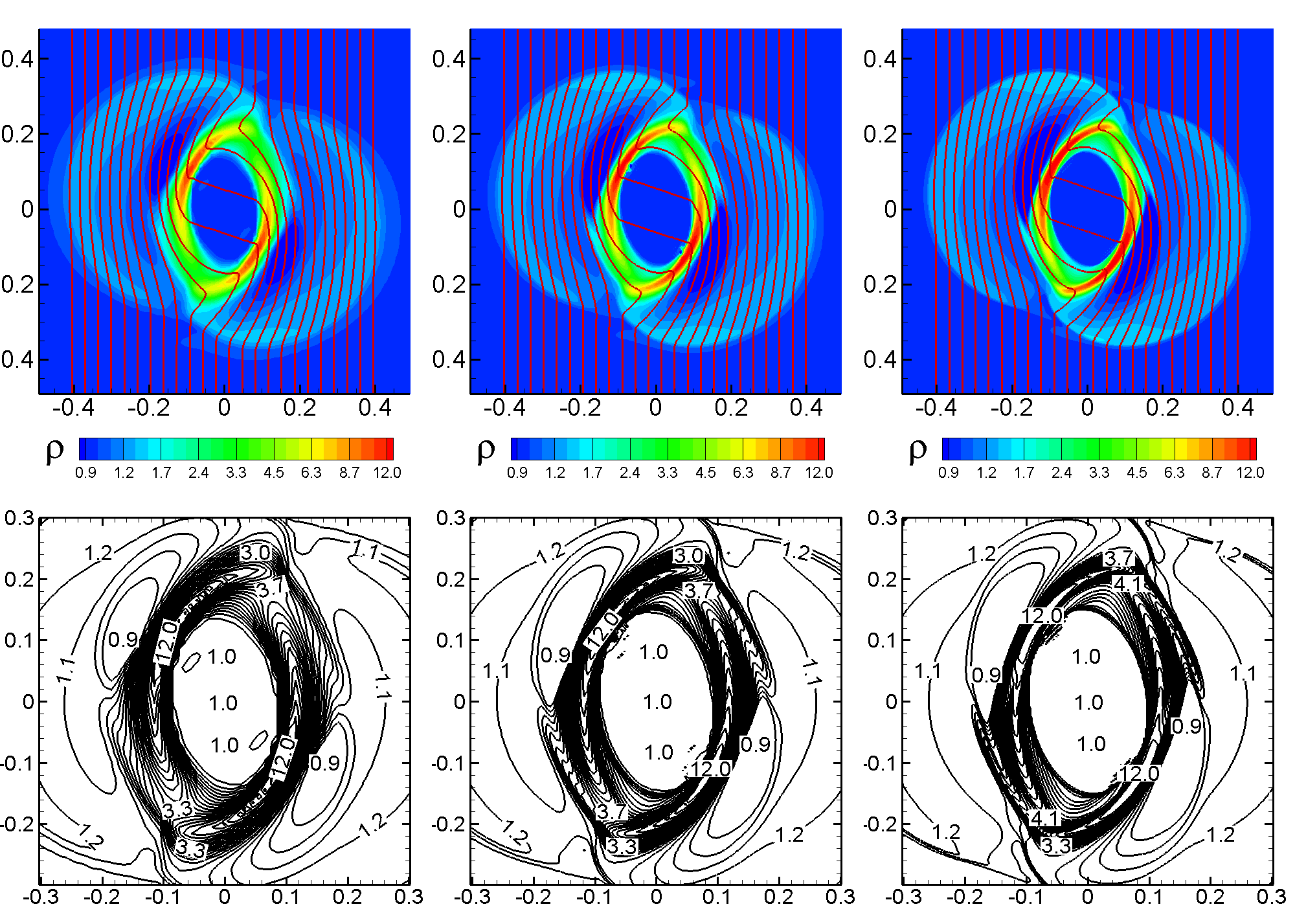}
\caption{Test of the ideal MHD block of the code with the ``rotor problem" test at different grid resolutions.
{\it Top Panels:}
Density distribution (color background) and field lines (solid lines)
with grid resolution $100\times 100$ (left panel), $200\times 200$ (middle panel)
and $400\times 400$ (right panel). {\it Bottom Panels:} The density contour
lines.}\label{ideal-6}
\end{figure*}

\subsection{Test of the ideal MHD block}

To check the MHD block of the code we performed the standard ``rotor problem"
test.
 This test has been used by a number of authors for testing  MHD solvers
including the
energy equation (e.g., Balsara \& Spicer 1999).
 We use this test to
check our ideal MHD block of the code (viscosity and diffusivity are switched off) with an isentropic
equation $dS/dt=0$ instead of the energy equation.
 In this situation the shock waves
do not have physical sense.
   However, the goal of this test
is not to test the physics of shock waves, but instead to demonstrate the ability of the
numerical algorithm to solve the 2D adiabatic MHD equations.

We solve the MHD equations numerically in the region $-0.5<x<0.5$, $-0.5<y<0.5$ in
a Cartesian geometry.
At the beginning of simulations,
$t=0$, the pressure in the region is constant, $p=1$, and the magnetic field is homogeneous,
$B_x=0$, $B_y=5$.
   In the center there is a circle with radius $r_0=0.1$
(radius $r=\sqrt{x^2+y^2}$) where the density of matter is $\rho_0=10$
and the matter rotates as a solid body with angular velocity $\omega_0=20$.     At $r>r_1=0.115$,
the density is $\rho_1=1$ and the matter is at rest.
  In the ring $r_0<r<r_1$, the density and velocity
are linearly interpolated between those at $r=r_0$ and $r=r_1$.

The equations of ideal adiabatic MHD were solved
with the Godunov-type scheme used in the ``main" code. In the test we used a homogeneous
grid with step $\Delta x = \Delta y = 1/N$, where $N=100, 200, 400$. The time-step
is chosen from the condition $\Delta t=0.4 \Delta x/v_{max}$, where $v_{max}$ is the maximum velocity
of propagation of the perturbations.
The results of the simulations are shown in  \fig{ideal-6}.
One can see that the density and the field line distribution is very similar in all three cases,
while simulations at the highest grid resolutions give almost identical results and the convergence
of the results is evident.
The bottom panels of   \fig{ideal-6} show selected streamlines with numbers
which confirm the similarity and convergence of the results.
The test had been performed on a
homogeneous grid, while  our simulations were done in spherical coordinates with a high grid
resolution near the star, and much coarser resolution at the outer boundary of the simulation region.
 In this paper we mainly investigate the launching of jets and winds from the disk, and therefore
we need to have adequate grid resolution
in the region where matter is launched from the disk into the winds, that is,
at radii $r\approx 2-3$ for the conical winds, and
$r\approx 3-5$ for the propeller-driven winds.
  The grid resolution at these radii
corresponds approximately to the homogeneous grid with the lowest grid resolution,
$100\times100$.
   The above test shows that
the grid resolution in this region is sufficiently good for investigation of the physics of outflows in this region.
It is clear that the grid resolution at  larger distances, and in
particular close to the outer boundary is not very high. However, this region
does not influence  the physics of the process.
  In this region matter flows either inward
due to the viscosity (in the disk), or outward as winds (in the corona).
   We expect
larger simulation errors in these regions, but these do not change the main result: the launching of
winds from the disk-magnetosphere boundary.
   The magnetic force accelerating matter into
the jet component has higher accuracy close to the star and less accuracy
at  larger distances. However, Fig. 11d shows that the main acceleration into the wind occurs close to the star where the grid resolution is good.

We are planning future simulations with
higher resolutions and in  larger regions with the recently MPI-parallelized version of our code.

\begin{figure*}
\centering
\includegraphics[width=6in]{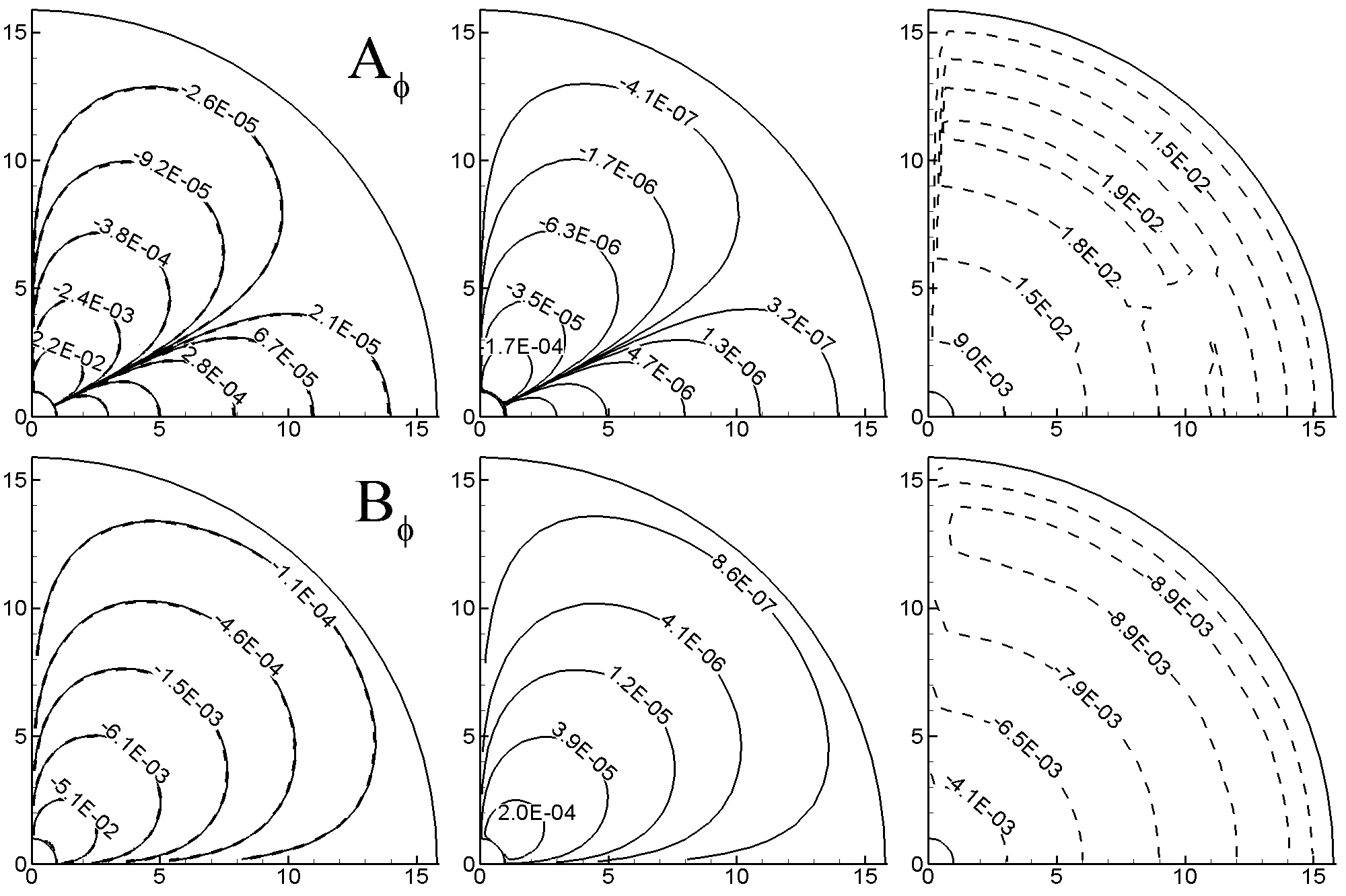}
\caption{Test of the diffusivity block of the code on the grid $51\times31$.
The top panels show analysis of the vector-potential, $A_\phi$.
The left hand panel shows the simulated value $A_\phi^{num}$ (solid lines) and the exact solution $A_\phi^{exact}$
(dashed lines). The middle panel shows the absolute error $\Delta A_\phi=A_\phi^{num}-A_\phi^{exact}$, and the right
hand panel shows the relative error, $\Delta A_\phi/A_\phi^{exact}$.
The bottom panels show similar analysis but for the azimuthal component of the magnetic field, $B_\phi$.}
\label{dif-6}
\end{figure*}

\begin{table}
\centering
\begin{tabular}{l@{\extracolsep{0.2em}}l@{}lll}
\hline
&   Grids                     & $\max|\Delta A_\phi|$     & $\max|\Delta B_\phi|$  \\
\hline
\multicolumn{2}{l}{$51\times 31$}     & $1.32\e{-3}$      & $7.11\e{-4}$    \\
\multicolumn{2}{l}{$75\times 45$}     & $6.20\e{-4}$      & $3.24\e{-4}$    \\
\multicolumn{2}{l}{$101\times 61$}    & $3.40\e{-4}$      & $1.75\e{-4}$     \\
\hline
\end{tabular}
\caption{The maximum absolute error in calculations of the $A_\phi$ and $B_\phi$
obtained in the diffusivity tests at different grid resolutions.} \label{tab:dif}
\end{table}

\subsection{Test of the diffusivity block}

Here we test the diffusivity block of the code.
For this we ``switch off" the hydrodynamic fluxes and the corresponding right-hand side terms
in equations A5 and A6 and integrate these equations numerically in the region  $r_0<r<r_1$,
$0<\theta<\pi/2$. In the case of the constant diffusivity coefficient (we take $\eta=1$)
equations A5 and A6 coincide, and so the problems for the azimuthal
component of the vector-potential and for the azimuthal magnetic field
differ only in the boundary conditions at the equator (at $\theta=\pi/2$) and the
initial conditions. That is,
$$
\frac{\partial{A_\phi}}{\partial{\theta}}{\bigg |}_{\theta=\pi/2}=0 ,~~
B_\phi{\bigg |}_{\theta=\pi/2} = 0.
$$
On the symmetry axis ($\theta=0$) we have:
$$
A_\phi |_{\theta=0} = 0,~~
B_\phi |_{\theta=0} = 0.
$$
We can find some particular solutions of equations (A5) and (A6) by the
separation of variables method. For testing we choose the following solutions:
\begin{equation}
\label{eq17}
A_\phi=\frac{C \sin\theta}{r^2}\bigg(1-\frac{5 \sin2\theta}{4}\bigg )
\bigg(1+\frac{10\eta(t_0-t)}{r^2}\bigg ),
\end{equation}
\begin{equation}
\label{eq18}
B_\phi=\frac{C \sin\theta \cos\theta}{r}\bigg(1+\frac{6\eta (t_0-t)}{r^2}\bigg ),
\end{equation}
where $C$ and $t_0$ are constants.
Equations for the $\phi$-components of the vector-potential
and magnetic field were integrated numerically using our 2.5D code
(with the ``switched-off" hydro fluxes and zero right-hand-side terms).
We used the same grid geometry as in the main simulations; that is, spherical coordinates
with an equidistant grid in the meridional direction and expanding grid in the
radial direction, where the expansion factor is determined by the fact that we keep all
sides of each grid cell to be approximately equal.
For the test we used the grid corresponding to the main simulation runs in the conical wind,
$N_r=51$ and $N_\theta=31$, and also we used finer grids, $75\times 45$ and  $101\times 61$.
We took the inner
and outer radii of the simulation region to be the same as in the conical wind case, $r_0=1$,
$r_1=15.9$.
For these boundaries we set $A_\phi$ and $B_\phi$ to be equal to the
analytical values determined by equations \ref{eq17} and \ref{eq18}. The initial conditions corresponded
to the exact solutions (\ref{eq17} and \ref{eq18}) at $t=0$.

The constants $C$ and $t_0$ in \ref{eq17} and \ref{eq18} were chosen so that the
solutions at the inner boundary are of the order of unity. We used $C=0.01$, $t_0=50$.
We integrated the equations up to  $t_1=2t_0=100$.

\fig{dif-6} (top panels) shows the result of the diffusivity test for the vector-potential, $A_\phi$.
The solid line in the left panel shows the numerical solution, $A_{\phi}^{num}$, which is obtained
by integrating equation A5, at time $t=100$. The
dashed line shows the exact solution, $A_{\phi}^{exact}$, obtained by integrating
the exact solution \ref{eq17}. One can see that the solutions are very close
to each other, and the dashed line is barely visible. The middle panel shows the absolute error
$\Delta A_\phi={A_{\phi}^{num} - A_{\phi}^{exact}}$. One can see
that the error is very small everywhere, taking into account the fact that the maximum value of the function,
$A_\phi\approx 1$. The error increases towards the star. However, this is because the value of the
function $A_\phi$ strongly increases towards the star. The right hand panel shows that the relative error,
$(A_{\phi}^{num} - A_{\phi}^{exact})/A_{\phi}^{exact}$, decreases towards the star. One can see that
the relative error in calculation of the diffusivity is also small and is of the order of $(1-2)\%$.
\fig{dif-6} (bottom panels) shows similar analysis for the $B_\phi$ component, where equations (A6)
and \ref{eq18} have been solved for the numerical and exact solutions.

We performed similar simulations and analysis at finer grid
resolutions, $75\times 45$ and  $101\times 61$. Table 2 shows the
maximum absolute error, $\max{|A_{\phi}^{num} -
A_{\phi}^{exact}|}$. One can see that at the finer grids,
$75\times 45$ and $101\times 61$, this error is $2.1$ and $3.9$
times smaller compared to the coarsest grid, $51\times 31$. This
confirms the convergence of the numerical solution towards the
analytical solution. Similar comparisons for the $B_\phi$
component give factors $2.2$ and $4.1$, which also show
convergence.

A complete description of our methods and tests will be given in a
separate paper.

\section{Parameter ranges of conical winds}

To investigate the dependence on different parameters we took the main
case and varied one parameter at a time.
We performed several sets of runs: (1) with a fixed diffusivity
and different viscosity coefficients; (2) with a fixed viscosity and
different diffusivity coefficients; (3) with different
rotation periods of the star $P_*$;
(4) with different coronal densities.

\subsection{Dependence on viscosity at fixed
diffusivity}

\begin{figure}
\centering
\includegraphics[width=3.4in]{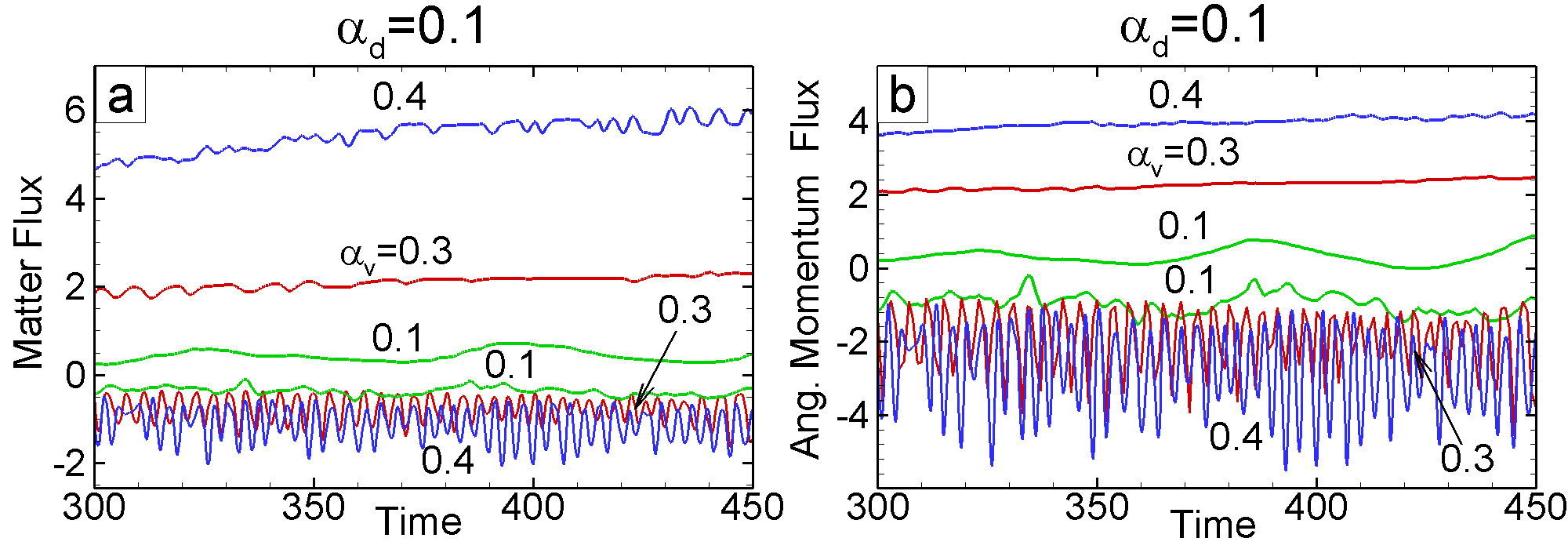}
\caption{Matter and angular momentum fluxes onto the star and into the wind for fixed diffusivity
$\alpha_d=0.1$ but different coefficients of viscosity, $\alpha_v$.}\label{dif01-x-2}
\end{figure}

We fixed the diffusivity at $\alpha_d=0.1$ and varied the viscosity
coefficient in
the range $\alpha_v = 0.01 - 1$. \fig{dif01-x-2} shows
matter fluxes onto the star and into the conical winds through the surface $r=6$.
We observed that for small
viscosity, $\alpha_v < 0.1$, the magnetic field of the dipole
diffuses through the inner regions of the incoming disk and an X-type
configuration does not form.
No conical winds appear in this
case.
We conclude that formation of conical winds requires
$\alpha_v \gtrsim \alpha_d$, that is, Pr$_m \gtrsim 1$.
Next we increased $\alpha_v$ and observed that an X-type
configuration formed and conical winds were generated.
We observed that the accretion rate to the star increases with
$\alpha_v$, while the mass outflow rate into the conical winds
increases but only slowly.
For $\alpha_v=0.1$, the matter fluxes onto the star and into the wind are small
and approximately equal.  For $\alpha_v=0.3$ and  $\alpha_v=0.4$, the wind
carries about $30\%$ and $20\%$ of mass respectively.
The angular momentum carried to the star also strongly increases with
$\alpha_v$. In all cases the star spins up, because the magnetospheric
radius, $r_m\approx 1.2$ is smaller than corotation radius
$r_{cor}=3$.
That is, the incoming matter brings positive angular momentum
onto the star.
The conical wind carries angular momentum away from the
disk.    Notice, however, that this is only a small part of the total angular momentum
of the disk as  shown above.

\begin{figure}
\centering
\includegraphics[width=3.4in]{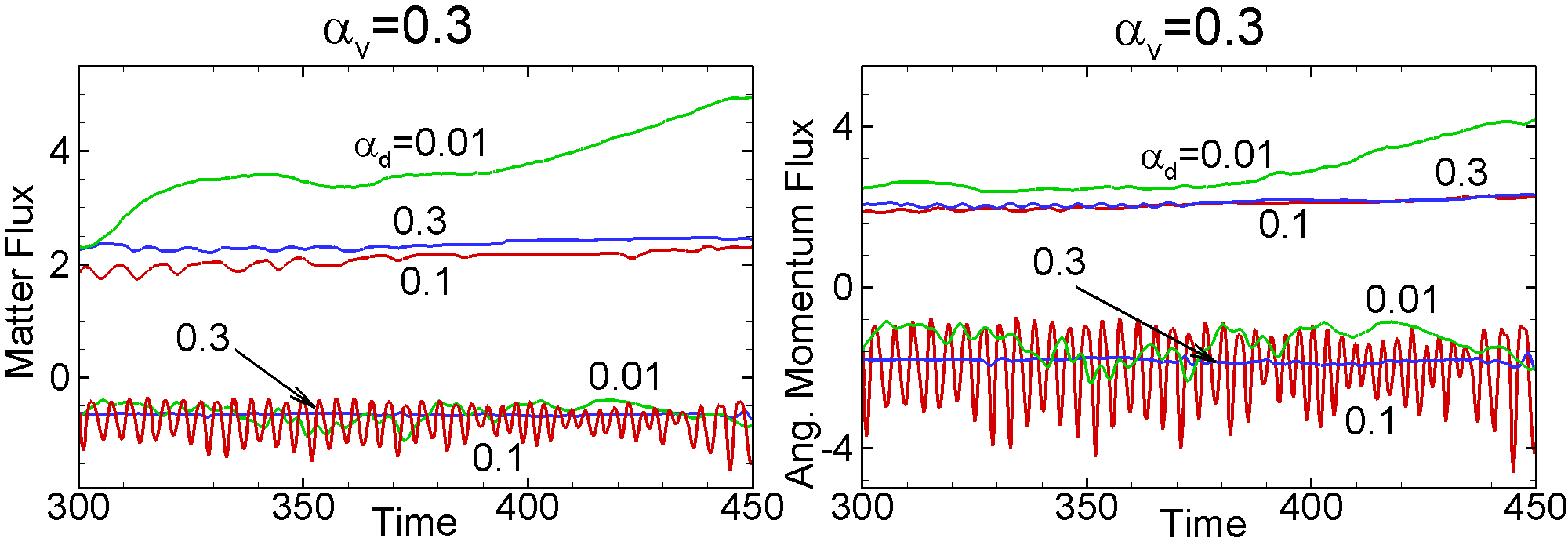}
\caption{Matter and angular momentum fluxes onto the star and into the wind for fixed viscosity
$\alpha_v=0.3$ but different coefficients of diffusivity, $\alpha_d$.}\label{vis03-x-2}
\end{figure}

\subsection{Dependence on diffusivity at fixed viscosity}

In the next set of runs we fixed the viscosity at $\alpha_v=0.3$ and
varied the diffusivity: from 0.01 to 1. \fig{vis03-x-2} shows the integrated
matter fluxes onto the star
and into the wind at different diffusivities. We observed
that no conical winds were
formed for $\alpha_d \gtrsim \alpha_v$.
At relatively high diffusivity,
$\alpha_d=0.1,~ 0.3$, about $30\%$ of the incoming matter flows into the conical wind.
  For $\alpha_d=0.1$ (Pr$_m=3$), the matter flux into the conical wind oscillates,
while at $\alpha_d=0.3$ (Pr$_m=1$) no oscillations are
observed.
The conical wind also forms for very small diffusivity, $\alpha_d=0.01$,
but with slightly smaller matter flux into the
wind.
   Angular momentum fluxes to the star and into the
winds are approximately the same, excluding the case $\alpha_d=0.01$ where the flux to the star is
larger.

\subsection{Variation of star's period}

We varied the {\it period of the star (via the corotation radius)} taking
$r_{cor} = 3, ~10$ for slowly rotating stars and $r_{cor}=1.0,~1.5,~ 2$
for more rapidly rotating stars. In the case of very slow rotation, $r_{cor} = 10$,
conical winds form and the outflow rate into
the wind is similar to that in the main case
($r_{cor}=3$), although the accretion rate onto the star is somewhat larger.
In stars with higher spin, $r_{cor}=2$, the amplitude of variability
increases and the matter flux into the outflows increases up to $50\%$
of the accretion rate to the star.
   For even higher spin,
$r_{cor}=1.5$, the accretion rate to the star decreases  by a
factor of $5$ compared with the main case, while the outflow rate into the conical wind strongly increases.
 Thus when a star rotates more rapidly, the winds become more powerful,
 and the accretion rate to the star decreases, and therefore the situation becomes closer to the propeller regime.
     In rapidly rotating stars, the outbursts become episodic, and the interval between outbursts increases with spin.
In addition, the volume occupied by the  fast coronal component increases with spin: at $r_{cor}=10$ there is no
fast component in the corona, at
$r_{cor}=3$ it occupies some region above the conical winds (see \fig{rotation}{\it a}),
while at  $r_{cor}=2$ it occupies a much larger region (see \fig{rotation}{\it b}). At even higher spin this region occupies the whole simulation region as in the propeller regime.

\subsection{Variation of coronal density}

Outflow of matter into a wind occurs if the corona is not very
dense, and hence outflowing matter of the winds does not lose its
energy while propagating through the corona. In the main simulation
runs the initial density of the corona is $10^{4}$ times lower compared to the
disk density ( $\rho_c=10^{-3}$ versus $\rho_d=10$). To test the
dependence on the coronal density we decreased its density by a factor of
$3$. These simulations showed that the matter fluxes onto the star and into the winds are
not appreciably different from the main case.
We conclude that the coronal
 density used in the main case is sufficiently small not to affect  the outflows.

The situation is different in the coronal region.
For slowly rotating stars (the conical wind regime)
there are no forces which tend to drive matter along the axis.
Fast flow appears only in the part of corona where the stellar field lines are strongly inclined
(see \fig{con-numb}, \fig{rotation}).
The rest of the corona has very slow motion
towards or away from the star.
The top half of the coronal region is matter-dominated
(\fig{ang-x-6}{\it a,d}) and might be an obstacle
to a fast outflow.
We suggest that at lower coronal density
the fast coronal component might occupy a larger region.

In the propeller regime where the star rotates rapidly,
the magnetic force is larger and
drives the low-density coronal matter into the fast jet.
The corona is magnetically-dominated (see \fig{ang-p-6}{\it a,d})
during the whole simulation time.
   Initially, the density in the corona $\rho_c=10^{-4}$ is $10$ times lower than
in case of slowly rotating stars.
 However, the initial density distribution is important
 only at the beginning of the long
simulation runs.
  Later the density distribution is established by the outflow process.
The main process is inflation of the dipole field lines.
   Inflating field lines carry matter along both
disk and stellar field lines, and as a result, some matter penetrates into the corona.
    Strong disk oscillations and violent processes
of inflation during which the conical wind component
often changes its opening angle) lead to
the penetration of a small amount of
matter into the corona.
    The density increases away from the axis, where it is very small, towards the conical wind.
 In the case of a slowly rotating star,  the disk oscillations are weaker
and there are no violent inflation events.
     For this reason the axial region above the star  has a very low density.

\end{document}